\documentclass[12pt]{article}
\usepackage{amssymb,epsf,color}
\def\lsim{\mathrel{\rlap {\raise.5ex\hbox{$ < $}}
{\lower.5ex\hbox{$\sim$}}}}
\def\gsim{\mathrel{\rlap {\raise.5ex\hbox{$ > $}}
{\lower.5ex\hbox{$\sim$}}}} 
\def\sqr#1#2{{\vcenter{\vbox{\hrule height.#2pt

        \hbox{\vrule width.#2pt height#1pt \kern#1pt

           \vrule width.#2pt}

        \hrule height.#2pt}}}}

\def\lsim{{\displaystyle
{{\raise-8pt\hbox{$ <$}}
\atop{\raise5pt\hbox{$\sim$}}}}}
\def\gsim{{\displaystyle
{{\raise-8pt\hbox{$ >$}}
\atop{\raise5pt\hbox{$\sim$}}}}}
\def\slsim{{\displaystyle
{{\raise-8pt\hbox{$\scriptstyle <$}}
\atop{\raise5pt\hbox{$\scriptstyle \sim$}}}}}
\def\sgsim{{\displaystyle
{{\raise-8pt\hbox{$\scriptstyle  >$}}

\atop{\raise5pt\hbox{$\scriptstyle \sim$}}}}}

\newskip\humongous \humongous=0pt plus 1000pt minus 1000pt

\newcommand{\sumpf}[0]{\sum_{(H^{\rm f},G^{\rm f})}\! \! \! \!
{\raise
4pt
\hbox{$'$}}\,}

\newcommand{\sump}[0]{\sum_{(H,G)}\! \! {\raise 4pt \hbox{$'$}}\,}

\def\bs{\begin{subequations}}
\def\es{\end{subequations}}

\catcode`\@=11
\newcount\hour
\newcount\minute
\newtoks\amorpm

\hour=\time\divide\hour by60
\minute=\time{\multiply\hour by60 \global\advance\minute by-\hour}
\edef\standardtime{{\ifnum\hour<12 \global\amorpm={am}%
        \else\global\amorpm={pm}\advance\hour by-12 \fi

        \ifnum\hour=0 \hour=12 \fi
        \number\hour:\ifnum\minute<10 0\fi\number\minute\the\amorpm}}
\edef\militarytime{\number\hour:\ifnum\minute<10 0\fi\number\minute}
\def\draftlabel#1{{\@bsphack\if@filesw {\let\thepage\relax
   \xdef\@gtempa{\write\@auxout{\string
      \newlabel{#1}{{\@currentlabel}{\thepage}}}}}\@gtempa
   \if@nobreak \ifvmode\nobreak\fi\fi\fi\@esphack}
        \gdef\@eqnlabel{#1}}
\def\@eqnlabel{}
\def\@vacuum{}
\def\draftmarginnote#1{\marginpar{\raggedright\scriptsize\tt#1}}
\def\draft{\oddsidemargin -.2truein
        \def\@oddfoot{\sl preliminary draft \hfil
        \rm\thepage\hfil\sl\today\quad\militarytime}
        \let\@evenfoot\@oddfoot \overfullrule 3pt
        \let\label=\draftlabel
        \let\marginnote=\draftmarginnote
   \def\@eqnnum{(\theequation)\rlap{\kern\marginparsep\tt\@eqnlabel}%
\global\let\@eqnlabel\@vacuum}  }
\def\subequations{\refstepcounter{equation}%
  \edef\@savedequation{\the\c@equation}%
  \@stequation=\expandafter{\theequation}
  \edef\@savedtheequation{\the\@stequation}
  \edef\oldtheequation{\theequation}%
  \setcounter{equation}{0}%
  \def\theequation{\oldtheequation\alph{equation}}}

\def\endsubequations{\setcounter{equation}{\@savedequation}%
  \@stequation=\expandafter{\@savedtheequation}%
  \edef\theequation{\the\@stequation}\global\@ignoretrue
  \vspace*{-12pt} \\}

\def\bs{\begin{subequations}}
\def\es{\end{subequations}}
\relax
\def\thefootnote{\fnsymbol{footnote}}
\def\be{\begin{equation}}
\def\ee{\end{equation}}
\def\ba{\begin{eqnarray}}
\def\ea{\end{eqnarray}}

\def\ee{\end{equation}}
\def\bea{\begin{eqnarray}}
\def\eea{\end{eqnarray}}
\def\nn{\nonumber}

\newcommand{\uarrw}[0]{\mathrel{
{\raise.5ex\vbox{\hrule width 1cm}\hskip-6pt\rightarrow}}}
\def\thebibliography#1{%
\vskip 0.5cm \centerline{\bf References}
\list{%
[\arabic{enumi}]}{\settowidth\labelwidth{[#1]}
\leftmargin\labelwidth
\advance\leftmargin\labelsep
\usecounter{enumi}}
\def\newblock{\hskip .11em plus .33em minus .07em}
\sloppy\clubpenalty4000\widowpenalty4000
\sfcode`\.=1000\relax}

\renewcommand{\theequation}{\arabic{section}.\arabic{equation}}

\renewcommand{\section}{\setcounter{equation}{0}\@startsection%
{section}{1}{0mm}{-\baselineskip}{0.5\baselineskip}%
{\normalfont\normalsize\bfseries}}

\renewcommand{\subsection}{\@startsection%
{subsection}{2}{0mm}{-\baselineskip}{0.5\baselineskip}%
{\normalfont\normalsize\slshape}}

\renewcommand{\subsubsection}{\@startsection%
{subsubsection}{2}{0mm}{-\baselineskip}{0.5\baselineskip}%
{\normalfont\normalsize\slshape}}

\begin{document}
%
%
\renewcommand{\theequation}{\arabic{section}.\arabic{equation}}
\begin{titlepage}
\begin{flushright}
\end{flushright}
\begin{centering}
\vspace{1.0in}
\boldmath

{ \large \bf Combinatorics, observables, and String Theory}

\unboldmath
\vspace{1.5 cm}

{\bf Andrea Gregori}$^{\dagger}$ \\
\medskip
\vspace{3.2cm}
{\bf Abstract} \\
\end{centering} 
\vspace{.2in}
We investigate the most general phase space of configurations, consisting 
of all possible ways of assigning elementary attributes, ``energies'',
to elementary positions, ``cells''. We discuss how this space possesses 
structures that can be approximated by a quantum-relativistic physical
scenario. In particular, we discuss how the Heisenberg's 
Uncertainty Principle and a universe with a three-dimensional space 
arise, and what kind of mechanics rules it. String Theory shows up as a 
complete representation of this structure in terms of time-dependent fields 
and particles.
Within this context, owing to the uniqueness of the underlying mathematical 
structure it represents, one can also prove the uniqueness of string theory.

\vspace{6cm}

\hrule width 6.7cm
\noindent
$^{\dagger}$e-mail: agregori@libero.it

\end{titlepage}
\newpage
\setcounter{footnote}{0}
\renewcommand{\thefootnote}{\arabic{footnote}}

\tableofcontents

\vspace{1.5cm}

\noindent

\section{Introduction}
\label{intro}

The search for a unified description of 
quantum mechanics and general relativity, within a theory that should
possibly describe also the evolution of the universe, is one of the long 
standing and debated open problems of modern theoretical physics. 
The hope is that, once such a theory has been found, it will
open us a new perspective from which to approach,
if not really answer, 
the fundamental question behind all that, that is ``why the universe is what
it is''. On the other hand, it is not automatical that, once such a unified
theory has been found, it gives us also more insight on the reasons
why the theory is what it is, namely, \emph{why} it has to be precisely 
that one, and why no other choice could work. But perhaps it is precisely
going first through this question that it is possible to make progress in
trying to solve the starting problem, namely the one of unifying quantum 
mechanics and relativity. Indeed, after all we don't know why do we need
quantum mechanics, and relativity, or, equivalently, why the speed
of light is a universal constant, or why there is the Heisenberg Uncertainty.
We simply know that, in a certain regime, Quantum Mechanics and Relativity
work well in describing physical phenomena.

In this work, we approach the problem from a different perspective.
The question we start with can be formulated as follows: is it possible
that the physical world, as we see it, doesn't proceed from a
``selection'' principle, whatever
this can be, but it is just the collection of all the
possible ``configurations'', intended in the most
general meaning? May the history of the Universe be viewed somehow as a
path through these configurations, and what we call time ordering an
ordering through the inclusion of sets, so that the universe at a
certain time is characterized by its containing as subsets all previous
configurations, whereas configurations which are not 
contained belong to the future of the Universe? What is the
meaning of ``configuration'', and
how are then characterized configurations, in order to say which one is
contained and which not? How do they contribute to make up what we observe?

Let us consider the most general 
possible phase space of ``spaces of codes of information''.
By this we mean products of spaces carrying strings of information
of the type ``1'' or ``0''. If we interpret 
these as occupation numbers for
cells that may bear or not a unit of energy, we can view the set of these
codes as the set of assignments of a map $\Psi$ from a space of unit
energy cells to a discrete target vector space, that can be of any 
dimensionality. 
If we appropriately introduce units of length and energy, we may ask
what is the geometry of any of these spaces. 
Once provided with this interpretation, it is clear that the problem of
classifying all possible information codes can be viewed as a classification
of the possible geometries of space, of any possible dimension.
If we consider the set of all
these spaces, i.e. the set of all maps, $\left\{ \Psi \right\}$,  
that we call the phase space of all maps, we may also ask whether 
some geometries occur more or less often in this phase space. 
In particular, we may ask this question about $\left\{ \Psi (E) \right\}$,
the set of all maps which assign a finite amount of energy
units, $N \equiv E$. The frequency by which these spaces occur
depends on the combinatorics of the energy assignments 
\footnote{In order to unambiguously define these frequencies, 
it is necessary to make a ``regularization''
of the phase space by imposing to work at finite volume. This 
condition can then be relaxed once a regularization-independent
prescription for the computation of observables is introduced.}. 
Indeed, it turns out that not only 
there are configurations which occur more often than other ones, but
that there are no two configurations with the same weight.
If we call $\left\{ \Psi (E) \right\}$ the ``universe'' at ``energy'' $E$,
we can see that we can assign a time ordering in a natural way, because
$\left\{ \Psi (E^{\prime}) \right\}$ ``contains''
$\left\{ \Psi (E) \right\}$ if
$E^{\prime } > E$, in the sense that 
$\forall \Psi \in \left\{ \Psi (E) \right\}$ $\exists \Psi^{\prime}
\in \left\{ \Psi (E^{\prime}) \right\}$ such that
$\Psi \subsetneqq \Psi^{\prime}$. $E$ plays therefore the role of 
a time parameter, that we can call the age of the universe, ${\cal T}$. 
Our fundamental assumption is that, at any time $E$, 
there is no ``selected'' geometry of the universe: the universe as it
appears is given by the superposition of all possible geometries.
Namely, we assume that
the partition function of the universe, i.e. the function through which
all observables are computed, is given by:
\be
{\cal Z} (E) \, = \, \sum_{\Psi(E)} {\rm e}^{S(\Psi(E))} \, ,
\label{zsum1}
\ee 
where $S(\Psi)$ is the entropy of the configuration $\Psi$ in the phase space
$\left\{  \Psi \right\}$, related to the weight of occupation in the
phase space $W(\Psi)$ in the usual way: $S = \log W$. Rather evidently,
the sum is dominated by the configurations of
highest entropy.

The most recurrent geometries 
of this universe turn out to be those corresponding to three dimensions.
Not only, but the very dominant configuration is the one corresponding
to a three-sphere of radius $R$ proportional to $E$. 
That is, a black hole-like 
universe in which the energy density 
is $\sim 1 / E^2 \propto 1 / R^2$~\footnote{The radius of the black hole
is the radius of the three-ball enclosed by the horizon surface. The radius of the three
sphere does not coincide with the radius of the ball; they are anyway proportional
to each other. How,
and in which sense, a sphere can have, like a ball, 
a boundary, which functions
as horizon, is a rather non-trivial fact that we are going to discuss in detail.}.
But the most striking feature is that all the other configurations 
summed up contribute
for a correction to the total energy of the universe of the 
order of $\Delta E \sim 1 / {\cal T}$. 
This is rather reminiscent of the inequality at the base of the
Heisenberg Uncertainty Principle on which quantum mechanics is based on:
${\cal T}$, the age/radius up to the horizon of observation,
can also be written as $\Delta t$, the interval of time during which
the universe of radius $E$ has been produced. That means, the universe is
mostly a classical space, plus a
``smearing'' that quantitatively
corresponds to the Heisenberg Uncertainty, $\Delta E \sim 1 / \Delta t$.
This argument can be refined and applied to any observable one may
define: all what we observe is given by a superposition of
configurations and whatever value of observable quantity we can
measure is smeared around, is given with a certain fuzziness, which
corresponds to the Heisenberg's inequality.
Indeed, a more detailed inspection
of the geometries that arise in this scenario, the way ``energy clusters''
arise, their possible interpretation in terms of matter, particles etc.
allows to conclude that \ref{zsum1} formally implies a quantum
scenario, in which the Heisenberg Uncertainty
receives a \emph{new interpretation}.
The Heisenberg uncertainty relation arises here as a
way of accounting not
simply for our ignorance about the observables, but for the 
ill-definedness of these quantities in themselves: all the observables
that we may refer to a three-dimensional world, together with
the three-dimensional space itself, exist only
as ``large scale'' effects. Beyond a certain degree of accuracy
they can neither be measured nor be defined. The space itself, with
a well defined dimension and geometry, 
cannot be defined beyond a certain degree of accuracy either. 
This is due to the fact that
the universe is not just given by one configuration, the dominant one,
but by the superposition of all possible configurations, an infinite
number, among which many (an infinite number) don't even correspond to a
three dimensional geometry.
The physical reality is the 
superposition all possible configurations, weighted as in \ref{zsum1}.

It is also possible to show that the speed of expansion of the 
geometry of the dominant
configuration of the universe,
i.e. the speed of expansion of the radius of the three-dimensional black hole, 
that by convention and choice of
units we can call ``$c$'', is also the
maximal speed of propagation of \underline{coherent}, i.e.
non-dispersive, information. This can be shown
to correspond to the $v=c$ bound of the speed of light.
Here it is
essential that we are talking of \emph{coherent information}, as tachyonic
configurations also exist and contribute to \ref{zsum1}: their contribution
is collected under the Heisenberg uncertainty.

One may also show that
the geometry of geodesics in this space corresponds to
the one generated by the energy distribution. This means that this
framework ``embeds'' in itself Special and General Relativity.

The dynamics implied by (\ref{zsum1}) is neither deterministic in the
ordinary sense of causal evolution, nor
probabilistic. At any age the universe is the superposition
of all possible configurations, weighted by their ``combinatorial''
entropy in the phase space. According to our definition of time and time
ordering,
at any time the actual superposition of configurations does not depend on
the superposition at a previous time, because the actual and the previous one
trivially are 
the superposition of all the possible configurations at their time. 
Nevertheless,
on the large scale the flow of mean values through time
can be approximated by a smooth evolution that we can, up to a certain extent,
parametrize through evolution equations. As it is not possible
to exactly perform the sum of infinite terms of \ref{zsum1},
and it does not even make sense, because an infinite number of less
entropic configurations don't even correspond to a description of
the world in terms of three dimensions,
it turns out to be convenient to
accept for practical purposes
a certain amount of unpredictability, introduce probability
amplitudes and work in terms of the rules of quantum mechanics. These
appear as precisely tuned to embed the uncertainty that
we formally identified with the Heisenberg Uncertainty
into a viable framework, which allows some
control of the unknown, by endowing the uncertainty with a probabilistic
interpretation.
Within this theoretical framework, we can therefore give an
argument for the \underline{necessity} of a quantum description of
the world: quantization appears to be a useful
way of parametrizing the fact of being the observed reality a
superposition of an infinite number of configurations.
Once endowed with this interpretation,
this scenario provides us with a theoretical framework that
unifies quantum mechanics and relativity in a description that,
basically, is neither of them: in this perspective, they turn out to be only
approximations, valid in a certain limit, of a more comprehensive
formulation.

The scenario implied by \ref{ZPsi} is 
\underline{highly predictive}, in that there is basically no free
parameter, except for the only running quantity, the age of the
universe, in terms of which everything is computed.
Out of the dominant configuration, a three-sphere,
the contribution given by the other configurations to (\ref{zsum1})
is responsible for the introduction of ``inhomogeneities'' in the universe. 
These are what gives rise to a varied spectrum of energy clusters, 
that we interpret as matter and fields evolving and interacting
during a time evolution set by the $E$--time-ordering. All of them fall
within the corrections to the dominant geometry implied by the Heisenberg's 
inequality. For instance, matter clusters constitute
local deviations of the mean energy/curvature of order 
$\Delta E \sim 1 / \Delta t $, where $\Delta t$ is the typical
time extension
(or, appropriately converted through the speed of light, the space extension)
of the cluster, and so on.

In this framework, String Theory arises as a consistent quantum theory
of gravity and interacting fields and particles, 
which constitutes a useful mapping of the
combinatorial problem of ``distribution of energy
along a target space'' into a continuum space.
To this purpose, one must think at string theory as defined in an
always compact, although arbitrarily extended, space.
By ``String Theory'' we mean here the collection of all supersymmetric string
constructions, which are supposed to be particular realizations
of a unique theory 
underlying all the different string constructions.
In this sense, when we speak of 
a string configuration, this has to be intended as a (generally
non-perturbative) configuration of which the possible 
perturbative constructions made in terms of heterotic, type II, type I
string, represent ``slices'', dual aspects of the same object.
Owing to quantization, and therefore to the embedding of the Heisenberg's
Uncertainties, the space of all possible string configurations
``covers'' all the cases of the
combinatorial formulation, of which it provides a representation in terms of a
probabilistic scenario, useful for practical computations. 
Indeed, in this theoretical framework
precisely the ``uniqueness'' of the combinatorial
scenario (because of its being absolutely general), and the fact of being
the collection of string constructions a faithful and complete
representation in terms
of fields and particles of this absolutely general structure, 
allow to view in a different light
the problem of the ``uniqueness of string theory'', namely the
fact that all perturbative superstring constructions should be part of a unique
theory.

In order to be a representation of the combinatorial scenario, 
as it happens for the universe coming out of
the geometry of codes, also the physical string vacuum must not 
follow a selection rule. 
In correspondence to the phase space of all the energy-combinatorial
configurations, it is possible to introduce the phase space of
all string configurations, and the corresponding partition function
for the universe at any age.
Since we work on the continuum, instead of a sum the
partition function of the string phase space will be an integral:
\be
{\cal Z}_{\rm string} \, = \, \int {\cal D} \psi \, {\rm e}^{S (\psi)} \, .   
\label{zint1}
\ee
One can show that, in order to correctly
reproduce the conditions of the combinatorial problem,
the ordering must be taken through the degree of symmetry 
and the volume of the compact space
these configurations are based on.

The detailed analysis of the string configuration of highest
entropy and the corresponding spectrum of particles and fields
will be discussed in Ref.~\cite{npstrings-2011}(see also \cite{spi}), 
together with a discussion
of various cosmological bounds (Oklo bound, nucleosynthesis bound), etc.
In this paper we discuss
the theoretical grounds of this approach, revisiting and completing
the content of
Refs.~\cite{assiom} and \cite{rel}.
As in Ref.~\cite{assiom},
we start our analysis in section~\ref{setup} by investigating the
combinatorics of the ``distribution of binary attributes'', and discuss how,
and in which sense, certain structures dominate. This allows to see an
``order'' in this ``darkness''. We discuss how a ``geometry'' shows up,
and how geometric inhomogeneities, that we can interpret as the discrete 
version of ``wave packets'',
arise. We recover in this way, through a completely different approach,
all the known concepts of particles and masses. In the ``phase space''
constituted by all possible configurations
we introduce the ``time ordering'' based on the inclusion of configurations,
and discuss what is the dimension and geometry which is
statistically dominant. 
In section~\ref{UncP} we discuss how the Uncertainty Principle shows up
in this framework, and what is its interpretation.
We devote section~\ref{detprob} to a discussion
of the issues of causality and in what limit ``quantum mechanics'' arises
in this framework.  
In section~\ref{relativity} we draw on the arguments of Ref.~\cite{rel}
to discuss how this scenario implies also Relativity.

We pass then (section \ref{stringT}) to discuss what is the role played
by string theory in this scenario:
in which sense and up to what extent it provides an approximation
to the description of the combinatoric/geometric scenario, of which 
Quantum (Super) String Theory constitutes an implementation in 
the framework of a continuum (differentiable) space.
String Theory is consistent only
in a finite number of dimensions. Therefore, it would seem that it represents
only a subset of the configurations of the combinatorial approach,
a subspace of the full phase space. However, through the implementation
of quantization, and therefore of the Heisenberg's Uncertainty Principle,
it considers also the neglected configurations of the phase space, covering
them under the uncertainty which is ``built in'' in its basic definition.
In other words, it comes already endowed with a 
``fuzziness'' that incorporates in its range the contribution
of all the other possible configurations. It is precisely due to this completeness,
ensured by canonical quantization, that String Theory can be seen to
constitute a unique theory, of which the various perturbative constructions
constitute dual slices. Canonical quantization can be also shown to
be directly related to the dimensionality of space-time; it is precisely
upon quantization that string theory is forced to a critical dimension, which
implies as most entropic compactification a configuration with
four space-time dimensions. At the end of the section we discuss 
then how the integral~\ref{zint1} can be viewed as the analogous of the
Feynman path integral for string configurations. 
This supports the 
idea that \ref{zsum1} constitutes the natural extension of
quantum field theory to gravity.
The concept of ``weighted sum over all paths''
is substituted by a weighted sum over all energy/space configurations. 
The traditional question about ``how to find the right string vacuum''
is then surpassed in a way that looks very natural
for a quantum scenario: the concept of ``right solution'' 
is a classical concept, as is the idea of ``trajectory'', compared
to the path integral. The physical configuration 
takes all the possibilities into account.
As much as the usual path integral contains all the quantum corrections
to a classical trajectory, similarly here in the functional~\ref{zsum1}
the sum over all configurations accounts for the corrections
to the classical, geometric vacuum.

In section~\ref{dimspt} we discuss how the Universe, as it appears 
to an observer, builds up. 
In particular, we discuss what is the meaning of boundary and
horizon in such a spheric geometry, and how 
an understanding of these properties is only possible outside of
the domain, and the properties, of classical geometry:
all oddities and paradoxes find their explanation only within
what we call quantum geometry.  
We conclude with some comments about how fundamental
is a description of the world in terms of discrete (binary) codes.
We argue that real numbers (the continuum) doesn't add any
information to a description made in terms of binary codes, which
therefore seems to be the most fundamental description of nature.
But our investigation, and the approach we are proposing,
leads us to dare asking
another question, about what is after all the world we experience.
We are used to order our observations according to phenomena that
take place in what we call space-time. An experiment, or, better,
an observation (through an experiment), basically consists in realizing that
something has changed: our ``eyes'' have been affected by something, that
we call ``light'', that has changed their configuration (molecular, atomic 
configuration). This light may carry information about changes in our
environment, that we refer either to gravitational phenomena, or 
to electromagnetic ones, and so on...
In order to explain them we introduce energies, momenta, ``forces'', i.e.
interactions, and therefore we speak in terms of masses, couplings etc...
However, all in all, what all these concepts refer to is a change in
the ``geometry'' of our environment, a change that ``propagates'' to us,
and eventually results in a change in our brain, the ``observer''.
But what is after
all geometry, other than a way of saying that, by moving along
a path in space, we will encounter or not some modifications? 
Assigning a ``geometry''
is a way of parametrizing modifications.
Is it possible then to invert the logical ordering from reality to
its description?  
Namely, can we argue that
what we interpret as energy,
or geometry, is simply a code of information? Something happens, i.e.
time passes, when the combinatorial of possible codes changes. 
Viewed in this way, it is not a matter of mapping
physical degrees of freedom to a language of abstract codes, but 
the other way around, namely: perhaps
the deepest reality \emph{is} ``information'', that we
arrange in terms of geometries, energies, particles, fields, and
interactions. 
Consider the most general and generic code. 
At the end of this paper, we argue that any code
can be expressed as a binary code. 
In this new point of view, the universe is
the collection of all possible codes. 
In order to ``see'' the universe,
we must \emph{interpret} these codes in terms of maps, 
from a space of ``energies'' to a target space, that take the ``shape''
of what we observe as the physical reality. From this
point of view, information is not just something that transmits knowledge
about what exists, but it is itself the essence of what exists, and the 
rationale of the universe is precisely that it ultimately
is the whole of rationale.

For a detailed analysis of the spectrum of the theory implied
by \ref{zsum1} and \ref{zint1}, and the phenomenological implications,
we refer the reader to \cite{npstrings-2011}, \cite{spc-gregori}
\cite{blackholes-g}, and~\cite{paleo}. In particular, Refs.~\cite{spc-gregori}
and~\cite{paleo} show how this theoretical framework, being on its ground a 
new approach to quantum mechanics and phenomenology, does not simply provide us
with possible answers to problems which are traditionally referred to quantum gravity and string theory, but opens new perspectives about problems apparently
pertaining to other domains of physics, such
as (high temperature) superconductivity and evolutionary biology.

\section{The general set up}
\label{setup}

Consider the system constituted by the following two ''cells'':
\be
\epsfxsize=3cm
\epsfbox{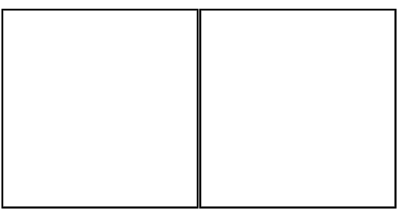}
\label{2cells1}
\ee
Let's assume that the only degrees of freedom this system possesses are that
each one of the two cells can independently be white or black.
We have the following possible configurations:
\be
\epsfxsize=3cm
\epsfbox{2cells1.eps}
\label{2cells2}
\ee
\be
\epsfxsize=3cm
\epsfbox{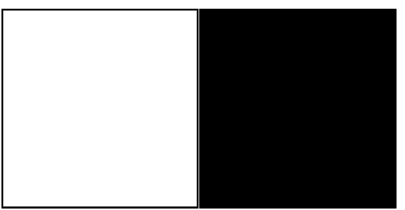}
\label{2cells3}
\ee
\be
\epsfxsize=3cm
\epsfbox{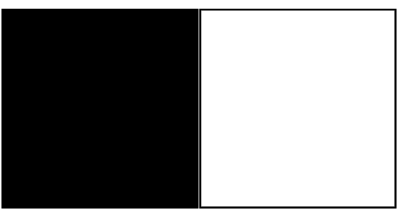}
\label{2cells4}
\ee
\be
\epsfxsize=3cm
\epsfbox{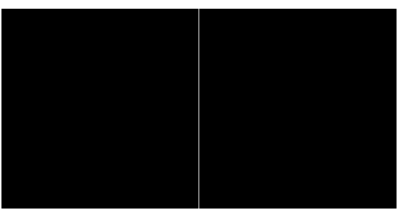}
\label{2cells5}
\ee
This is the ``phase space'' of our system.
The configuration ``one cell white, one cell black'' is realized two
times, while the configuration ``two cells white'' and ``two cells black''
are realized each one just once. 
Let's now abstract from the practical fact that these pictures 
appear inserted in a page, 
in which the presence of a written text clearly selects an orientation.
When considered as a ``universe'', something standing alone in its own,
configuration \ref{2cells3} and \ref{2cells4} are equivalent.
In the average, for an observer possessing the same ``symmetry'' of
this system (we will come back later to the subtleties of
the presence of an observer), the ``universe'' will appear 
something like the following:   
\be
\epsfxsize=3cm
\epsfbox{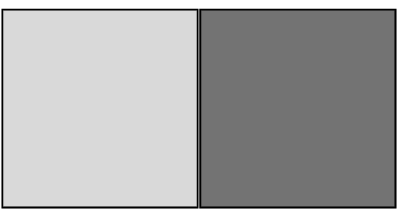}
\label{2cells6}
\ee
or, equivalently, the following: 
\be
\epsfxsize=3cm
\epsfbox{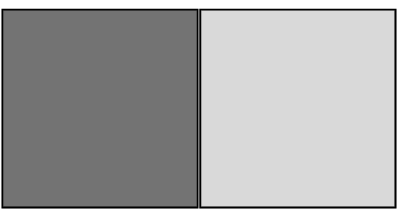}
\label{2cells7}
\ee
namely, the ``sum'':
\be
\epsfxsize=3cm
\epsfbox{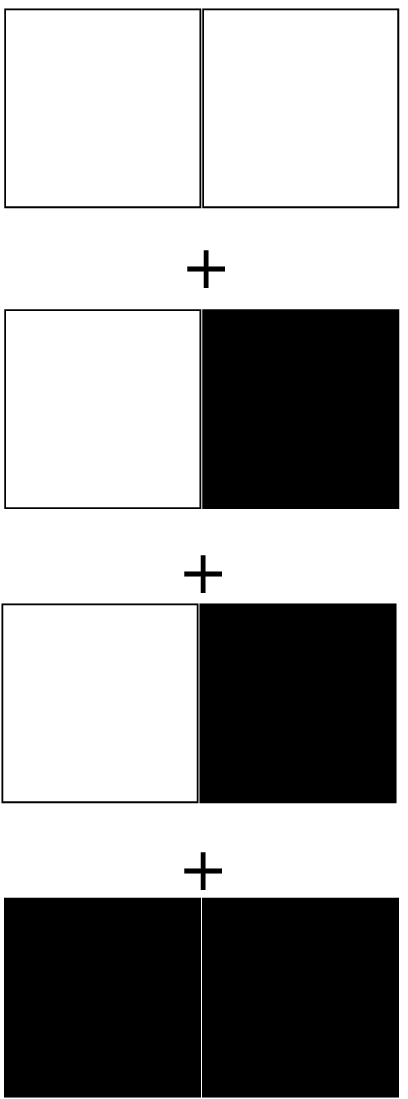}
\label{2cells8}
\ee
or equivalently the sum:
\be
\epsfxsize=3cm
\epsfbox{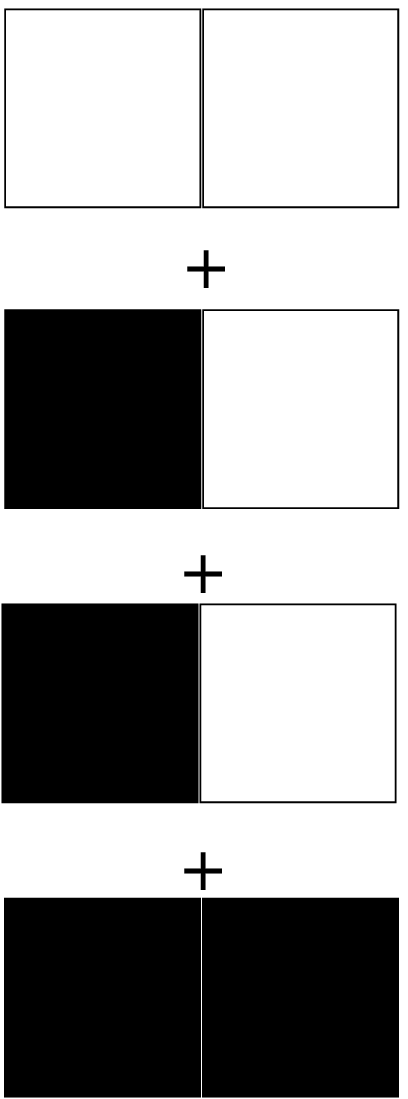}
\label{2cells9}
\ee
Notice that the observer ``doesn't know'' that we have rotated
the second and third term, because he possesses the same
symmetries of the system, and therefore is not able to distinguish
the two cases by comparing the orientation with, say, the
orientation of the characters of the text. What he sees, is
a universe consisting of two cells which appear slightly differentiated, 
one ``light gray'', the other ``dark gray''.

The system just described can be viewed as a two-dimensional space,
in which one coordinate specifies the position of a cell
along the ``space'',
and the other coordinate the attribute of each position, namely, the color.
Our two-dimensional ``phase space'' is 
made by $2 ({\rm space}) \times 2 ({\rm colors})$ cells.
By definition the volume occupied in the phase
space by each configuration (two white; two black; one white one black)
is proportional to the logarithm of its entropy. 
The highest occupation corresponds
to the configuration with highest entropy. The effective appearance,
one light-gray one dark-gray, \ref{2cells6} or \ref{2cells7}, 
mostly resembles the highest entropy configuration.
  
Let's now consider in general cells and colors. The colors are attributes we 
can assign to the cells, which represent the positions in our space.
On the other hand, these ``degrees of freedom'' can themselves be viewed as 
coordinates. Indeed, if in our space with $m ({\rm space}) \times n 
({\rm colors})$ we have $n > m$, then we have more degrees of freedom
than places to allocate them. In this case, it is more appropriate
to invert the interpretation, and speak of $n$ places to which to assign
the $m$ cells. The colors become the space and the cells their ``attributes''.
Therefore, in the following we consider always $n \leq m$.

\subsection{Distributing degrees of freedom}
\label{ddf}

Consider now a generic ``multi-dimensional'' space, consisting of 
$M_1^{p_1} \times \ldots \times M_i^{p_i} \ldots \times M_n^{p_n}$ 
``elementary cells''. Since an elementary, ``unit'' 
cell is basically a-dimensional, it makes sense to measure the volume
of this $p$-dimensional space, $p = \sum^n_i p_i$, in terms of unit cells:
$V = M_1^{p_i} \times \ldots \times M_n^{p_n} \stackrel{\rm def}{\equiv} 
M^p $. Although with the same volume, from the point of 
view of the combinatorics of cells and attributes this space is deeply
different from a one-dimensional space with $M^p$ cells. 
However, independently on the dimensionality, 
to such a space we can in any case assign, in the sense of ``distribute'',
$N \leq M^p$ ``elementary'' attributes. Indeed, in order to preserve the
basic interpretation of the ``$N$'' coordinate as ``attributes'' and
the ``$M$'' degrees of freedom as ``space'' coordinates, to which attributes
are assigned, it is necessary that $N \leq M_n$, $\forall \, n$ 
\footnote{In the case 
$N > M_n$ for some $n$, we must interchange the interpretation of 
the $N$ as attributes and instead consider them as a space coordinate, whereas
it is $M_n$ that are going to be seen as a coordinate of attributes.}.   
What are these attributes? Cells, simply cells:
our space doesn't know about ``colors'', it is simply a mathematical
structure of cells, and cells that we attribute in certain positions
to cells. By doing so, we are constructing a discrete 
``function'' $y = f (\vec{x})$, where 
$y$ runs in the ``attributes'' and $\vec{x} \in \{ M^{\otimes p} \}$
belongs to our $p$-dimensional space. 
We \underline{define} the phase space as the space of the assignments, the
``maps'':
\be
N \, \to \, \prod_i \otimes M_i^{\otimes p_i} \, , ~~~~~ M_i 
\, \geq \, N \, . 
\ee 
For large $M_i$ and $N$, we can approximate
the discrete degrees of freedom with continuous coordinates: $M_i \to r_i$,
$N \to R$.
We have therefore a $p$-dimensional space with volume $\prod r_i^{p_i}$, and
a continuous map $\vec{x} \in \{ \vec{r}^{\vec p} \} 
\stackrel{f}{\to} y \in \{ R \}$,
where $y$ spans the space up to $R = \prod r_i^{p_i} \equiv r^p$ and no more. 
In the following we will always consider $M_i \gg N$, while keeping $V = M^p$
finite. This has to considered as a regularization condition,
to be eventually relaxed by letting $V \to \infty$.

An important observation is that
\emph{there do not exist two configurations with the same entropy}:
if they have the same entropy, they are perceived as the same configuration.
\label{noneq}
The reason is that we have a combinatoric problem, and,
at fixed $N$, the volume of occupation
in the phase space is related to the symmetry group of the configuration.
In practice, we classify configurations through combinatorics: 
a configuration corresponds to a certain combinatoric group. Now,
discrete groups with the same volume, i.e. the same number of elements,
are homeomorphic. This means that they describe
the same configuration. Configurations and entropies are therefore
in bijection with discrete groups, and this removes the degeneracy.
Different entropy = different occupation volume = different volume of the
symmetry group; in practice  this means that we have a different configuration.

We ask now: what is the most realized configuration, namely, are there
special combinatorics in such a phase space that single out ``preferred''
structures, in the same sense as in our ``two-cells $\times$ two colors''   
example we found that the system in the average appears 
``light-gray--dark-gray''?
The most entropic configurations are the 
``maximally symmetric'' ones, i.e.
those that look like spheres in the above sense.

\subsection{Entropy of spheres}
\label{eSp}

Let us now consider distributing the $N$ energy attributes
along a $p$-sphere of radius $m$. We ask what are
the most entropic ways of occupy $N$ of the $\sim m^p$ cells of the
sphere~\footnote{For simplicity we neglect numerical coefficients: we are
interested here in the scaling, for large $N$ and $m$.}.
For any dimension, the most symmetric configuration is of course the one 
in which one
fulfills the volume, i.e. $N \sim m^p$. However, we are bound to the constraint
$N \leq m$ for any coordinate, otherwise we loose the interpretation 
at the ground of the whole construction, namely of $N$
as the coordinate of attributes, and $m$ as the target of the assignment. 
$N \sim m^p$ means $m \sim \sqrt[p]{N}$,
which implies $m < N$. The highest entropy we can attain is therefore
obtained with the largest possible value of $N$ as compared to $m$, i.e.
$N = m$, where once again the equality is intended up to an appropriate,
$p$-dependent coefficient. 
Let us start by considering the entropy of a three-sphere.
The weight in the phase space
will be given by the 
number of times such a sphere can be formed by moving along
the symmetries of its geometry, times
the number of choices of the position of, say, its center, in the whole space.
Since we eventually are going to take the limit $V \to \infty$,
we don't consider here this second contribution, which
is going to produce an infinite factor, equal for each kind of geometry, for 
any finite amount of total energy $N$. We will therefore concentrate here
on the first contribution, the one that from three-sphere and other geometries.
To this purpose, we solve
the ``differential equation'' (more properly, a finite difference equation)
of the increase in the combinatoric when passing from $m$ to $m + 1$.
Owing to the multiplicative structure of the phase space 
(composition of probabilities),
expanding by one unit the radius, or equivalently the scale of all 
the coordinates, means
that we add to the possibilities to form the configuration for
any dimension of the sphere some more $\sim m+1$ times
(that we can also approximate with $m$, because we work at large $m$) 
the probability of one cell
times the weight of the configuration of the remaining $m$ 
(respectively $m-1$) cells.
But this is not all the story: since distributing $N$ energy cells along a 
volume
scaling as $\sim m^3$, $m \geq N$ means that our distribution does not
fulfill the space, the actual symmetry group of the distribution will be
a subgroup of the whole group of the pure "geometric" symmetry: moving
along this space by an amount of space shorter than the distance between
cells occupied by an energy unit will not be a symmetry, because one
moves to a "hole" of energy. It is easy to realize that
in such a "sparse" space, the effective symmetry group will have a 
volume that stays 
to the volume of a fulfilled space in the same ratio as the respective energy
densities. Taking into account all these effects, we obtain the following 
scaling:
\be
W(m+1)_3 \, \sim \, W(m)_3 \times   (m+1)^3 \times {N \over m^3} 
\times {m \over N} \, . 
\label{Wmm0}
\ee 
The last factor expresses the density of a circle,  
whereas the
factor ${N \over m^3}$ is the density of the three-sphere. 
In order to make the origin of the various terms more clear, in these 
expressions
we did not use explicitly the fact that actually $N$ is going
to be eventually identified with $m$. 
Indeed, in \ref{Wmm0} there should be one more factor:
when we pass from radius $m$ to $m+1$ while keeping $N$ fixed, 
the configuration becomes less dense, and we loose a symmetry factor
of the order of the ratio of the two densities:
$[m / (m+1 )] ^3 \sim 1 + {\cal O}(1 / m)$.
Expanding $W(m+1)$ on the left hand side
of \ref{Wmm0} as $W(m) \, + \, \Delta W(m)$, and neglecting on the r.h.s. 
corrections of order $1 / m$,
we can write it as:
\be
{\Delta W(m)_3 \over W(m)_3 } \, \simeq \, m \, . 
\label{dwwm3}
\ee
Since we are interested in the behavior at large $m$, we can approximate it
with a continuous variable, $m \to x$, $x$, and approximate the finite
difference equation with a differential one. Upon integration, we obtain:
\be
S_{3} \, \propto \, \ln W(m)_3 \, \sim \, {1 \over 2} \, m^{2} \, ,
\label{Spm3}
\ee
where it is intended that $N = m$. Without this identification, the factor
$(m/ N)$ in \ref{Wmm0} would not be the density of a 1-sphere.
Under this condition, the energy density of the three-sphere
scales as $1 / N^2$,
and we obtain an equivalence between energy density and curvature $R$:
\be
\rho_3 (N) \, \sim \, {1 \over N^2} ~  \cong ~ {1 \over r^2} \, 
\sim \,  R_{(3)} \, .
\label{r3N}
\ee 
This is basically the Einstein's equation relating the curvature of space
to the tensor expressing the energy density. 
Indeed, here this relation can be assumed to be the physical description of a sphere
in three dimension. We can certainly think to formally distribute the $N$ energy units
along any kind of space with any kind of geometry, but what makes a curved
space \emph{physically} distinguishable from a flat one, and a particular geometry from another one? 
Geometries are characterized by the curvature, but how does one observer 
measure the curvature? The coordinates $m$
of the target space have no meaning without energy units distributed
along them. The geometry is decided by the way we assign the $N$
occupation positions. 
Here therefore we \emph{assume} that measuring
the curvature of space is nothing else than measuring the energy density.
For the time being, let us just take the equivalence between energy density and curvature as purely formal;
we will see in the next sections that this, with our definition of energy, will also imply 
that physical particles move along geodesics of the so characterized space, 
precisely as one expects from the Einstein's equations. 
We will come back to these
issues in section~\ref{relativity}.
In a generic dimension $p \geq 2$ the condition for having the geometry of
a sphere reads \footnote{We recall that we omit here $p$-dependent numerical coefficients
which characterize the specific normalization of the curvature of a sphere
in $p$ dimensions, because we are interested in the scaling at generic $N$, and $m$,
in particular in the scaling at large $N$.}:
\be
\rho_p (E) \, \sim \, {N \over m^p} ~ \cong ~ {1 \over m^2} \, . 
\label{rpN}
\ee
In dimension $p \geq 3$ it is solved by:
\be
m \, \sim \,  N^{1 \over p-2} ~ < \, N \, , ~~~ p \geq 3 \, .
\label{mphigher}
\ee
In two dimensions, \ref{rpN} implies $N = 1$ (up to some numerical coefficient). This means that, although it is technically possible to distribute $N > 1$
energy units along a two-sphere of radius $m > 1$, from a physical point of view
these configurations do not describe a sphere. This may sound strange, because
we can think about a huge number of spheric surfaces existing in our physical
world, and therefore we may have the impression that attempting to
give a characterization of the physical world in the way we are here doing
already fails in this simple case. The point is that all the two spheres of our
physical experience do not exist as two-dimensional spaces alone, but only as embedded
in a three-dimensional physical space. i.e. as subspaces of a three-dimensional
space.
In dimensions higher than three, the equivalent of \ref{Wmm0} reads:
\be
W(m+1)_p \, \sim \, W(m)_p \times   (m+1)^p \times {N \over m^p} 
\times {m \over N} \, . 
\label{Wm+1pN}
\ee
The last term on the r.h.s. is actually one, because it was only formally written
as $N / m$ to keep trace of the origin of the various terms. Indeed, it
indicates the density of a fulfilling space, to which the scaling of the weight of any
dimension must be normalized.
Inserting the condition for the $p$-sphere, equation~\ref{rpN}, we obtain:
\be
W(m+1)_p \, \sim \, W(m)_p \times   (m+1)^p \times {1 \over m^2} \, , 
\label{Wm+1p}
\ee
which leads to the following finite difference equation:
\be
{\Delta W (m)_p \over W(m)_p} ~ \approx ~
m^{p-2} \, .
\label{Wmp-2}
\ee
This expression obviously reduces to~\ref{dwwm3} for $p = 3$.
Proceeding as before, by transforming the finite difference equation
into a differential one, and integrating, we obtain:
\be
S_{( p \geq 2)} \, \propto \, \ln W(m) \, \sim \, {1 \over p-1} \, m^{p-1} \, , ~~~~
p \geq 3 \, .
\label{Spm}
\ee
This is the typical scaling law of the entropy of 
a $p$-dimensional black hole (see for instance \cite{Rabinowitz:2001ag}).
For $p=2$, if we start
from~\ref{Wm+1pN}, without imposing the condition~\ref{rpN}
of the sphere, we obtain, upon integration:
\be
S_{( 2)} \, \sim \, N^2 \, ,
\label{Sp2mN}
\ee 
formally equivalent to the entropy of a sphere in three dimensions.
However, the fact that the condition of the sphere \ref{rpN} implies
$N = 1$ means that a homogeneous distribution of the $N$ energy
units corresponds to a staple of $N$ two-spheres.
Indeed, if we use \ref{Wm+1p} and \ref{Wmp-2},
for which the condition $N = 1$ is intended, we obtain:
\be
S_{( 2)} \, \sim \, m \, .
\label{Sp2m}
\ee
For a radius $m = N$, this gives $1 / N$ of the result~\ref{Sp2mN},
confirming the interpretation of this space as the superposition of $N$
spheres. From a physical point of view, we have therefore $N$ times
the repetition of the same space, whose true entropy is not
$N^2$ but simply $N$. As we will see in the next sections,
such a kind of geometries correspond to what we will interpret as
quantum corrections to the geometry of the universe.
In the case of $p = 1$, from a purely formal point of view the condition
of the sphere~\ref{rpN} would imply $N = 1/m$. Inserted in~\ref{Wm+1pN}
and integrated as before, it gives:
\be
S_{( 1)} \, \propto \, \ln W(m) \, \sim \, \ln m \, , ~~~~~~
p = 1 \, .
\label{Sp1m}
\ee
Indeed, in the case of the one-sphere, i.e. the circle,
one does not speak of Riemann curvature, proportional to $1 / r^2$,
but simply of inverse of the radius of curvature, $1 / r$.
It is on the other hand clear that the most entropic
configuration of the one-dimensional space is obtained by a complete
fulfilling of space with energy units, $N=m$, and that the weight in the
phase space of this configuration is simply:
\be
W(N)_1 \, \sim \, N \, ,
\ee
in agreement with~\ref{Sp1m}~\footnote{We always factor out the group of permutations, which brings a volume factor $N!$ common to any
configuration of $N$ energy cells.}.
For the spheres in higher dimension, 
from expression~\ref{Spm} and~\ref{mphigher}we derive:
\be
S_{(p \geq 3)} \vert_N \, \sim \, {1 \over p-1} \, 
m^{p-1} \, \sim \, {1 \over p-1} \, N^{p-1 \over p -2} \, .
\ee 
For large $p$ the weights tend therefore to a $p$-independent value:
\be
W(N)_p ~ \stackrel{p \ggg 3}{\longrightarrow} ~ \approx \, {\rm e}^{N} \, , 
\label{Wlargep}
\ee
and their ratios tend to a constant. As a function
of $N$ they are exponentially suppressed
as compared to the three-dimensional sphere.
The scaling of the effective entropy as a function of $N$ 
allows us to conclude that:
\begin{itemize}
\item \emph{\underline{At any energy $N$, the 
most entropic configuration is the one corresponding to the geo-}}

\emph{\underline{metry of a 
three-sphere. Its relative entropy scales as $S \sim N^2$}}. 
\end{itemize}

\noindent
Spheres in different dimension have an unfavored ratio
entropy/energy. Three dimensions are then statistically
``selected out'' as the dominant space dimensionality.

\subsection{The ``time'' ordering}
\label{timev}

Consider the set $\Phi (N) \equiv \{ \Psi (N) \}$ 
of all configurations
at any dimensionality $p$ and volume $V \gg N$ ($V \to \infty$ at fixed $N$).
A property of $\Phi (N)$ is that, if $M < N$ $\forall \Psi(M) \in \Phi (M)$
$\exists \Psi^{\prime} (N) \in \Phi (M)$ such that $\Psi^{\prime} (N)
\supsetneq \Psi (M) $, something that, with an abuse of language, we
write as:  
$\Phi(N) \supset \Phi(M)$, $\forall ~ M < N$. 
It is therefore natural to
introduce now an ordering in the whole phase space, that
we call a ``time-ordering'', through the identification 
of $N$ with the time coordinate: $N \leftrightarrow t$. 
We call ``history of the Universe'' the ``path'' $N \to \Phi (N)$
\footnote{Notice that $\Phi(N)$, the ``phase space at time $N$'', 
includes also tachyonic configurations.}. 
This ordering turns out to quite naturally correspond to our
every day concept of time ordering. 
In our normal experience, the reason why
we perceive a history basically consisting in a progress toward
increasing time lies on the fact that higher times bear the ``memory'' 
of the past, lower times.
The opposite is not true,  because ``future'' configurations are not contained
in those at lower, i.e. earlier, times. But in order to
be able to say that an event $B$ is the follow up of $A$, $A \neq B$ 
(time flow from $A \to B$), at the time we observe $B$ we need to
also know $A$. This precisely means $A \in \Phi(N_A)$ \underline{and}
$A \in \Phi(N_B)$, which implies $\Phi(N_A) \subset \Phi(N_B)$ in the
sense we specified above.
Time reversal is not a symmetry of the system \footnote{Only by restricting to
some subsets of physical phenomena one can approximate the description 
with a model symmetric under reversal of the time coordinate, at the price
of neglecting what happens to the environment.}.

\subsection{How do inhomogeneities arise}
\label{inho}

We have seen that the dominant geometry, the spheric geometry, corresponds to a
homogeneous distribution of cells along the positions of the space, that 
we illustrate in figure~\ref{inhomo-1},
\be
\epsfxsize=4cm
\epsfbox{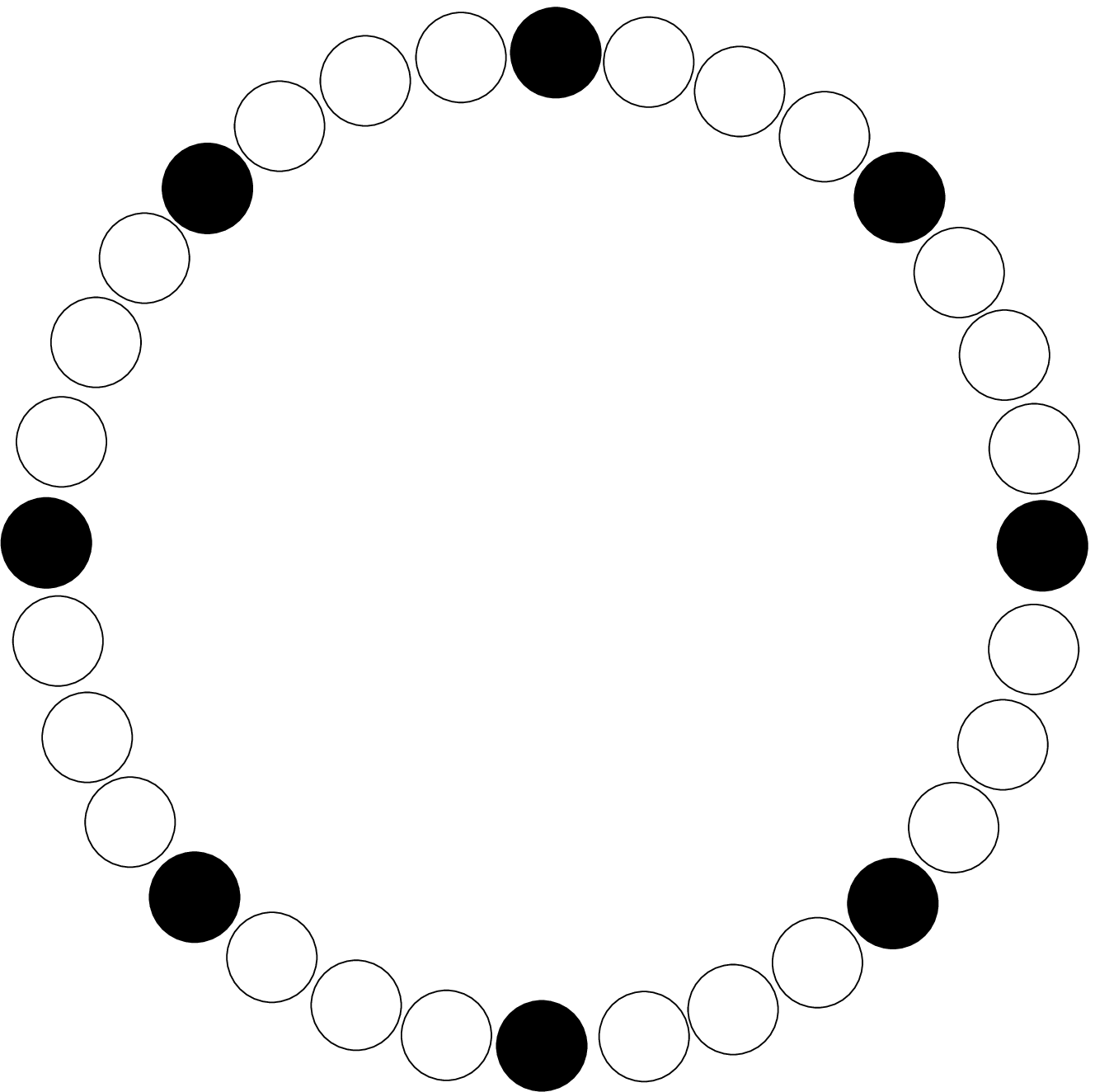}
\label{inhomo-1}
\ee
where we mark in black the occupied cells.
However, also the following configurations have spheric symmetry:
\be
\epsfxsize=3cm
\epsfbox{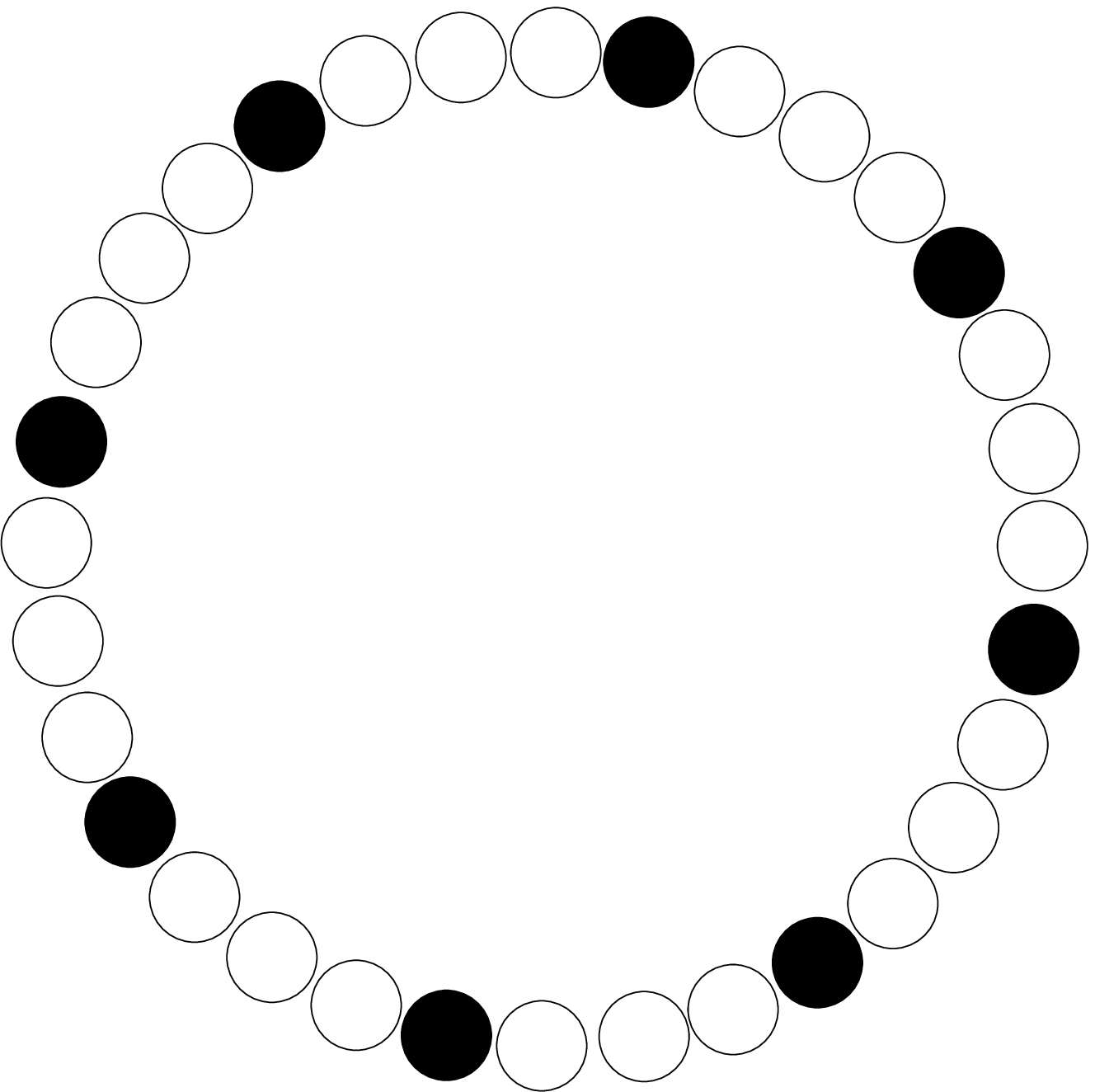}
~~~~~~~~~~~
\epsfxsize=3cm
\epsfbox{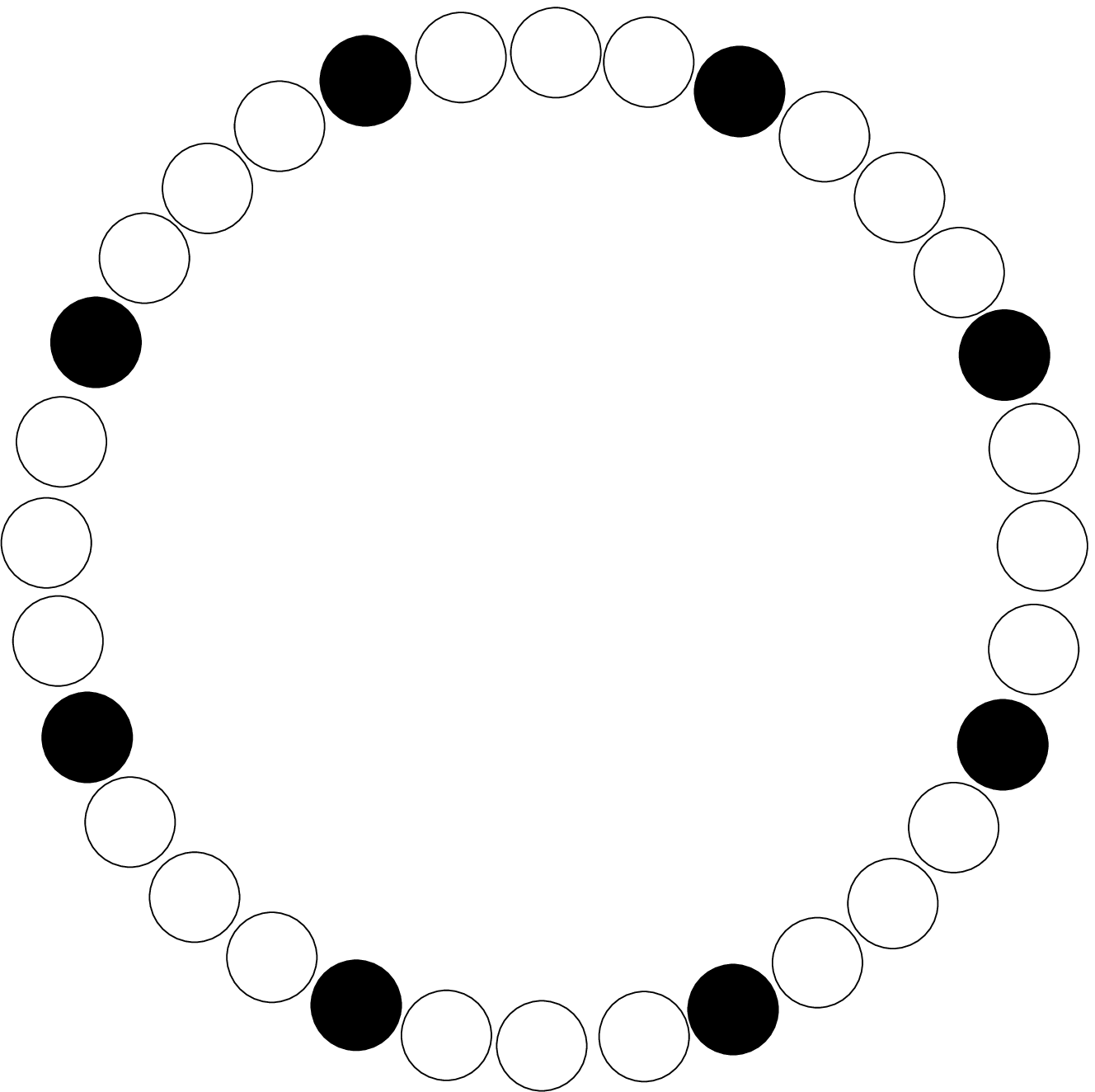}
~~~~~~~~~~~
\epsfxsize=3cm
\epsfbox{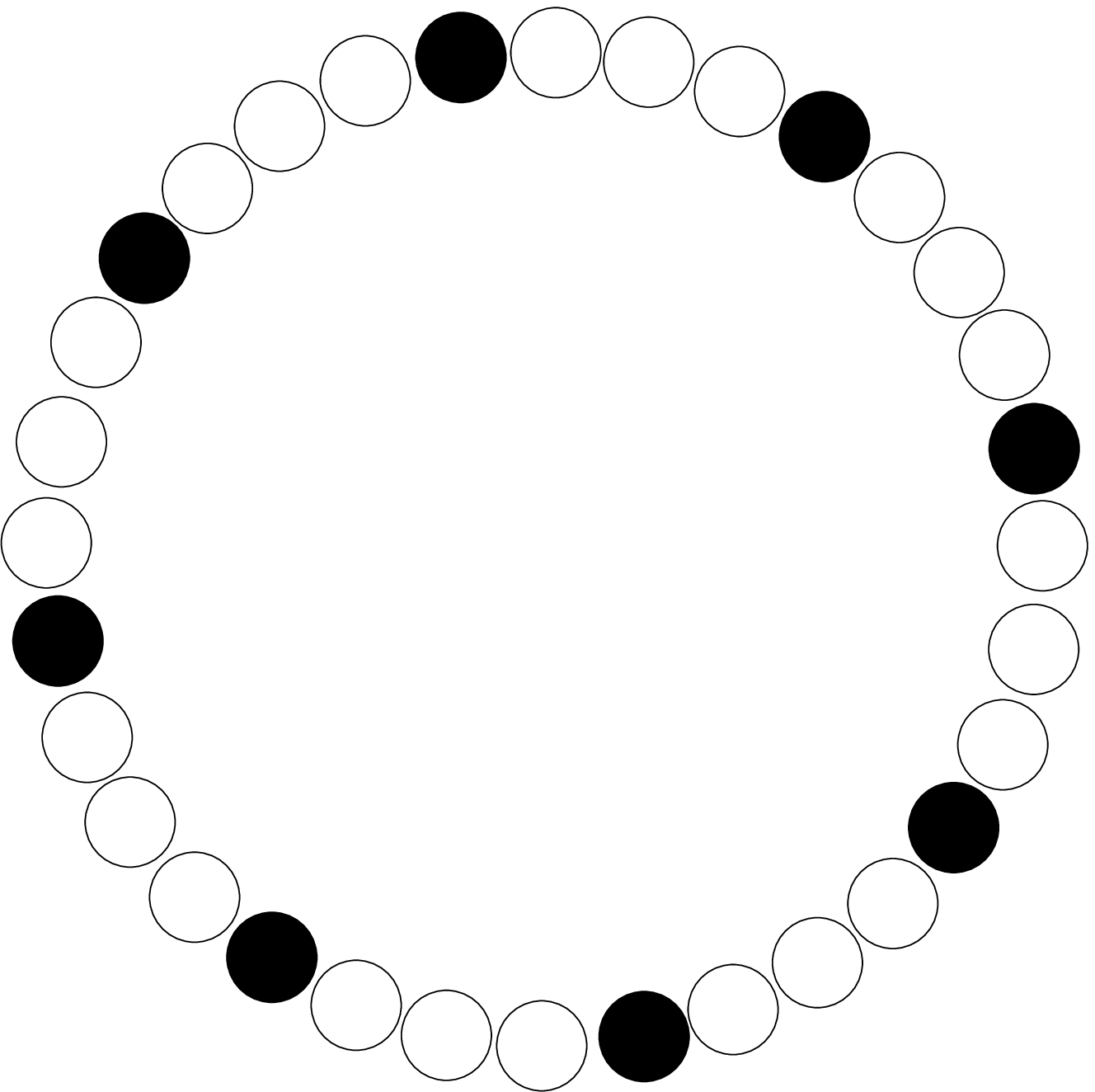}
\label{inhomo-4}
\ee
They are obtained from the previous one by shifting clockwise by one position 
the occupied cell.
One would think that they should sum up to an apparent averaged
distribution like the following:
\be
\epsfxsize=4cm
\epsfbox{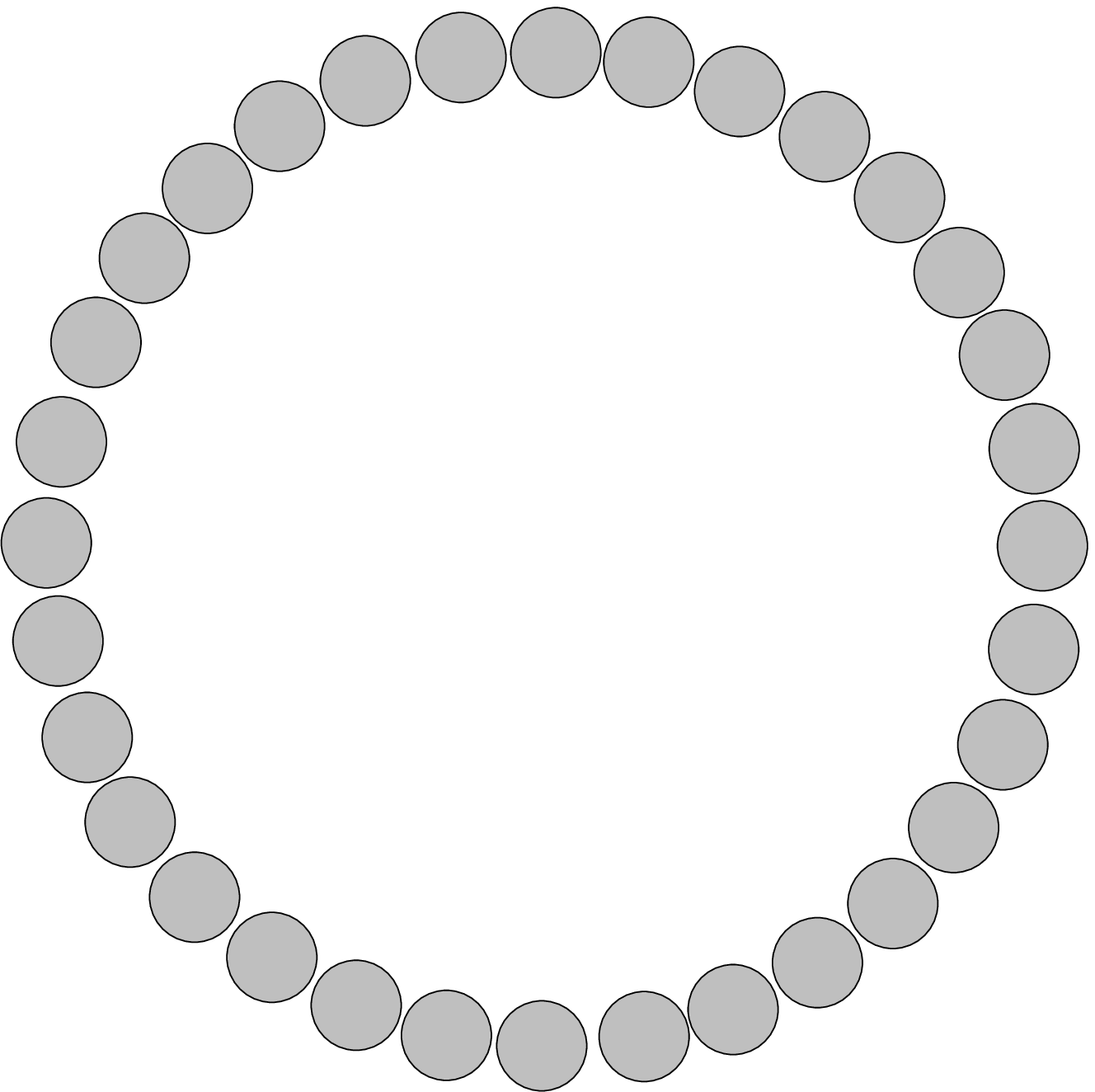}
\label{inhomo-5}
\ee 
This is not true: the Universe will indeed look like in 
figure~\ref{inhomo-5}, however this will be the ``smeared out'' result
of the configuration~\ref{inhomo-1}. As long as there are no reference points
in the space, which is an absolute space, all the above configurations
are indeed the same configuration, because nobody can tell in which sense
a configuration differs from the other one: ``shifted clockwise''
or ``counterclockwise'' with respect to what?
We will discuss later how the presence of an
observer by definition breaks some symmetries. Let's see here how
inhomogeneities (and therefore also configurations that we call
``observers'') do arise. Configurations with almost maximal, 
although non-maximal entropy, correspond to a slight breaking of the 
homogeneity of space. 
For instance, the following configuration, in which only one cell is shifted
in position, while all the other ones remain as in \ref{inhomo-1}:
\be
\epsfxsize=4cm
\epsfbox{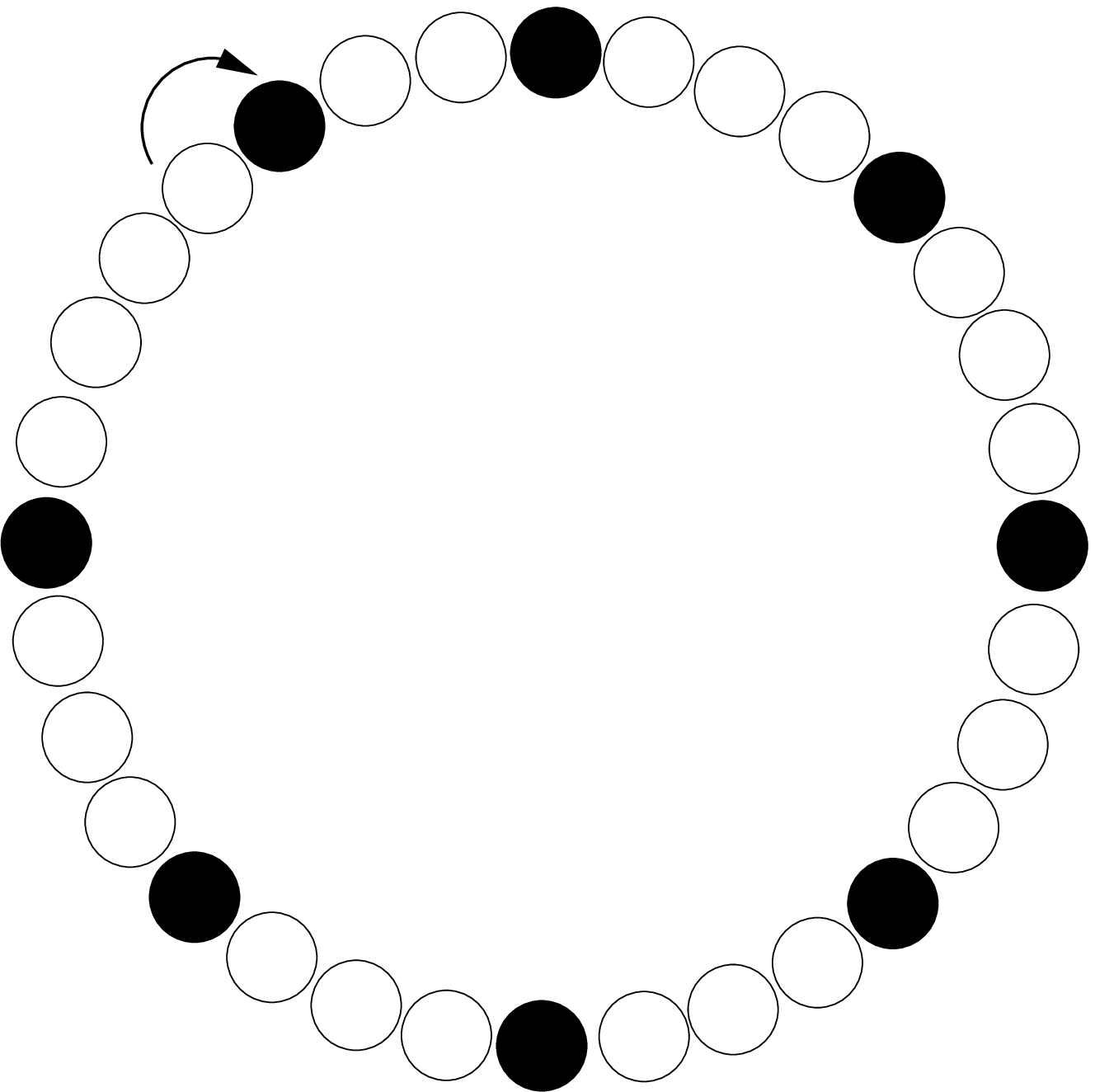}
\label{inhomo-6}
\ee 
This configuration will have a lower weight as compared to the fully symmetric
one. In the average, including also this one, the universe will
appear more or less as follows:
\be
\epsfxsize=4cm
\epsfbox{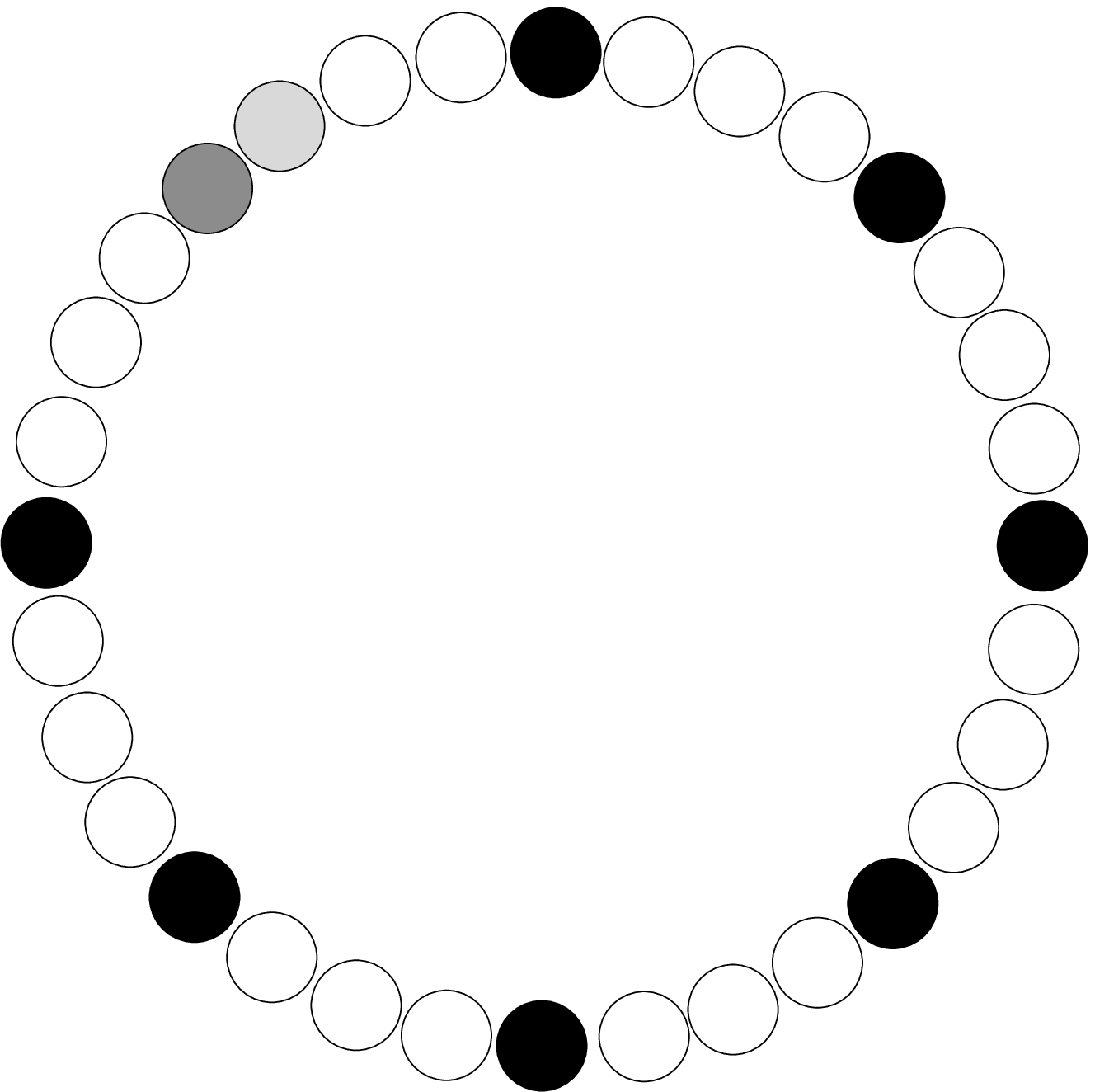}
\label{inhomo-7}
\ee 
where we have distinguished with a different tone of gray
the two resulting adjacent occupied cells, as a result of the
different occupation weight.
For the same reason as before, we don't have to consider summing 
over all the rotated configurations, in which the inhomogeneity
appears shifted by 4 cells, because all these are indeed the very same 
configuration as \ref{inhomo-6}. This is therefore the way inhomogeneities
build up in our space, in which the ``pure'' spheric geometry is
only the dominant aspect. We will discuss in section~\ref{sumgeo} how heavy
is the contribution of non-maximal configurations, and therefore what is
the order of inhomogeneity they introduce in the space. 

\subsection{The observer}
\label{obsv}

An observer is a subset of space, a ``local inhomogeneity''
(if one thinks a bit about,
this is what after all a person or a device is:
a particular configuration of a portion of space-time!). 
Wherever it is placed, the observer breaks the homogeneity of space.
As such, it \emph{defines} a privileged point, the point of observation.
Indeed, in this theoretical frame, everything is referred to the observer, 
which in this way defines the ``center of the universe''.
The observer is only sensitive to 
its own configuration. He, or it, ``learns'' about the full space only through
its own configurations. For instance, he can perceive that the configurations
of space of which he is built up change with time, and \emph{interprets}
this changes as due to the interaction with an environment.

It is not hard to recognize that these properties basically correspond
to the usual notion of observer.
There is no ``instantaneous'' knowledge: we know about objects placed at a 
certain distance from us only through interactions, light or gravitational
rays, that modify our configuration. But we know that, for instance,
light rays are light rays, because we compare configurations through a certain
interval of time, and we see that these change as according to
an oscillating ``wave'' that ``hits'' our cells. When we talk about energies,
we talk about frequencies. We cannot talk of periods and 
frequencies if we cannot compare configurations at different times.

\subsection{Mean values and observables}
\label{vev}

The mean value of any (observable) quantity ${\cal O}$
at any time ${\cal T} \sim N$ is the sum of the contributions to ${\cal O}$
over all configurations $\Psi$, weighted according to their volume of 
occupation in the phase space:
\be
< {\cal O} > \; \propto \, \sum_{\Psi ({\cal T})} 
W (\Psi)\, {\cal O}(\Psi) 
\, .
\ee
We have written the symbol $\propto$ instead of $=$ because,
as it is, the sum on the r.h.s. is not normalized. 
The weights don't sum up to 1, and not even do they sum up to a finite
number: in the infinite volume limit, they all diverge
\footnote{As long as the volume, i.e. the total number of cells of
the target space, for any dimension, is finite, there is only a finite 
number of ways one can distribute energy units. Moreover, also the possible
dimensionality of space are finite, bound by $D = V$,
because it does not make sense
to speak of a space direction with less than one space cell.
In the infinite volume limit, both the number of possibilities
for the assignment of energy, and the
number of possible dimensions, become infinite.}. However, as we 
discussed in section~\ref{ddf}, what matters is their 
relative ratio, which is finite because the infinite volume factor
is factored out. In order to normalize mean values,
we introduce a functional that works as ``partition
function'', or ``generating function'' of the Universe:
\be
{\cal Z} \, \stackrel{\rm def}{=} \sum_{\Psi ({\cal T})} 
W(\psi)   
\, = \, \sum_{\Psi ({\cal T})} 
{\rm e}^{S(\Psi)} \, .
\label{ZPsi}
\ee
The sum has to be intended as always performed at finite volume.
In order to define mean values and observables,
we must in fact always think in terms of finite space volume, 
a regularization condition to be eventually relaxed. 
The mean value of an observable can then be written as:
\be
< {\cal O} > \, \stackrel{\rm def}{\equiv} \, {1 \over {\cal Z}}
\sum_{\Psi ({\cal T})} W (\Psi)\, {\cal O}(\Psi) \, .
\label{meanO}
\ee
Mean values therefore are not defined in an absolute way, 
but through an averaging procedure in which
the weight is normalized to the total weight of all the configurations,
at any finite space volume $V$.

From the property stated at page~\pageref{noneq} 
that at any time ${\cal T} \sim N$
there do not exist two inequivalent
configurations with the same entropy, and from the fact that less
entropic configurations possess a lower degree of symmetry, we can already
state that:
\begin{itemize}
\item 
\emph{At any time ${\cal T}$ the average appearance of the universe is that
of a space in which \underline{all} \underline{the symmetries are broken.}}  
\end{itemize}
\noindent
The amount of the breaking, depending on the weight of non-symmetric
configurations as compared to the maximally symmetric one, involves
a relation between the energy (i.e. the deformations of the geometry) and the time
spread/space length, of the space-time deformation, as it will be
discussed in the next section.

\subsection{Summing up geometries}
\label{sumgeo}

We may now ask what a ``universe'' given by the collection of
all configurations at a given time $N$ looks like to an observer.
Indeed, a physical observer will be part of the universe, and as such
correspond to a set of configurations that identify a preferred point, 
something less symmetric and homogeneous than a sphere. 
However, for the time being, let us just assume that the observer
looks at the universe from the point of view of the most entropic 
configuration, namely it lives in three dimensions, and interprets 
the contribution of any configuration in terms
of three dimensions. This means that he will not perceive the universe
as a superposition of spaces with different dimensionality, but will
measure quantities, such as for instance energy densities,
referring them to properties of the three dimensional space, although
the contribution to the amount of energy may come also
from configurations of different dimension (higher or lower than three)
\footnote{These concept are not unfamiliar
to string theory, which implies a similar
interpretation of the three-dimensional world.}.

From this point of view, let us see how the contribution  
to the average energy density of space
of all configurations which are not the three-sphere is perceived. 
In other words, we must see how do the $p \neq 3$ configurations
project onto three dimensions.
The average density should be given by:
\be
\langle \rho(E)  \rangle ~ = ~
{\sum_{\Psi(N)} W(\Psi(N)) \rho(E)_{\Psi (N)} 
\over
\sum_{\Psi(N)} W(\Psi(N)) } \, .
\ee
We will first consider the contribution of spheres.
To the purpose, it is useful to keep in mind that at fixed $N$ (i.e. fixed time) higher
dimensional spheres become the more and more ``concentrated'' around
the (higher-dimensional) origin, and the weights tend
to a $p$-independent value for large $p$ (see \ref{mphigher} and \ref{Wlargep}).
When referred to three dimensions,
the energy density of a $p$ sphere, $p > 3$, is
$1 / N^{p-1}$, so that, when integrated over the volume, which
scales as $\sim N^p$, it gives a total energy $\sim N$.
There is however an extra factor $N^3 / N^{p} $ due to the fact that
we have to re-normalize volumes to spread all the higher-dimensional energy
distribution along a three-dimensional space.
All in all, this gives a factor $1 / N^{2(p-2)}$ in front of
the intrinsic weight of the $p$-spheres. Since the latter depend
in a complicated exponential form on $P$ and $N$, it is not possible
to obtain an expression of the mean value of the energy distribution
in closed form. However, as long as we are interested in just giving 
an approximate estimate, we can make several simplifications.
A first thing to consider is that, as we already remarked, at finite
$N$, the number of possible dimensions is finite, because
it does not make sense to distribute less than one 
unit of energy along a dimension: as a matter of fact such a space would
not possess this dimension. Therefore, $p \leq N$.
In the physically relevant cases $N \ggg 1$, and
we have anyway a sum over a huge number of terms, so that we can approximate
all the weights but the three dimensional one by their asymptotic value,
$W \sim \exp N$. This considerably simplifies our computation, because
with these approximations we have:
\be
\langle \rho(E)_N  \rangle ~ \approx ~
{1 \over {\rm e}^{N^2} \, + \, N \, {\rm e}^N} \times
\left[ {1 \over N^2} \, {\rm e}^{N^2} \, + \, \sum_{p > 3} 
{1 \over N^{2(p-2)}} \, {\rm e}^{N} \right] \, ,
\ee 
that, in the further approximation that $\exp N^2 \gg N \exp N$,
so that $\exp N^2 \, + \,  N \exp N \, \approx \, \exp N^2$,
we can write as:
\be
\langle \rho(E)_N  \rangle ~ \approx ~
{1 \over N^2} ~ + ~ {\rm e}^{- N} \, \left[ {1 \over 1 - {1 \over N^2}}
\right] ~ \approx ~ {1 \over N^2} ~ + ~ {\cal O} \left( {\rm e}^{- N} \right)
\, .
\label{rhoE3}
\ee 
We consider now the contribution
of configurations different from the spheres.
Let us first concentrate on the dimension $D = 3$, which is the most relevant one.
The simplest deformation of a 3-sphere consists in moving just one energy unit
one step away from its position on the sphere. Owing to this move, we 
break part of the symmetry. Further breaking is produced by moving more
units of energy, and by larger displacements. Indeed, it is in this way that
inhomogeneities in the geometry arise.
Our problem is to estimate the amount of reduction of the weight as compared
to the sphere. Let us consider displacing just one unit of energy.
We can consider that the overall symmetry group of the sphere is
so distributed that the local contribution is proportional to the density 
of the sphere, $1 / N^2$. Displacing one unit energy cell should then reduce 
the overall weight by a factor $\sim (1 - {1 / N^2})$. Displacing the unit by two steps 
would lead to a further suppression of order
$1 / N^2$. Displacing more units may lead to partial symmetry restoration
among the displaced cells. Even in the presence of
partial symmetry restorations the suppression factor due to the
displacement of $n$ units remains of order
$\approx n^2 / N^{2n}$ (the suppression factor divided
by the density of a sphere made of $n$ units) as long as $n \ll N$. The maximal effective
value $n$ can attain in the presence of maximal
symmetry among the displaced points is of course $N/2$, beyond
which we fall onto already considered configurations. 
This means that summing up all the contributions
leads to a correction which is of the order of the sum of an (almost) 
geometric series of ratio $1 / N^2$. 
Similar arguments can be applied to $D \neq 3$, to conclude that
expression \ref{rhoE3} receives all in all a correction of order $1 / N^2$.
This result is remarkable. As we will discuss in the following along this
paper, the main contribution to the geometry of the universe, the one
given by the most entropic configuration, can be viewed as the
classical, purely geometrical contribution, whereas those given by 
the other, less entropic geometries, can be considered contributions to the
"quantum geometry" of the universe. In Ref.~\cite{npstrings-2011}
we will discuss how the classical part of the curvature can be
referred to the cosmological constant, while the other terms to
the contribution due to matter and radiation. In particular, we will recover
the basic equivalence of the order of magnitude of these contributions, as
the consequence of a non-completely broken
symmetry of the quantum theory which is going to represent
our combinatorial construction in terms of quantum fields and particles. 
From \ref{rhoE3} we can therefore see that not only the three-dimensional
term dominates over all other ones, but that it is
reasonable to assume that 
\emph{the universe looks mostly like three-dimensional}, indeed
mostly like a three-sphere.
This property becomes stronger and stronger as time goes by 
(increasing $N$).
From the fact that the maximal entropy is the one of
three spheres, and scales as $S_{(3)} \sim N^2$, we derive also that
the ratio of the overall weight of the configurations at time $N-1$,
normalized to the weight at time $N$, is of the order:
\be
W(N-1) \, \approx \, W(N) \, {\rm e}^{-2N} \, .
\ee
At any time, the contribution of past times is therefore negligible as compared
to the one of the configurations at the actual time. The suppression factor
is such that the entire set of three-spheres at past times sums up 
to a weight of the order of $W(N-1)$:
\be
\sum_{n =1}^{N-1} W(n) ~ \approx ~ \sum {1 \over ({\rm e}^2)^n} ~ \sim ~
{\cal O}(1) \, .
\label{sumnN}
\ee

\vspace{.3cm}

We want to estimate now the overall contribution to the partition function
due to all the configurations, as
compared to the one of the configuration of maximal entropy.
We can view the whole spectrum of configurations as obtained by
moving energy units, and thereby deforming parts of the symmetry, starting
from the most symmetric (and entropic) configuration. In this way,
not only we cover all possible configurations in three dimensions, but we
can also walk through dimensions:
since we are basically working with space cells, 
it makes sense to think 
of moving and deforming also through different dimensions of space. 
In order to account for the contribution
to the partition function of all
the deformations of the most entropic
geometry, we can think of a series of steps, in which we move from
the spheric geometry one, two, three, and so on, units of symmetry.
At large $N$, we can approximate sums with integrals, and
account for the contribution to 
the ``partition function'' \ref {ZPsi} of all the configurations
by integrating 
over all the possible values of entropy, decreasing from the maximal one. 
In the approximation of variables on the continuum, symmetry groups
are promoted to Lie groups, and moving positions
and degrees of freedom is a ``point-wise'' operation that can be viewed
as taking place on the algebra, not on the group elements. Therefore,
the measure of the integral is such that we sum over incremental
steps on the exponent, that is on the logarithm of the weight, the entropy.
At large, asymptotically infinite
volume of space, volume factors due to the sum over all possible
positions at which the configurations can be placed (e.g. where a sphere is centered
in the target space) can be considered universal, 
in the sense that relevant deviations due to border 
effects concern only configurations
very sparse in space, and therefore remote in the phase space. 
With good approximation we can therefore factor out from all the weights
a common volume factor, and assume that the maximal entropy is volume-independent,
and corresponds to the
entropy of a three-sphere, as given in~\ref{Spm}, namely $S_{\rm max} = S_0 =
\exp N^2$. We can therefore write:
\be
{\cal Z} ~~ \gsim ~~  
\int^{S_0}_0 d L \; 
{\rm e}^{S_0 \left(1 - L \right) } \, .
\label{Zall1}
\ee
The domain of integration is only formal, in the sense that, as the entropy
approaches zero, it is no more allowed to neglect the volume factors depending
on the size of the overall volume. Indeed, if on one hand for any finite $N$
there are only a finite number of dimensions, the number of possible 
configurations in infinite, because on a target space of infinite extension, no matter of its 
dimension, the $N$ units
of energy can be arranged in a infinite number of different configurations. 
\ref{Zall1} has not to be taken as a rigorous expression,
but as an approximate way of accounting for the order
of magnitude of the contribution of the infinity of configurations.
Integrating~\ref{Zall1}, we obtain:
\be
{\cal Z} ~~ \approx ~~ {\rm e}^{S_0} \left(1 \, + \, {1 \over S_0}  \right)
\, .
\label{Zallint1}
\ee  
The result would however not change if, instead of considering the integration on
just one degree of freedom, parametrized by one coordinate, $L$, we would
integrate over a huge (infinite) number of variables, each one contributing
independently to the reduction of entropy, as in:
\be
{\cal Z} ~~ \approx ~~ \sum^N_{n=1} 
\int d^n L \; 
{\rm e}^{S_0 \left[1 -  (L_1 + \ldots + L_n) \right]} \, ,
\label{Zall}
\ee
In the second case, \ref{Zall}, we would have:  
\be
{\cal Z} ~~ \approx ~~ {\rm e}^{S_0} \sum_n {1 \over S_0^n } \, = 
\, {\rm e}^{S_0} \left( 1 + {1 \over S_0 -1} \right) \, ,
\label{Zallint}
\ee  
anyway of the same order as\ref{Zallint1}. Together with \ref{sumnN},
this tells us also that instead of \ref{zsum1} we could as well define the
partition function of the universe at "time" ${\cal E}$ by the sum over all
the configurations at past time/energy $E$ up to ${\cal E}$:
\be
{\cal Z}_{\cal E} ~ = ~ \sum_{\psi(E \leq {\cal E})} {\rm e}^{S (\psi)} \, .
\label{zsum}
\ee

\subsection{``Wave packets''}
\label{wavep}

Let's suppose there is a set of configurations of space that differ 
for the position of one energy cell, in such a way that 
the unit-energy cell is ``confined'' to a take a place in 
a subregion of the whole space. Namely, we have a sub-volume $\tilde{V}$ 
of the space with unit energy, or energy density $1 / \tilde{V}$. 
For $N$ large enough as compared to $\tilde{V}$,  
we must expect that all these configurations have almost the same weight.
Let's suppose for simplicity
that the subregion of space extends only in one direction, so that we work
with a one-dimensional problem: $\tilde{V} = r$. 
The ``average energy'' of this 
region of length $n \sim r $,  averaged over this subset
of configurations, is:
\be
E \, = \, {1 \over  n} \, = \, {1 \over  r} \, .
\label{Er}
\ee      
This is somehow a familiar expression: if we call this subregion   
a ``wave-packet'' everybody will recognize that this is nothing else than
the minimal energy according to the Heisenberg's Uncertainty Principle.
Each cell of space is ``black'' or ``white'', 
but in the average the region is ``gray'', 
the lighter gray the more is the ``packet'' spread out
in space (or ``time'', a concept to which we will come soon).
If we interpret this as the mass of a particle present
in a certain region of space, we can say that the particle is more heavy
the more it is ``concentrated'', ``localized'' in space. Light particles are 
``smeared-out mass-1 particles''.

\subsection{Masses}
\label{masses}

As discussed in section \ref{wavep}, the energies of the inhomogeneities , the
``energy packets'', are inversely 
proportional to their spreading in space: $E \sim 1 / r$.
Indeed, this is strictly so only in the case the configurations constituting
the wave packet exhaust the full spectrum of configurations, Namely, 
let's suppose we have a wave packet spread over 10 cells.
If we have 10 configurations contributing to the ``universe'',  in each
one of which nine cells corresponding to this set are ``empty'',
i.e. of zero energy, and one occupied, with the occupied cells 
occurring of course in a different position 
for each configuration, then we can rigorously say that the energy of the
wave packet is 1/10.
However, at any $N$ the universe consists of an infinite number of 
configurations, which contribute to ``soften'' (or strengthen)
the weight of the wave packet.
A priori, the energy of this wave packet could therefore be lower (higher)
than 1/10.
We are therefore faced with an uncertainty in the value of the mass/energy
of this packet, due to the lack of knowledge of the full spectrum
of configurations. As we already mentioned in section~\ref{eSp}, and will
discuss more in detail in section~\ref{UncP}, this uncertainty is at most
of the order of the mass/energy itself. For the moment, let's therefore accept
that such ``energy packets'' can be introduced, with a precision/stability 
of this order.
According to our definition of time, the volume of space increases with
time. Indeed, it mostly increases as the cubic power of time
(in the already explained sense that the most entropic configuration 
behaves in the average like a three-sphere), while the total energy increases
linearly with time. The energy of the universe therefore ``rarefies''
during the evolution ($\rho(E) \sim 1 / N^2  \sim 1 / {\cal T}^2$).
It is reasonable to expect
that also the distributions in some sense ``rarefy'' and
spread out in space. Namely, that also the sub-volume $\tilde{V}$ in which
the unit-energy cell is confined, and represents
an excitation of energy $1 / \tilde{V}$, spreads out as time goes by.

If the rate of increase of this volume is $d r / dt = 1$, namely, if
at any unit step of increase of time 
${\cal T} \sim N \to {\cal T} + \delta {\cal T} \sim N + 1$
we have a unit-cell increase of space: 
$r \sim n \to r + \delta r \sim n + 1$, the energy of this excitation
``spreads out'' at the same speed of expansion of the universe.
This is what we interpret as the propagation of the fundamental
excitation of a \underline{massless} field.

If the region where the unit-energy cell is confined
expands at a lower rate, $dr / dt < 1$, we have, within
the full space of a configuration, a reference frame
which allows us to ``localize'' the region, because we can
remark the difference between its expansion and the expansion of the full
space itself. We perceive therefore this excitation
as ``localized'' in space; its energy, its lowest energy, is always
higher than the energy of a corresponding massless excitation.
In terms of field theory, this is interpreted as the propagation
of a \underline{massive} excitation.

Real objects in general consist of a superposition of ``waves'', 
or excitations, and possess energies higher than the fundamental one.
Nevertheless, the difference between what we call massive and massless
objects lies precisely in the rate of expansion of the region of space
in which their energy is ``confined''.
The appearance of unit-energy cells at larger distance would be 
interpreted as ``disconnected'', belonging to another excitation,
another physical phenomenon; a discontinuity consisting in a ``jump'' by
one (or more) positions in this increasing one-dimensional ``chess-board''
implying a non-minimal jump in entropy. A systematic expansion of the
region at a higher speed is on the other hand what we call a 
\underline{tachyon}.
A tachyon is a (local) configuration of geometry that ``belongs
to the future''. In order for an observer
to interpret the configuration as coming from the future, 
the latter must corresponds to an energy density lower than the present one.
Indeed, also this kind of configurations contribute to the mean values
of the observables. Their contribution is however highly suppressed, as we
will see in section~\ref{UncP}.

\section{The Uncertainty Principle}
\label{UncP}

According to \ref{meanO}, quantities which are observable by an
observer living in three dimensions do not receive contribution
only from the configurations of extremal or near to extremal entropy: 
all the possible configurations at a certain time
contribute. Their value bears therefore a ``built-in'' uncertainty,
due to the fact that, beyond a certain approximation, 
experiments in themselves cannot be defined as physical quantities of a 
three-dimensional world.

In section~\ref{timev} we have established the correspondence 
between the ``energy'' $N$ and the ``time'' coordinate that orders the history
of our ``universe''. Since the distribution of the $N$ degrees of 
freedom basically determines the curvature of space, it is quite right 
to identify it with our concept of energy, as we intend it after the 
Einstein's General Relativity equations. 
However, this may not coincide with the
\emph{operational} way we define energy, related to the way we measure it.
Indeed, as it is, $N$ simply reflects the ``time'' coordinate, and coincides 
with the global energy of the universe, proportional to the time. 
From a practical point of view, what we
measure are curvatures, i.e. (local) modifications of the geometry, and we 
refer them to an ``energy content''. An exact measurement of energy
therefore means that we exactly measure the geometry and its 
variations/modifications within a certain interval of time. 
On the other hand, 
we have also discussed that, even at large $N$, 
not all the configurations of the universe at time $N$ admit an interpretation
in terms of geometry, as we normally intend it. The universe as we see it
is the result of a superposition in which also very singular configurations 
contribute, in general uninterpretable within the usual conceptual framework 
of particles, or wave-packets, and so on.    
When we measure an energy, or equivalently a ``geometric curvature'', 
we refer therefore
to an average and approximated concept, for which we consider only a subset of
all the configurations of the universe. Now, we have seen that the 
larger is the ``time'' $N$, the higher is the dominance of the most probable
configuration over the other ones, and therefore more picked is the average,
the ``mean value'' of geometry. The error in the evaluation of the
energy content will therefore be the more reduced, 
the larger is the time spread we consider, 
because relatively lower becomes the weight of the configurations we ignore.
From \ref{Zallint} we can have an idea of what is the order of the 
uncertainty in the evaluation of energy.
According to \ref{Zall}
and \ref{Zallint}, the mean value of the total energy, receiving contribution 
also from all the other configurations, results to be ``smeared'' by an amount:
\be
< E  > ~ \approx ~ E_{S_0} \, + \, E_{S_0} \times \, {\cal O} 
\left( 1 / S_0  \right) \, .
\label{Emean} 
\ee  
That means, inserting $S_0 \approx N^2 \equiv t^2 \sim \, E^2_{S_0}$:
\be
< E  > ~ \approx ~ E_{S_0} \, + \, \Delta E_{S_0} ~ \approx ~ 
E_{S_0} \, + \, {\cal O} \left( {1 \over t} \right) 
\, .
\label{deltaEmean} 
\ee  
Consider now a subregion of the universe, of extension 
$\Delta t$ \footnote{We didn't yet introduce units distinguishing between 
space and time. In the usual language we could consider this region as being
of ``light-extension'' $\Delta x = c \Delta t$.}. 
Whatever exists in it, namely,
whatever differentiates this region from the uniform spherical ground geometry
of the universe, must correspond to a superposition of 
configurations of non-maximal entropy. 
From our considerations of above, we can derive that it is not possible to
know the energy of this subregion with an uncertainty lower than the inverse
of its extension. In fact, let's 
see what is the amount of the contribution to this energy given by the
sea of non-maximal, even ``un-defined'' configurations. As discussed,
these include higher and lower space dimensionalities, and any other kind
of differently interpretable combinatorics. 
The mean energy will be given as in \ref{Emean}.
However, this time the maximal entropy
$\tilde{S}_0(\Delta t)$ of this subsystem will be lower than the upper bound
constituted by the maximal possible entropy of a region enclosed in a time 
$\Delta t$, namely the one of a three-sphere of radius $\Delta t$:
\be
\tilde{S}_0(\Delta t) ~ < ~ \left[ \Delta t \right]^2 \, ,
\label{Stilde}
\ee 
and the correction corresponding to the
second term in the r.h.s. of \ref{deltaEmean} will just constitute a lower 
bound to the energy uncertainty \footnote{The maximal energy can be 
$E \sim \Delta t$ even for a class of non-maximal-entropy, non-spheric 
configurations.}:
\be
\Delta E ~~ \gsim ~~ {{\Delta t} \over S_0 (\Delta t)} \, ~~
\approx \, {1 \over \Delta t } \, . 
\label{Etuncertainty}
\ee
In other words, \emph{no region of extension $\Delta t$
can be said with certainty to possess an energy lower than $1/ \Delta t$}.
When we say that we have measured
a mass/energy of a particle, we mean that we have measured 
an average fluctuation of the configuration of the universe around the 
observer, during a certain time interval. 
This measurement is basically a process that takes place along the time 
coordinate. 
As also discussed also Ref.~\cite{spi}, during the time of the ``experiment'', 
$\Delta t$, a small ``universe'' opens up for this particle. Namely, what we 
are probing are the configurations of a space region created in a time 
$\Delta t$. According to \ref{Etuncertainty}, the particle
possesses therefore a ``ground''
indeterminacy in its energy:
\be
\Delta E \, \Delta t ~\gsim ~ 1\, . 
\label{HUP}
\ee
As a bound, this looks quite like the time-energy Heisenberg
uncertainty relation. From an historical point of view, 
we are used to see
the Heisenberg inequality as a ground relation of Quantum Mechanics, ``tuned'' 
by the value of $\hbar$. Here it appears instead as a ``macroscopic relation'',
and any relation to the true Heisenberg's uncertainty looks only formal. 
Indeed, as I did already mention, we have not yet introduced units in which
to measure, and therefore physically distinguish, space and time, and
energy from time, and therefore also momentum. Here we have for the
moment only cells and distributions of cells. However, 
one can already look through where we are getting to: it is not difficult to
recognize that the
whole contruction provides us with the basic formal structures we need
in order to describe our world. Endowing it
with a concrete physical meaning will just be a matter of appropriately
interpreting these structures.
In particular, the introduction of $\hbar$ will just be a matter of
introducing units enabling to measure energies in terms of time (see discussion
in section~\ref{stringT}).

In the case we consider the whole Universe itself,
expression \ref{Zallint}
tells us that the terms neglected in the partition function,
due to our ignorance of the ``sea'' of all the possible configurations
at any fixed time, contribute to an ``uncertainty''
in the total energy of the same order as the inverse of the age of the 
Universe:
\be
\Delta E_{\rm tot} \, \sim \, {\cal O} \left( 1 \over {\cal T}  \right)
\, .
\ee
Namely, an uncertainty of the same order as the imprecision due
to the bound on the size of the minimal energy steps at time ${\cal T}$.

The quantity
$1/ S_0 \sim 1 / {\cal T}^2$ basically corresponds to 
the parameter usually called ``cosmological constant'', 
that in this scenario is not constant. 
The cosmological constant therefore 
not only is related to the size of the energy/matter density of the 
universe (see Ref.~\cite{spi}), setting thereby the minimal measurable ``step''
of the actual universe, related to the Uncertainty Principle \footnote{See also
Refs.~\cite{estring,lambda}.},
but also corresponds to a bound on the effective precision of calculation
of the predictions of this theoretical scenario.
Theoretical and experimental uncertainties are therefore of the same order.
There is nothing to be surprised that things are like that: this is 
the statement that the limit/bound to an experimental access to the
universe as we know it corresponds to the limit within which such a
universe is in itself defined. Beyond this threshold, there is a ``sea''
of configurations in which  
i) the dimensionality of space is not fixed;
ii) interactions are not defined, iii) there are tachyonic contributions,
causality does not exist etc...
beyond this threshold there is a sea of...uninterpretable 
combinatorics.

\begin{itemize}
\item \emph{It is not possible to go beyond the Uncertainty 
Principle's bound with the precision in the measurements, because this bound 
corresponds to the precision with which the quantities to be measured 
themselves are defined}.   
\end{itemize}

\section{Deterministic or probabilistic physics?}
\label{detprob}

We have seen that masses and energies are
obtained from the superposition, with different weight, of configurations
attributing unit-energy cells to different positions, that concur
to build up what we usually call a ``wave packet''. Unit energies appear
therefore ``smeared out'' over extended space/time regions.
The relation between energies and space extensions is of the type
of the Heisenberg's uncertainty.
Strictly speaking, in our case there is no uncertainty: in themselves,
all the configurations of the superposition
are something well defined and, in principle,
determinable. There is however also a true
uncertainty: in sections \ref{eSp} and \ref{masses} 
we have seen that to the appearance of the universe, and therefore to the 
``mean value'' of observables, contribute also higher and lower than three
dimensional space configurations, as well as tachyonic ones. In 
section~\ref{UncP} we have also seen how,
at any ``time'' $N$, all ``non-maximal''
configurations sum up to contribute to the geometry of space by an
amount of the order of the Heisenberg's Uncertainty. 
This is more like what we intend
as a real uncertainty, because it involves the very possibility of
defining observables and interpret observations according to 
geometry, fields and particles.
The usual quantum mechanics relates on the other hand the concept
of uncertainty with the one of probability: the ``waves'' (the set of
simple-geometry configurations which are used as bricks for building the
physical objects) are interpreted as ``probability waves'',
the decay amplitudes are ``probability amplitudes'', which allow to
state the probability of obtaining a certain result when making
a certain experiment. In our scenario, there seems to be no room
for such a kind of ``playing dice'': everything looks well determined.
Where does this aspect come from, if any, namely
where does the ``probabilistic'' nature
of the equations of motion originates from and what is its meaning in our
framework?

\subsection{A ``\sl{Gedankenexperiment}''}
\label{gedank}

Let's consider a simple, concrete example of such a situation.
Let's consider the case of a particle (an ``electron'')
that scatters through a double slit. This is perhaps the 
example in which classical/quantum effects manifest their peculiarities
in the most emblematic way, and where at best the deterministic vs. 
probabilistic nature of time evolution can be discussed.

\begin{figure}
\centerline{
\epsfxsize=4cm
\epsfbox{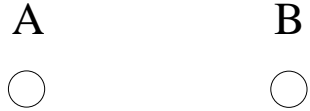}
}
\vspace{0.3cm}
\caption{A and B indicate two points of space-time, 
symmetric under reflection or 180$^0$ rotation.
They may represent the positions on a target place where light, 
or an electron beam, scattering through a double slit, can hit.}     
\label{exmp1}
\end{figure}

\begin{figure}
\centerline{
\epsfxsize=4cm
\epsfbox{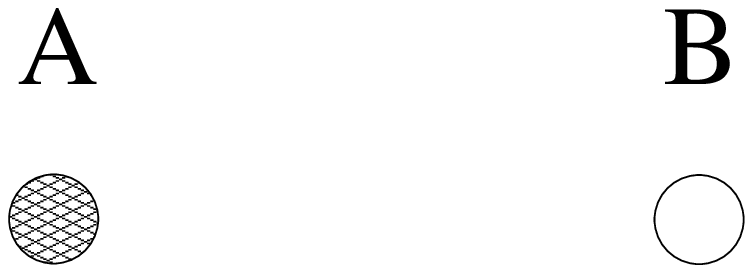}
}
\vspace{0.3cm}
\caption{Point A is marked by some property that distinguishes it from point
B.}     
\label{exmp2}
\vspace{0.5cm}
\centerline{
\epsfxsize=4cm
\epsfbox{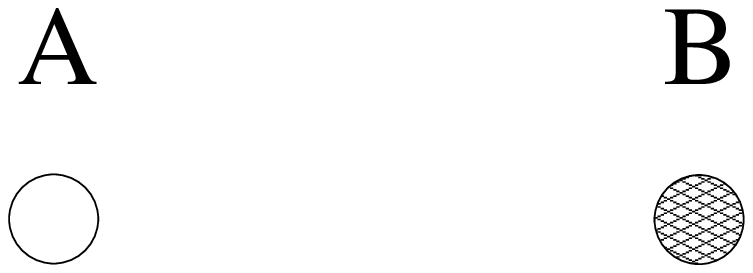}
}
\vspace{0.3cm}
\caption{The situation symmetric to figure~\ref{exmp2}.}     
\label{exmp3}
\end{figure}

As is known, it is possible to carry out 
the experiment by letting the electrons
to pass through the slit only one at once. In this case, each electron
hits the plate in an unpredictable position, but in a way that as time
goes by and more and
more electrons pass through the double slit, they build up the 
interference pattern typical of a light beam. This fact is therefore advocated 
as an example of probabilistic dynamics: we have a problem with a symmetry
(the circular and radial symmetry of the target plate, 
the symmetry between the two holes of the intermediate plate, etc...);
from an ideal point of view, in the ideal, abstract world in which formulae
and equations live, the dynamics of the single
scattering looks therefore absolutely unpredictable, although in the whole
probabilistic, statistically predictable \footnote{The 
probabilistic/statistical interpretation comes
together with a full bunch of related problems. For instance, the fact that
if a priori the probability of the points of the target plate 
corresponding to the 
interference pattern to be hit has a circular symmetry, as a matter of fact
once the first electron has hit the plate, there must be a higher probability 
to be hit for the remaining points, otherwise the interference pattern would
come out asymmetrical. These are subtleties that can be theoretically solved
for practical, experimental purposes in various ways, 
but the basic of the question remains, 
and continues to induce theorists and philosophers to come back to the 
problem and propose
new ways out (for instance K. Popper and his ``world of propensities'').}.    
Let's see how this problem looks in our theoretical framework.
Schematically, the key ingredients of the situation can be 
summarized in figure~\ref{exmp1}. 
This is an example of ``degenerate vacuum''
of the type we want to discuss. Points A and B are absolutely 
indistinguishable, and, from an ideal point of view, we can perform a 180$^0$
rotation and obtain exactly the same physical situation. 
As long as this symmetry exists, namely, as long as the \emph{whole universe},
including the observer, is symmetric under this operation, there is
no way to distinguish these two situations, the configuration and
the rotated one: they appear as only one configuration, weighting 
twice as much. Think now that A and B represent 
two radially symmetric points in the target plate of the double slit 
experiment. Let's
mark the point A as the point where the first electron 
hits. We represent the situation in which we have distinguished the properties
of point A from point B by shadowing the circle A, figure~\ref{exmp2}.
Figure~\ref{exmp3} would have been an equivalent choice. 
Indeed, since \emph{everything else} in the universe
is symmetric under 180$^0$ rotation,
figure~\ref{exmp2} and \ref{exmp3} represent \emph{the same} vacuum,
because \emph{nothing} enables to distinguish between figure~\ref{exmp2}
and figure~\ref{exmp3}. 

\begin{figure}
\centerline{
\epsfxsize=4cm
\epsfbox{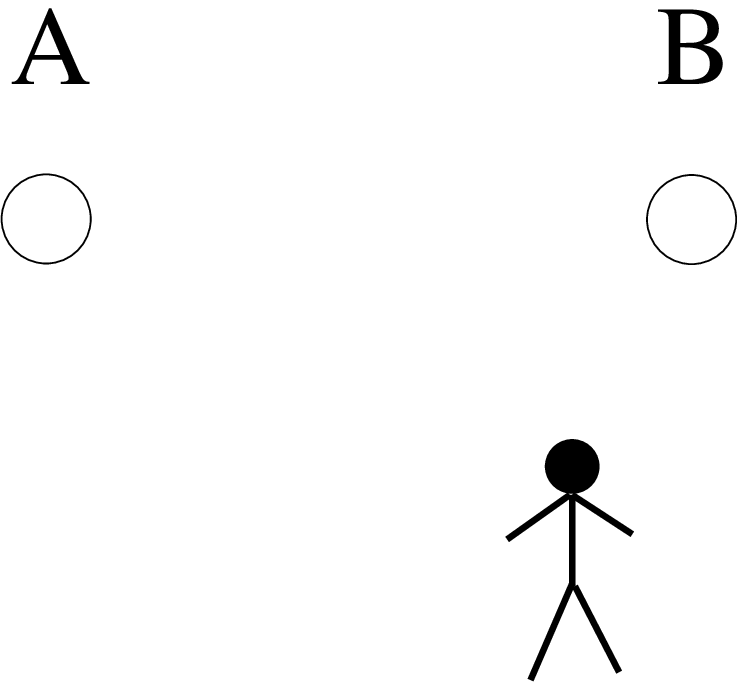}
}
\vspace{0.3cm}
\caption{The presence/existence of the observer
breaks the symmetry of the physical configuration
under 180$^0$ rotation.}     
\label{exmp4}
\centerline{
\epsfxsize=4cm
\epsfbox{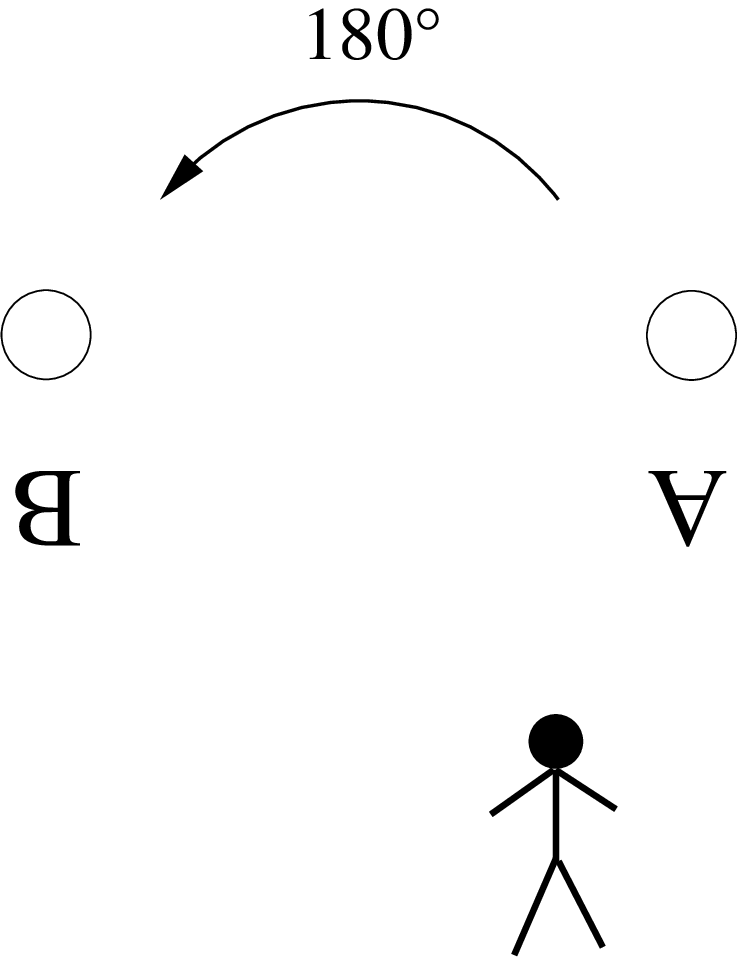}
}
\vspace{0.3cm}
\caption{The observer does not rotate. Now the rotated situation
is not equivalent to the previous one.}     
\label{exmp5}
\end{figure}

\begin{figure}
\centerline{
\epsfxsize=4cm
\epsfbox{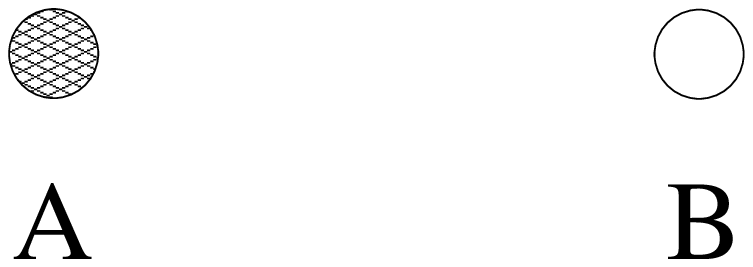}
}
\vspace{0.3cm}
\caption{The presence of an observer able
to detect a motion according to a group action is something that breaks 
the symmetry of the universe under this group, otherwise the
action would not be detectable. Here we represent the observer as
something that distinguishes A from B.}     
\label{exmp6}
\end{figure}

\begin{figure}
\centerline{
\epsfxsize=8cm
\epsfbox{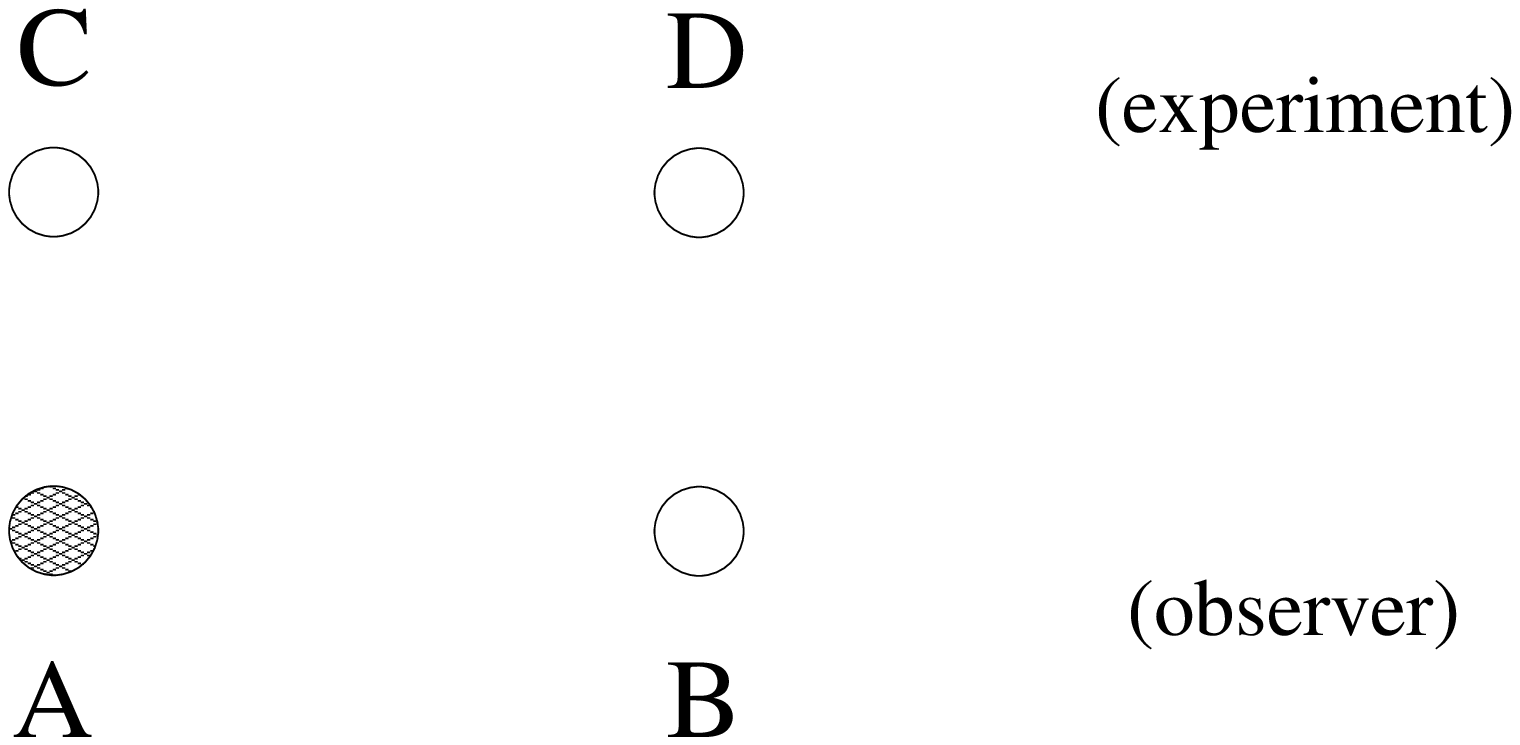}
}
\vspace{0.3cm}
\caption{In the presence of an observer, here represented by the points
A and B, even with a ``symmetric'' system, the points C, D, the universe
is no more symmetric. Points C and D can be identified, by saying
that C is the one closer to A, D the one closer to B.}     
\label{exmp7}
\end{figure}

\begin{figure}
\centerline{
\epsfxsize=8cm
\epsfbox{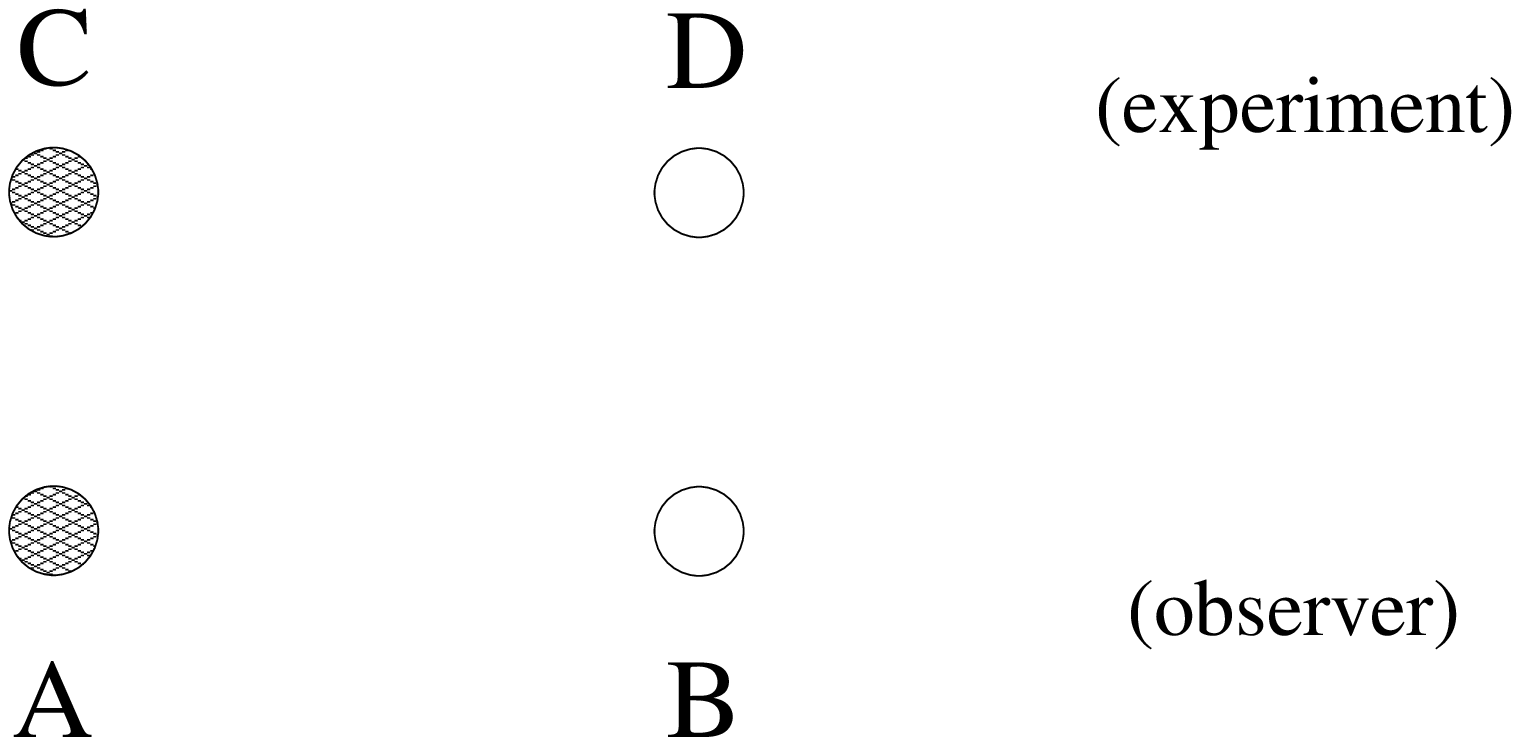}
}
\vspace{0.3cm}
\caption{The analogous of figure~\ref{exmp2} in the presence of an 
observer.}     
\label{exmp8}
\vspace{1cm}
\centerline{
\epsfxsize=8cm
\epsfbox{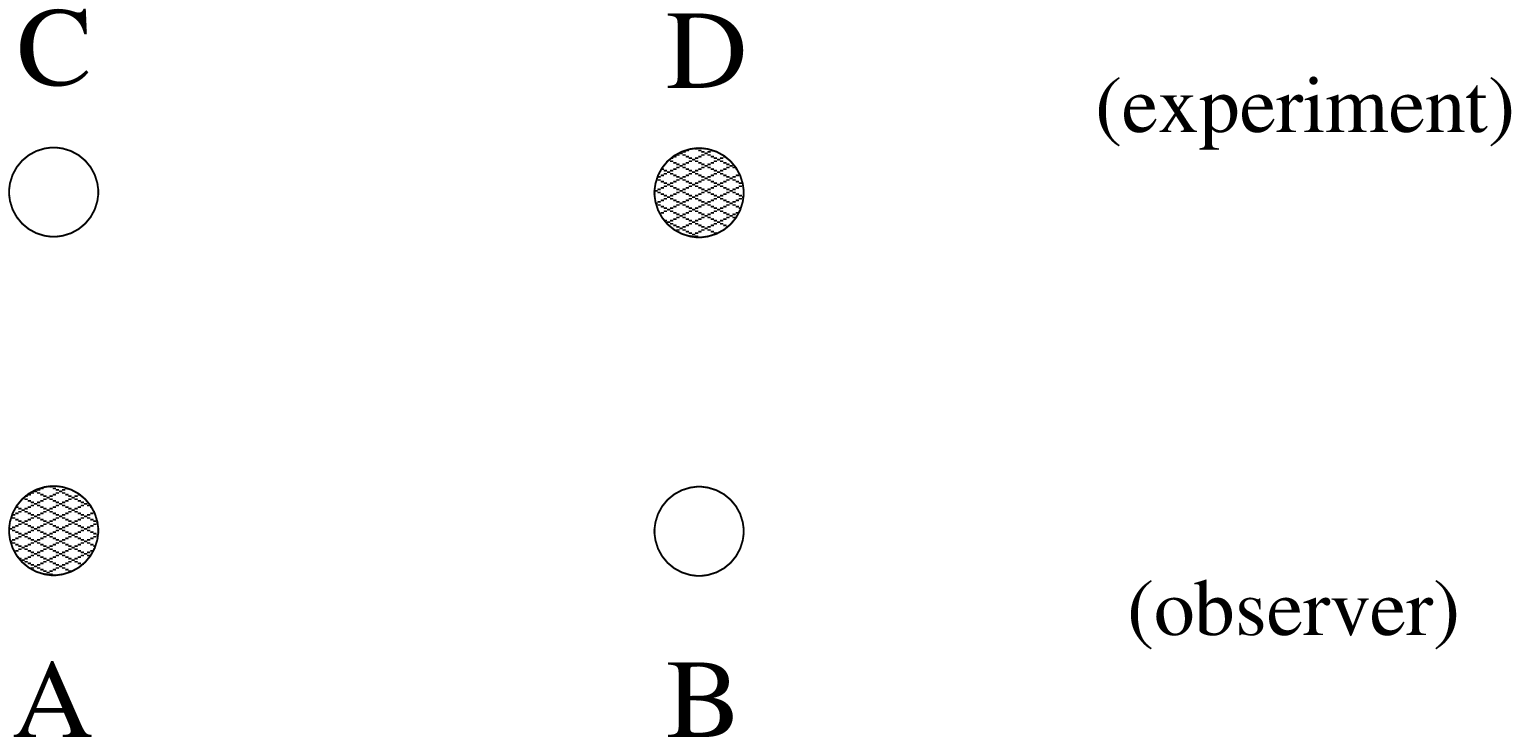}
}
\vspace{0.3cm}
\caption{The analogous of figure~\ref{exmp3} in the presence of an 
observer.}     
\label{exmp9}
\end{figure}

As we discussed in section~\ref{vev}, in our framework in the
universe all symmetries are broken. This matches with the fact that  
in \underline{any real experiment}, the environment doesn't possess
the ideal symmetry of our {\sl Gedankenexperiment}. For instance,
the target plate \emph{in} the environment, 
\emph{and the environment itself}, don't possess
a symmetry under rotation by 180$^0$: the presence of an ``observer''
allows to distinguish the two situations, as illustrated in figures~\ref{exmp4}
and \ref{exmp5}. 
There is therefore a choice which 
corresponds to the maximum of entropy.  
The \underline{real} situation can be schematically depicted 
as follows. The ``empty space'' is something like in figure~\ref{exmp6},
in which the two dots, distinguished by the shadowing, represent
the observer, i.e. not only ``the person who observes'', but more crucially 
``the object (person or device) which can distinguish between configurations''.
Now we add the experiment, figure~\ref{exmp7}.
In this case, the previous figures~\ref{exmp2} and \ref{exmp3}
correspond to figures~\ref{exmp8} and \ref{exmp9}.
It should be clear that entropy in the configuration of 
figure~\ref{exmp8} is not the same
as in the configuration of figure~\ref{exmp9}.
This means that the observer ``breaks the symmetries'' in the universe, 
it \emph{decides} that \emph{this one}, namely figure~\ref{exmp6}, 
is the actual configuration of the universe, i.e. the one contributing
with the highest weight to the appearance of the universe, 
while the one obtained by exchanging A and B is not.  

The observer is itself part of the universe, and the symmetric situation
of the ideal problem of the double slit is only an abstraction.
In our approach, it is the very presence of an observer, i.e. of an
asymmetrical configuration of space geometry, 
what removes the degeneracy of the 
physical configurations, thereby solving the paradox of equivalent 
probabilities of ordinary quantum mechanics. In this
perspective there are indeed no ``probabilities'' at all: the universe
is the superposition of configurations in the same sense as wave packets
are superpositions of elementary (e.g. plane) waves; real waves,
not ``probability wave functions''.
This means also that mean values, given by \ref{meanO}, are 
sufficiently ``picked'', so that
the universe doesn't look so ``fuzzy'', as it would if rather
different configurations contributed with a similar weight. 
Indeed, the fuzziness due to
a small change in the configuration, leading to a smearing out
of the energy/curvature distribution around a space region, 
corresponds to the Heisenberg's uncertainty, section~\ref{UncP}.
The two points on the target plate correspond to
a deeply distinguished asset of the energy distribution, the curvature
of space, whose distinction
is well above the Heisenberg's uncertainty.

When objects, i.e. special configurations of space
and curvature, are disentangled beyond the ``Heisenberg's scale'',
``randomness'' and
``unpredictability'' are rather a matter of the infinite number
of variables/degrees of freedom which concur to
determine a configuration, i.e., seen from a dynamical point of view,
``the path of mean configurations'', their time evolution.
In itself, this universe is though deterministic. Or, to better say,
``determined''. ``Determined'' is a better expression, 
because the universe at time
$N^{\prime} \sim {\cal T}^{\prime} = {\cal T} + \delta {\cal T} \sim N +1$ 
cannot be obtained by running forward the configurations at time 
$N \sim {\cal T}$.
The universe at time ${\cal T} + \delta {\cal T}$ is not the 
``continuation'', obtained through equations of motion,
of the configuration at time ${\cal T}$; it is given
by the weighted sum of all the configurations at time
${\cal T} + \delta {\cal T}$, as the universe at time ${\cal T}$ was
given by the weighted sum of all the configurations at time ${\cal T}$. 
In the large $N$ limit,
we can speak of ``continuous time evolution'' only 
in the sense that for a small change of time, the dominant
configurations correspond to distributions
of geometries that don't differ that much from those at previous time.
With a certain approximation we can therefore speak of evolution in 
the ordinary sense of (differential, or difference) time equations. 
Strictly speaking, however, initial conditions don't determine the future.

Being able to predict the details of an event, such
as for instance the precise position each electron will hit on the plate,
and in which sequence, requires to know the function
``entropy'' for an infinite number of configurations,
corresponding to any space dimensionality at fixed ${\cal T} \approx N$,
for any time ${\cal T}$ the experiment runs on.
Clearly, no computer or human being can do that. 
If on the other hand we content ourselves with an approximate predictive 
power, we can roughly reduce physical situations to certain ideal schemes, 
such as for instance ``the symmetric double slit'' problem. 
Of course, from a theoretical point of view we lose  
the possibility of predicting the position 
the first electron will hit the target (something anyway practically 
impossible to do), but we gain, at the
price of introducing symmetries and therefore also concepts like
``probability amplitudes'', the capability of predicting with a good
degree of precision the shape an entire beam of electrons will draw on the
plate. We give up with the ``shortest scale'', and we concern ourselves
only with an ``intermediate scale'', larger than the point-like one,
shorter than the full history of the universe itself. 
The interference pattern arises as the dominant mean configuration,
as seen through the rough lens of this ``intermediate'' scale.

\subsection{Going to the continuum}
\label{continuum}

For $N$ sufficiently large, it is not only possible but convenient to
map to a description on the continuum, because 
this not only makes things easier
from a computational point of view, but also better corresponds
to the way the physical world shows up to us, or, more precisely, to
the interpretation we are used to give of it. This however
does not mean that
the description on the continuum is the most fundamental one
of the physical world (see discussion of section~\ref{NorR}).

If we want to pass to a description in terms of continuous variables,
we must introduce a ``length'' to be used as a measure: 
in the continuum, lengths must be measured in terms of a given unit.
Differently from the discrete formulation, in which all quantities:
the extension of space, the amount of ``energy'', the ``time'', could be
measured in terms of ``number of cells'', in the continuum
we must a priori introduce a distinguished unit of measure for any 
type of measurable quantity. 
To start with, we must introduce a unit of length, that we call
$\ell$. This not only serves as a measure, 
but it can be chosen to coincide with
the elementary size, the radius of the unit cell. In this way, we introduce
what we call the ``Planck length'', $\ell_{\rm Pl}$.

Energies and momenta 
are conjugate to space lengths, relation~\ref{Er}, and the natural unit in 
which they are measured is the inverse of the Planck length. This leads
to the introduction of the Planck Mass $m_{\rm Pl}$ and the unit 
of conversion between the energy/momentum and space/time
scale, the Planck constant $\hbar$ according to the relation:
\be
\left[E,P   \right] \, \sim \, {1 \over \left[ R, t  \right]} ~ \leadsto
~ m_{\rm Pl } \, \stackrel{\rm def}{\equiv} \, 
{\hbar \over \ell_{\rm Pl}} \, . 
\ee
This corresponds to the usual relation between these quantities, apart
from the fact that here doesn't appear any power of the ``speed of light''.
In fact, till now we have paired the concepts of energy/momentum/mass because
we have not yet distinguished the unit of measure of time from the one of 
space. Indeed, were all the objects either massless, or permanently at rest,
this distinction would be unnecessary. We need to disentangle time from space
in order to measure the rate of expansion of objects, ``inhomogeneities''
in the average geometry of space, as compared to the rate of the expansion
of the space itself. As discussed,
non-trivial massive objects correspond to subregions that spread out
at particular rates, giving therefore rise to a full spectrum of 
non-trivial ``speeds''. We measure these speeds in terms of $c$,
the rate of expansion of the radius of the three-sphere with respect
to $N$, intended as the time. In section~\ref{speedlight} we will discuss
how this can be identified with the ``speed of light in the vacuum''.
Obviously, the formulation in terms of discrete numbers and combinatorics
corresponds to a choice of units 
for which all these ``fundamental constants'' are 1.   

From the perspective of a theory on the continuum,
in themselves these scales could be considered as free parameters.
One could think to be forced to introduce them as regulators,
but that in principle they are free to take any 
possible value. However, being $\ell$ the unit in which the length
of space is measured, by varying it one varies the ``unit of volume'', 
or equivalently ``the size of the point'', $v$. When considering the full
span of volumes $V$ we obtain a series of equivalent sets describing the same 
system, equivalent histories of the universe. 
Running in the set 
$\{V /v  \}$ by letting both $V$ and $v$ take any possible value results
in a redundancy reproducing an infinite number of times
the same situation. Similar arguments hold for the Planck constant
$\hbar$ and the speed of expansion $c$. In particular,
fixing the speed of expansion to a constant $c$
allows to establish a bijective map between the time $t$ and
the volume $V$ of the three-spheres.  Varying the map $t \to V $
through a change of $c$ would lead to an over-counting in the
``history of the universe''
\footnote{Also introducing a space dependence $c = c(\vec{X})$ would be 
a nonsense, because the functional \ref{ZPsi} always gives
the universe as it appears at the point of the observer, the 
``present-time point'', say $\vec{X}_0$. 
Saying that $c(\vec{X}^{\prime}) \neq c(\vec{X}_0) $ would be like
saying that volumes appear at $\vec{X}^{\prime}$ 
differently scaled than
how they appear at $\vec{X}_0$: this is a matter of properly
reducing observables to the point of the observer, through a rescaling.}. 
In summary, we would have \underline{classes} of universes, 
parametrized by the values of $c$ and $\ell_{\rm Pl}$. 
The real, effective phase space is therefore the coset:
\be
[{\rm eff.~phase~space}] \, = \, [{\rm phase~space}] 
\Big/ \{ c \} ,
\{ \ell_{\rm Pl} \} \, .
\ee 
When passing to the continuum, at large $N$, we must therefore look for
a mapping of the combinatoric problem to a description in terms of continuous
geometry, which i) contains as built-in the notion of minimal length,
finite speed of propagation of information, i.e. locality of physics,
in which ii) energies are related to space extensions through relations
such as \ref{Er}, and in which iii) the ``evolution'' is labeled through
a correspondence between configurations and a parameter that we call ``time''.
Through the relation between this parameter and the curvature, or the 
``radius'' of the maximally entropic configurations, this parameter too
can be viewed as a coordinate. Measuring also time in terms of unit cells,
the same units we use to measure the space,
corresponds to fixing the ``speed of expansion'' to 1 (later on we will
see how this can be seen as the ``speed of light'').

\section{Relativity}
\label{relativity}

As we discussed in sections~\ref{ddf} and~\ref{vev}, 
although the volume of the target spaces of the ``unit of energy cells'' 
is eventually considered infinite, $V \to \infty$,
at any finite time the dominant configuration of the universe 
corresponds to a three-sphere 
of radius $N \sim {\cal T}$. 
The configurations which correspond to a geometry
not bounded within a region of radius 
$N \sim {\cal T}$, nor three-dimensional,
contribute in the form of quantum perturbations: 
they all fall under the ``cover'' of the Uncertainty Principle, and are
related to what we interpret as the quantum nature of physical phenomena.
In other words, this means that at any time ${\cal T}$ we 
indeed \emph{do see} an infinitely-extended universe, 
but this can be reduced to the ordinary 
geometric interpretation of space only up to a distance ${\cal T}$.
\begin{itemize}
\item
\emph{The space ``outside'' the horizon is certainly infinitely 
extended, and somehow we see it, but it contributes to our perception and 
measurements only for an ``uncertainty'' of mean values, accounted for by the
Heisenberg's uncertainties}.
\end{itemize}
\noindent 
From a classical point of view, at any finite time
${\cal T}$ what we call ``space'' in the ordinary sense is of finite extension:
it exists only up to a radius $R \sim {\cal T} \sim N$.
This is for us the extension of the "classical" space.
In the following we want to see how in this space
Einstein's special (and general) relativity are implied as a particular limit.

\subsection{From the speed of expansion of the universe to a maximal speed for 
the propagation of information}
\label{speedlight}

The classical space corresponds to a universe of radius $\sim N$ at time $N$, 
with total energy
also $N$. It expands at speed $1$. Indeed, we can introduce
a factor of conversion from time to space, $c$, and say that, by choice
of units, we set the speed of expansion to be $c = 1$.  
We want to see how this is also the 
maximal speed for the propagation of information within the classical space.
It is important to stress that all this refers only to the classical space
as we have defined it, 
because only in this sense we can say that the universe is three dimensional:
the sum \ref{ZPsi} contains in fact also configurations in which higher 
speeds are 
allowed (we may call them ``tachyonic'' configurations), along with
configurations in which it is not even clear what is the meaning of
speed of propagating information in itself, as there is no recognizable 
information at all, at least in the sense we usually intend it.

Indeed, when we say we get information about, say, the motion of a particle,
or a photon,
we intend to speak of a non-dispersive wave packet, so that we can say
we observe a particle, or photon, that remains particle, or photon,
along its motion \footnote{Like a particle, also a physical photon, 
or any other field,
is not a pure plane wave but something localized, therefore a 
superposition of waves, a wave packet.}.
Let's consider the simplified case of a universe
at time $N$ containing only one such a wave packet \footnote{We may think to
concentrate onto only a portion of the universe, where only such a wave
packet is present.}, as illustrated in 
figure~\ref{gridN}, where it is represented by the shadowed cells, and
the space is reduced to two dimensions.
\be
\epsfxsize=6cm
\epsfbox{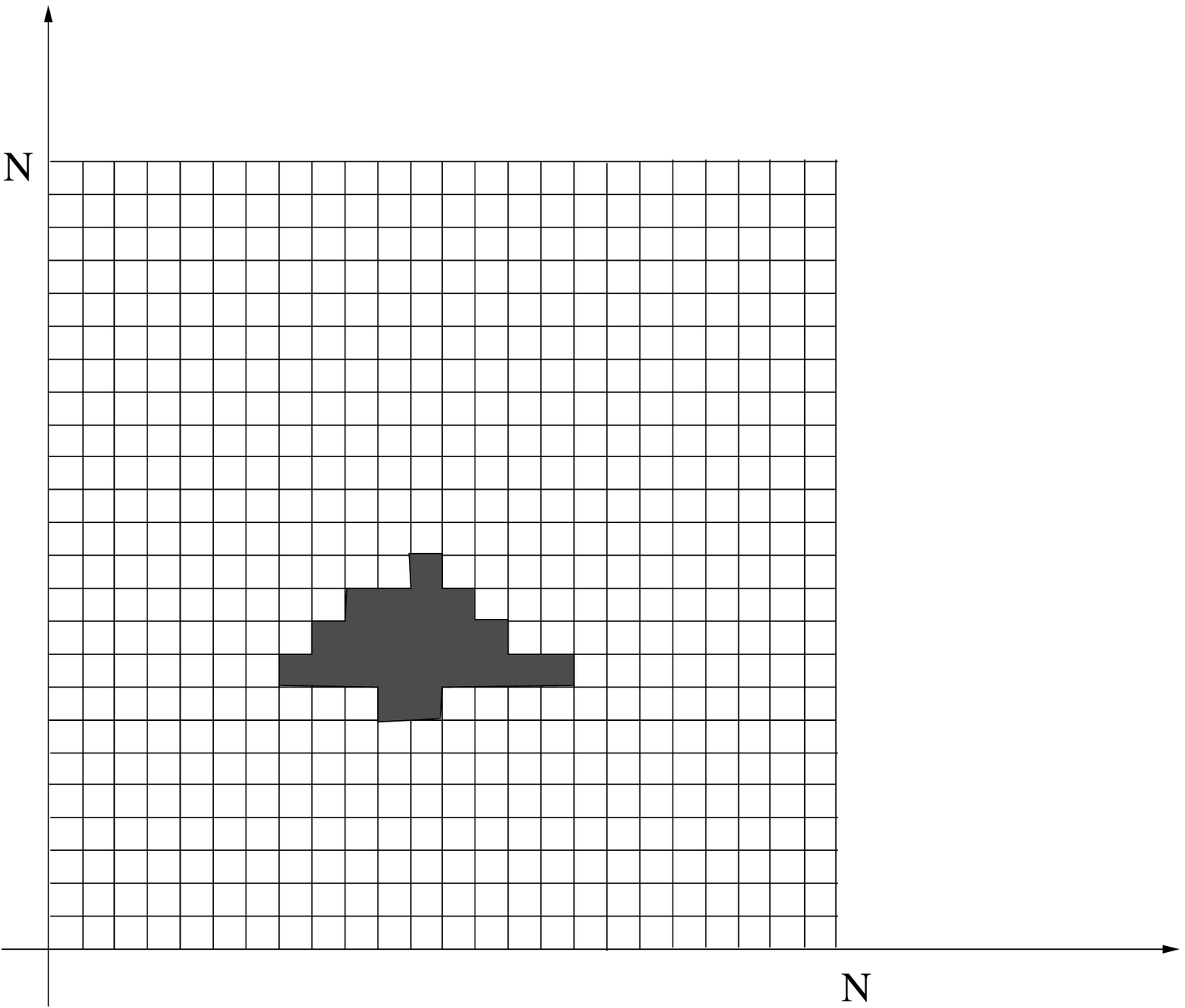}
\label{gridN}
\ee
Consider now the evolution at the subsequent instant of time,
namely after having progressed by a unit of time. 
Adding one point, $N \to N+1$, does 
produce an average geometry of a three 
sphere of radius $N+1$ instead of $N$. In the average, it is therefore like 
having added $4 \pi N^2$ ``points'', or unit cells. 
Remember that we work always with an 
infinite number of cells in an unspecified number of dimensions; when we talk 
of universe in three dimensions within a region of a certain radius, 
we just talk of the dominant geometry.
Let's suppose the position of the
wave packet jumps by steps (two cells) back, 
as illustrated in figure~\ref{gridN+1}. Namely, as time, and consequently also
the radius of the universe, progresses by one
unit, the packet moves at higher speed, jumping by two units: 
\be
\epsfxsize=6cm
\epsfbox{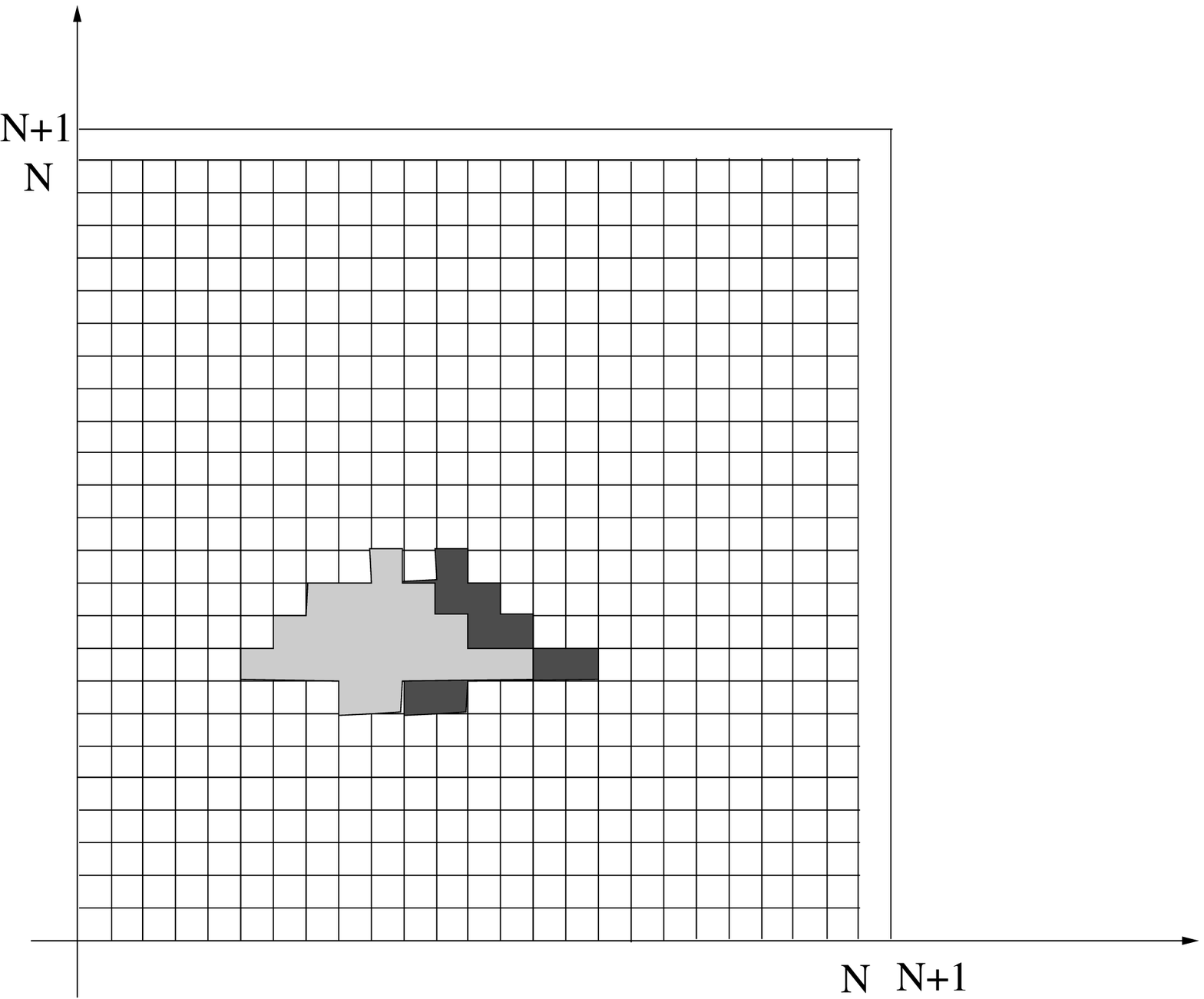}
\label{gridN+1}
\ee
Consider now the case in which the packet jumps by just one unit,
as in figure~\ref{gridN+1bis} here below: 
\be
\epsfxsize=6cm
\epsfbox{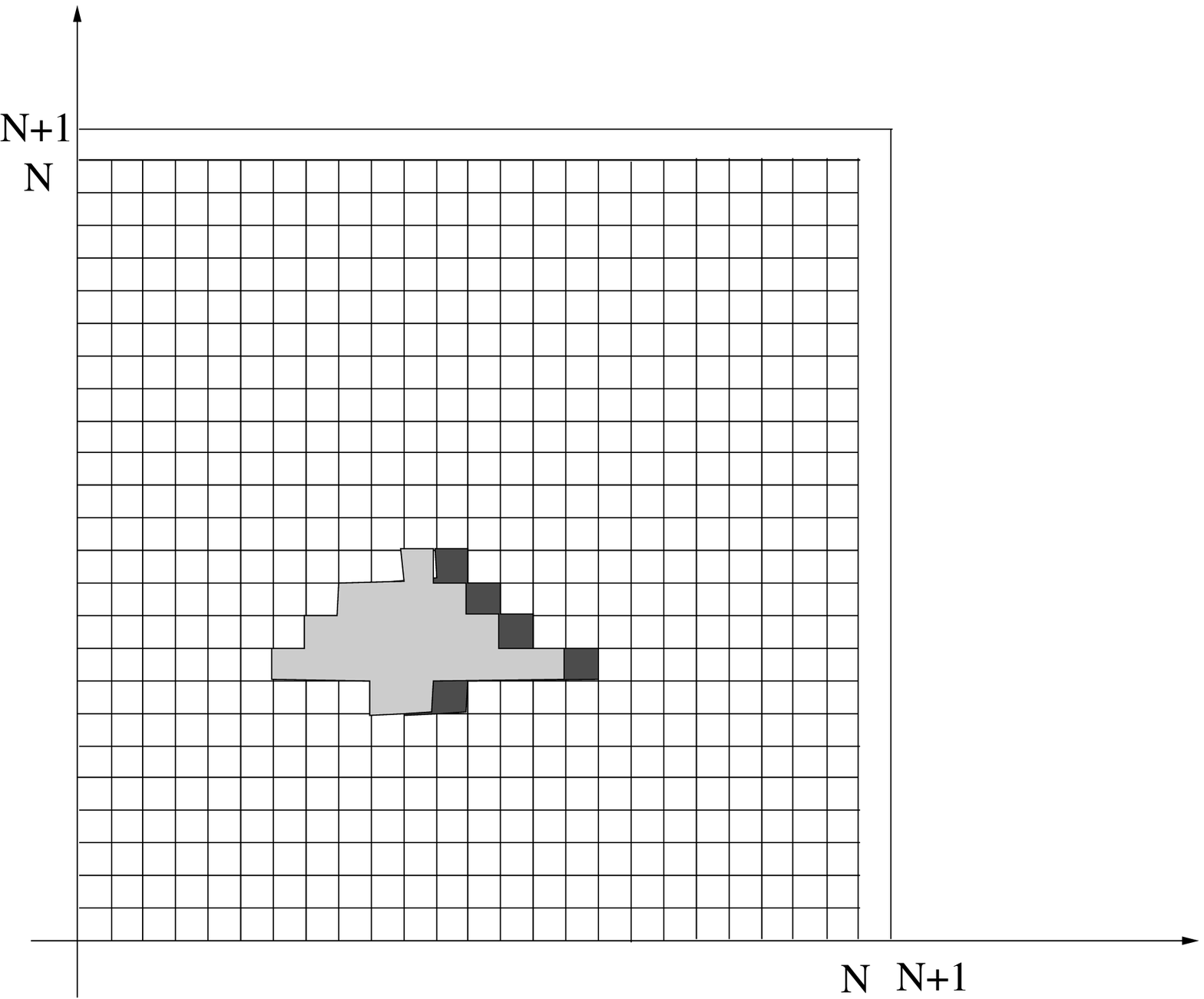}
\label{gridN+1bis}
\ee
The entropy of this latter configuration, intermediate between the first and 
second one, cannot be very different from the one of the second configuration,
figure~\ref{gridN+1}, in which the packet
jumps by two steps, because that was supposed to be the dominant configuration
at time $N+1$, and therefore the one of maximal entropy. Indeed, 
by ``continuity'' it must interpolate between step 2 and the configuration
at time $N$, that was also supposed to be a configuration of maximal
entropy. Therefore, the actual
appearance of the universe at time $N+1$ must be somehow a superposition
of the configurations 2 and 3, thereby contradicting our hypothesis that
the wave packet is non-dispersive~\footnote{If it was dispersive,
it would be something like a particle that, during its motion, ``dissolves'', 
and therefore we cannot anymore trace as a particle. It would be just
a ``vacuum fluctuation'' without true motion, something that does not carry
any information in the classical sense.}. 
Therefore, the wave packet cannot jump by two steps, 
and we conclude that the maximal speed allowed is that of expansion of the 
radius of the universe itself, namely, $c$.

According to this theoretical framework,  
the reason why we have a universal bound on the speed of light
is therefore that light carries what we call classical information. 
Information about whatever kind of event tells 
about a change of average entropy of the observed system, 
of the observer, and what surrounds and connects them too. 
The rate of transfer/propagation
of information is therefore strictly related to the rate of variation of
entropy. Variation of entropy is what gives the measure of time progress
in the universe. Any vector of information that ``jumps'' steps of the
evolution of the universe, going faster than its rate of entropy variation, 
becomes therefore dispersive, looses information
during its propagation. Light must therefore propagate at most at the rate of
expansion of space-time (i.e. of the universe itself). Namely, at the rate of  
the space/time conversion, $c$.

\subsection{The Lorentz boost}
\label{Lboost}

Let's now consider physical systems that can be identified as ``massive particles'',
i.e. localizable and which exist also ``at rest'', therefore travelling 
at speeds always lower than $c$. 
Since the phase space has a multiplicative structure, 
and entropy is the logarithm of 
the volume of occupation in this space, it is possible to separate for each 
such a system the entropy into the sum of an internal, ``rest'' entropy, and
an external, ``kinetic'' entropy. The first one refers to the structure
of the system in itself, that can be a point-like particle or an entire
laboratory \footnote{In our approach, there does not exist a strictly
point-like object. A point-like particle is an extended object of which
we neglect the geometric structure.}. 
The second one refers to the relation/interaction of this
system with the environment, the external world: its motion, the
accelerations and external forces it experiences, etc.      

\begin{figure}
\centerline{
\epsfxsize=7cm
\epsfbox{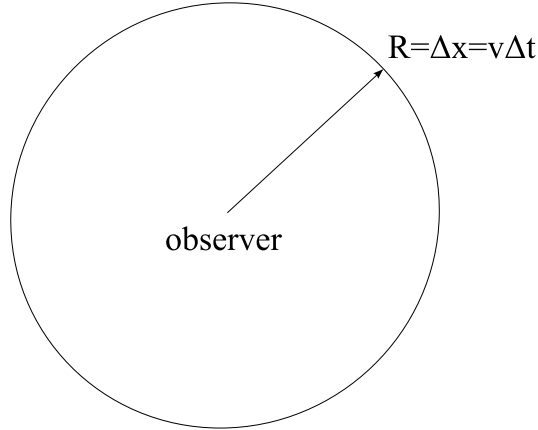}
}
\vspace{0.3cm}
\caption{During a time $\Delta t$, the pure motion ``creates'' a universe
with an horizon 
at distance $\Delta x = v \Delta t$ from the observer. As seen from the
rest frame, this part of the physical system does not exist. The ``classical''
entropy of this region is given by the one of its dominant configuration, i.e.
it corresponds to the entropy of a black hole of radius $\Delta x$.} 
\label{v-universe}
\end{figure}

Let us for a moment abstract from the fact that the actual
configuration of the universe implied by \ref{ZPsi} at any time 
describes a curved space. In other words, let's neglect the so called
``cosmological term''. This approximation can make sense at large $N$,
as is the case of the present-day physics, a fact that historically
allowed to introduce special relativity and Lorentz boosts before addressing 
the problem of the cosmological constant.
Let us also assume we can just focus our attention on two observers sitting
on two \underline{inertial} frames, $A$ and $A^{\prime}$, 
moving at relative speed $v$, neglecting everything else \footnote{In our 
theoretical framework, there is no ``external observer'': \ref{ZPsi} 
describes a universe ``on shell'', the totality of the physical world.}.  
For what above said, $v < 1$. 
An experiment is the measurement of some event that, owing to the
fact that happening of something means changing of entropy and therefore
is equivalent to a time progress, 
gives us the perception of having taken place during
a certain interval of time. Let us
consider an experiment, i.e. the detection of some event, taking place in
the co-moving frame of $A^{\prime}$, as reported by both
the observer at rest in $A$,
and the one at rest in $A^{\prime}$ (from now on we will
indicate with $A$, and $A^{\prime}$, indifferently the frame as well as
the respective observer). Let's assume
we can neglect the space distance separating the two observers, or suppose
there is no distance between them \footnote{In our scenario, 
huge (=cosmological) distances
have effect on the measurement of masses and couplings.}. 
For what above said, such a detection amounts in observing
the increase of entropy corresponding to the occurring 
of the event, as seen from $A$, 
and from $A^{\prime}$ itself. Since we are talking of the same event, the
\emph{overall} change of entropy will be the same for both 
$A$ and $A^{\prime}$.
One would think there is an ``absolute'' time
interval, related to the evolution of the universe corresponding to the
change of entropy due to the event under consideration. However, 
the story is rather different as soon as we consider \emph{time measurements}
of this event, as reported by the two observers, $A$ and $A^{\prime}$. 
The reason is that the two observers will in general attribute
in a different way what amount of entropy change has to be considered
a change of entropy of the ``internal'' system, and which amount
refers to an ``external'' change. Proper time measurements have to do with
the \emph{internal} change of entropy. 
For instance, consider the entropy of
all the configurations contributing to form, say, a clock. The part of phase
space describing the uniform motion of this clock will not be taken into
account by an observer moving together with the clock, as it will not even
be measurable. This part will however be considered by the other observer.    
Therefore, when reporting measurements of time intervals made by 
two clocks, one co-moving with $A$, and one seen
by $A$ to be at rest in $A^{\prime}$, owing to a different way of
attributing elements within the configurations building up the system,
between ``internal'' and 
``external'', we will have in general two different time
measurements.  
Let us indicate with $\Delta S$ the change of entropy as it is observed
by $A$. We can write:
\ba
\Delta S \, (\equiv \Delta S(A) ) 
& = & \Delta S ({\rm internal}\, = \, {\rm at \, rest}) 
\, + \, \Delta S ({\rm external})  \label{DeltaSS} \\
&& \nn \\
& = & \Delta S (A^{\prime}) \, + \, 
\Delta S_{\rm Kinetic}(A) \, ,  
\label{DeltaSSk}
\ea
with the identifications 
$\Delta S ({\rm internal}
\, = \, {\rm at \, rest}) \equiv \Delta S (A^{\prime})$ and
$\Delta S ({\rm external}) \equiv \Delta S_{\rm Kinetic}(A)$.
In section~\ref{eSp} we discussed how the entropy of a three sphere is 
proportional to $N^2 = E^2$. This is therefore also the entropy of 
the average, classical universe, that in the continuum limit, via the 
identification of total energy with time, can be written as: 
\be
S \; \propto \; \left( c {\cal T} \right)^2 \, ,
\label{ScT2}
\ee
where ${\cal T}$ is the age of the universe.
This relation matches with the Hawking's expression of the entropy of a 
black hole of radius $r = c {\cal T}$ \cite{Bardeen:1973gs,Bekenstein:1973ur}.
It is not necessary to write explicitly  the proportionality constant in 
(\ref{ScT2}),
because we are eventually interested only in ratios of entropies.
During the time of an event, $\Delta t$, the age of the universe passes
from ${\cal T}$ to ${\cal T} + \Delta t$, and
the variation of entropy, $\Delta S = S ({\cal T} + \Delta t)-S({\cal T})$,
is:
\be
\Delta S \; \propto \; \left( c \Delta t \right)^2 \, + \, 
c^2 {\cal T}^2 \left( {2 \Delta t \over {\cal T}}  \right) \, . 
\ee 
The first term corresponds to the entropy of a ``small universe'', 
the universe which is ``created'', or ``opens up'' around an observer
during the time of the experiment, and embraces within its horizon the entire 
causal region about the event. The second term is a 
``cosmological'' term, that couples the local physics to the history of the 
universe. The influence of this part of the
universe does not manifest itself through elementary, classical causality
relations within the duration of the event, but indirectly, through a (slow)
time variation of physical parameters such as masses and couplings,
(we refer to \cite{npstrings-2011} for a discussion of the time dependence 
of masses and couplings. See also~\cite{spi}).
In the approximation of our abstraction to the rather ideal case of
two inertial frames, we must neglect this part, concentrating the discussion
to the local physics. In this case, each experiment must be considered
as a ``universe'' in itself.
Let's indicate with $\Delta t$ the time interval as reported by $A$, and 
with $\Delta t^{\prime}$
the time interval reported by $A^{\prime}$.
In units for which $c=1$, and omitting the normalization
constant common to all the expressions like~\ref{ScT2} , 
we can therefore write:
\be
\Delta S(A) \rightarrow \langle \Delta S(A)  \rangle 
\approx (\Delta t)^2 \, ,
\label{dst}
\ee
whereas
\be
\Delta S (A^{\prime}) \rightarrow 
\langle \Delta S (A^{\prime})  \rangle
\approx (\Delta t^{\prime})^2 \, ,
\label{dstp}
\ee
and
\be
\Delta S_{\rm Kinetic} (A) ~ = ~ ( v \, \Delta t )^{2} \, . 
\label{vt2}
\ee
These expressions have the following interpretation. As seen from $A$,
the total increase of entropy corresponds to the black hole-like entropy
of a sphere of radius equivalent to the time duration of the 
experiment. Since $v=c=1$ is the maximal 
``classical'' speed of propagation of information, all the classical 
information about the system is contained within the horizon set by the 
radius $c \Delta t= \Delta t$. However, when $A$ attempts to refer this time 
measurement to what $A^{\prime}$ could observe, it knows that 
$A^{\prime}$ perceives itself at rest, and 
therefore it cannot include in the computation of entropy also the change in 
configuration due to its own motion (here it is essential that we consider 
inertial systems, i.e. constant motions). 
``$A$'' separates therefore its measurement 
into two parts, the ``internal one'', namely the one involving changes that
occur in the configuration as seen at rest by $A^{\prime}$ 
(a typical example is for instance a muon's decay at rest in
$A^{\prime}$), and a part accounting 
for the changes in the configuration due to the very being 
$A^{\prime}$ in motion at speed $v$.
If we subtract the internal changes, namely we think at the system at rest in
$A^{\prime}$ as at a point without meaningful physics 
apart from its motion in space \footnote{No internal physics means
that we also neglect the contribution to the energy/entropy due to the
mass.}, the entire information about the change of entropy is contained in 
the ``universe''
given by the sphere enclosing the region of its displacement, 
$v^2 (\Delta t)^2 ~ = ~ \Delta S_{\rm Kinetic} (A)$. In other words,
once subtracted the internal physics, the system behaves, from the point
of view of $A$, as a universe which expands at speed $v$, because the only
thing that happens is the displacement itself, of a point otherwise fixed
in the local universe (see figure~\ref{v-universe}). 
Inserting expressions \ref{dst}--\ref{vt2} in \ref{DeltaSSk} we obtain:
\be
(\Delta t)^2 ~ = ~ { (\Delta t^{\prime})^2 \over 1 - v^2 } \, , 
\ee
that is:
\be
\Delta t ~ = ~ { \Delta t^{\prime} \over \sqrt{1 - v^2}  } \, .
\ee
The time interval as measured by $A$ results to be longer by a factor 
$(\sqrt{1 - v^2})^{-1}$ than as measured by $A^{\prime}$.
We stress that, when we use expressions like ``as seen from'', ``it observes''
and alike, we intend them in an ideal sense, not in the concrete sense of
``detecting a light ray coming from what is observed''. 
We stress that in this argument the bound on the speed of information, 
and therefore of light, enters 
on when we write the variation of entropy of the ``local
universe'' as $\Delta S = (c \Delta t)^2$. If $c \to \infty$, namely, if within
a finite interval of time an infinitely extended causal region opens up around
the experiment, both $A$ and $A^{\prime}$ turn out to have access to the full
information, and therefore $\Delta t = \Delta t^{\prime}$.
This means that they observe
the same overall variation of entropy.

\subsubsection{the space boost}

In this framework we obtain in quite a natural way the Lorentz 
\underline{time} boost. 
The reason is that, for us, time evolution is directly related to entropy 
change, and we identify configurations (and geometries) through their entropy. 
The space length is somehow a derived quantity, 
and we expect also the space boost to be 
a secondary relation. Indeed, it can be easily derived from the time boost, 
once lengths and their measurements are properly defined. However, these 
quantities are less fundamental, because they are related to the classical 
concept of geometry. 
We could produce here an argument leading to the space boost. However, 
this would basically be a copy of the classical derivation within the 
framework of special relativity. The derivation of the time boost through 
entropy-based arguments opens 
instead new perspectives, allowing to better understand where relativity ends
and quantum physics starts. Or, to better say, it provides us with an 
embedding of this problem into a scenario that contains both these aspects,
relativity and quantization, as 
particular cases, to be dealt with as useful approximations.

\subsection{General time coordinate transformation}
\label{gtrans}

Lorentz boosts are only a particular case of a more general
transformation. They are valid when systems are not 
accelerated; in particular, when they are not subjected to a gravitational 
force. Traditionally, we know that 
the general coordinate transformation has to 
be found within the context of General Relativity; in that case the measure 
of time lengths 
is given by the time-time component of the metric tensor. In the absence of 
mixing with space boosts, i.e., with a diagonal metric, we have:
\be
(d s)^2 ~ = ~ g_{00} (dt)^2 \, . 
\ee 
As the metric depends on the matter/energy content through the Einstein's 
Equations:
\be
{\cal R}_{\mu \nu} - {1 \over 2} g_{\mu \nu} {\cal R} \, 
= \, 8 \pi G_N T_{\mu \nu} \, ,
\label{Eeq}
\ee
$g_{00}$ can be computed when we know the energy of the system. For instance,
in the case of a particle of mass $m$ moving at constant speed 
$\vec{v}$ (inertial motion),
the energy, the ``external'' energy, is the kinetic energy 
${1 \over 2} m v^2$, and we recover the
$v^2$-dependence of the Lorentz boost~\footnote{In the determination of the 
geometry, what matters here is not the full force experienced by the particle
but the field  in which the latter moves. The mass $m$ therefore drops out from
the expressions (see for instance~\cite{landau}).}.   

In the simple case of the previous section, we have considered 
the physical system of the wave packet as
decomposed into a part experiencing an ``internal'' physics, and a 
part which corresponds to the point of view
of the center of mass, that is a part 
in which the complex internal physics is dealt with as a point-like 
particle. The Lorentz boost has been derived as the consequence of a 
transformation of entropies. 
Indeed, our coordinate transformation is based on the same physical grounds
as the usual transformation of General Relativity, 
based on a metric derived from the energy tensor. 
Let us consider the transformation
from this point of view:
although imprecise, the approach through the linear approximation
helps to understand where things come from. In the linear
approximation, where one keeps only the first two terms of the expansion of the
square-root $\sqrt{1 - v^2/c^2}$, the Lorentz boost can be obtained from
an effective action in which in the Lagrangian appear the rest and the kinetic
energy. These terms correspond to the two terms on the r.h.s. of 
equation~\ref{DeltaSSk}. 
Entropy has in fact the dimension of an energy multiplied by a time 
\footnote{By definition, $d S = dE / T$, where $T$ is the temperature, and
remember that in the conversion of thermodynamic formulas, 
the temperature is the inverse of time.}. Approximately, we can write:
\be
\Delta S ~ \simeq ~
\Delta E \Delta t \, , 
\label{SEt}
\ee 
where $\Delta E$ is either the kinetic, or the rest energy.
The linear version of the Lorentz boost is obtained by inserting 
in~(\ref{SEt}) the expressions $\Delta E_{rest} = m$ and 
$\Delta E_{kinetic} = {1 \over 2} m v^2$. In this case, the linearization
of entropies lies in the fact that we consider the mass a constant, instead
of being the full energy of the 
``local universe'' contained in a sphere of 
radius $\Delta t$, i.e. the energy (mass) of a black hole of radius 
$\Delta t$: $m = \Delta E = \Delta t / 2$.

In our theoretical framework, the general expression of the time coordinate
transformation is:
\be
(\Delta t^{\prime})^2 ~ = ~ 
\langle \Delta S^{\prime}(t) \rangle \, - \, 
\langle \Delta S^{\prime}_{external}(t) \rangle \, .
\label{SpSS}
\ee
Here $\Delta S^{\prime}(t)$ is the total variation of entropy of the ``primed''
system as measured in the ``unprimed'' system of coordinates:
$\langle \Delta S^{\prime}(t) \rangle \, = \, (\Delta t)^2$.
We can therefore write expression~\ref{SpSS} as:
\be
(\Delta t^{\prime})^2 ~= ~ \left[ 1 - \mathcal{G}(t) \right] (\Delta t)^2 \, , 
\label{dtdt2}
\ee  
where:
\be
\mathcal{G}(t) ~ \stackrel{\mathrm{def}}{=} ~ 
{\Delta S^{\prime}_{external}(t) \over (\Delta t)^2}   \, . 
\label{Gdef}
\ee
With reference to the ordinary metric tensor $g_{\mu \nu}$, we have:
\be
\mathcal{G}(t) ~ = ~ 1 - g_{00}  \, .
\ee
$\Delta S^{\prime}_{external}(t)$ is the part of change of entropy of 
$A^{\prime}$ referred to by the observer $A$
as something that does not belong to the rest frame of $A^{\prime}$. It can be
the non accelerated motion of $A^{\prime}$, as in the previous example, or more
generally the presence of an external force that produces an acceleration.
Notice that the coordinate transformation \ref{dtdt2} starts with a constant
term,~1: this corresponds to the rest entropy term expressed in
the frame of the observer. For the observer, the new time metric is always
expressed in terms of a deviation from the identity.

By construction, \ref{Gdef} is the ratio between the metric in 
the system which is observed and the metric in the system of the observer.
From such a coordinate transformation we can pass to the metric of
space-time itself, provided we consider the coordinate transformation 
between the metric $g^{\prime}$ of a point in space-time,
and the metric of an observer which lies on 
a flat reference frame, whose metric is expressed in flat coordinates. 
We have then:
\be
1 \, - \, \mathcal{G}(t)  ~ = ~ { g_{00}^{(\prime)} \over g^{(0)}_{00} \, 
= \eta_{00} = 1} \, .
\ee
As soon as this has been clarified, we can drop out the denominator and we 
rename the primed metric as the metric tout court.

\subsection{General Relativity}
\label{genRel}

Once the measurement of lengths
is properly introduced, as derived from a measurement of configurations
along the history of the system, 
it is possible to extend the relations also to the transformation of
space lengths. This gives in general the components of the metric tensor
as functions of entropy and time. In classical terms, whenever this 
reduction is possible, this can be rephrased into a dependence on energy 
(energy density) and time. They give therefore a generalized, integrated version
of the Einstein's Equations.
Let's see this for the time component of the metric. We want
to show that the metric $g_{00}$ of the effective space-time corresponds
to the metric of the distribution of energy in the mean space, i.e., 
in the classical limit of effective three-dimensional space as it
arises from \ref{ZPsi}. This will mean
that the geometry of the motion of a particle within this space is the
geometry of the energy distribution. In particular, if the energy is 
distributed according to the geometry of a sphere, so it will be the
geometry of space-time in the sense of General Relativity.
To this regard, we must remember that: 

\noindent
${\rm i})$  All these arguments make only sense in the ``classical
limit'' of our scenario, namely only in an average sense, where 
the universe is dominated by a configuration that can be described in 
classical geometric terms. It is in this limit that the universe
appears as three dimensional. Configurations which
are in general non three-dimensional, non-geometric, possibly tachyonic, and,
in any case, configurations for which General Relativity and Einstein 
Equations don't apply, are covered under the ``un-sharping'' relations of the
Uncertainty Principle.
All of them are collectively treated as ``quantum effects'';

\noindent
${\rm ii})$
In the classical limit, \underline{nothing}
travels at a speed higher than $c$. As during an experiment
no information comes from outside the 
local horizon set by the duration of the experiment itself, to cause
some (classical) effects on it, any consideration
about the entropy of the configuration of the object under consideration
can be made ``local'' (tachyonic effects are taken into account
by quantization). That means, when we consider the motion of an
object along space we can just consider the local entropy, 
which depends on, and is determined by, 
the energy distribution around the object.

Having these considerations in mind,
let us consider the motion of a particle, or, more precisely, a 
non-dispersive wave-packet, in the mean, three-dimensional, classical space. 
Consider to perform a (generally point-wise) coordinate transformation to
a frame in which the metric of the energy distribution external 
to the system intrinsically building the wave packet in itself is flat,
or at least remains constant. As seen from this set of frames, along the motion
there is no change of the (local) entropy around the particle, and the right
hand side of \ref{Gdef} vanishes, implying that also the metric of the motion
itself remains constant (remember that \ref{Gdef} in
this case gives the \emph{ratio} between metrics at different points/times).
This means that the metric of the energy distribution and the metric of the motion
are the same, and proves the equivalence of \ref{SpSS} and \ref{Gdef}
with the Einstein's equations \ref{Eeq}.

If on the other hand we keep the frame of the observer fixed,
and we ask ourselves what will be the direction chosen by the particle
in order to decide the steps of its motion, the answer will be:
the particle ``decides'' stepwise to go in the direction that maximizes
the entropy around itself. Let us consider configurations in which 
the only property of particles is their mass
(no other charges), so that entropy is directly related to the ``energy
density'' of the wave packet. In this case,  
between the choice of moving toward 
another particle, or far away, the system will proceed in order to increase
the energy density around the particle. Namely, moving 
the particle toward, rather than away from, the other particle,
in order to include in its horizon also the new system. This is how 
gravitational attraction originates in this theoretical framework.

In order to deal with more complicated cases, such as those in which
particles have properties other than just their mass 
(electro-magnetic/weak/strong charge), we need a more detailed description of 
the phase space. In principle things are the same, but the appropriate
scenario in which all these aspects are taken into account is the one
in which these issues are phrased and addressed
within a context of (quantum) String Theory.
This analysis, first presented in Ref.~\cite{spi}, will be discussed 
in~Ref.~\cite{npstrings-2011} in an updated and corrected form.

\subsection{The metric around a black hole}
\label{mbkhl}

Let us consider once more the general expression relating the evolution of a system 
as is seen
by the system itself, indicated with $A^{\prime}$,
and by an external observer, $A$, expressions~\ref{DeltaSS} and \ref{DeltaSSk}.
In the large-scale, \emph{classical} limit, the variations of entropy
$\Delta S(A)$ and $\Delta S (A^{\prime})$ can be written in terms
of time intervals, as in \ref{dst} and \ref{dstp}, in which
$t$ and $t^{\prime}$ are respectively
the time as measured by the observer, and the proper time of the system
$A^{\prime}$.
In this case, as we have seen expression~\ref{DeltaSSk} can be written as
$
(\Delta t^{\prime})^2 =  (\Delta t)^2 \, - \, 
\langle \Delta S^{\prime}_{\rm external}(t) \rangle
$ (see expression~\ref{SpSS}),
and the temporal part of the metric is given by:
\be
g_{00} ~ = ~ 
{\langle \Delta S^{\prime}_{\rm external}(t) \rangle 
\over (\Delta t)^2} \, - \, 1   \, . 
\label{g00}
\ee
As long as we consider systems for which $g_{00}$ is far
from its extremal value, expression \ref{g00} constitutes a good approximation
of the time component of the metric. 
However, a black hole does not fall within the domain of this
approximation.
According to its very (classical) definition, 
the only part we can probe of a black hole is the surface at the horizon.
In the classical limit the metric at this surface vanishes:
$g_{00} \to 0$ (an object falling from outside toward the black hole
appears to take an infinite time in order to reach the surface).
This means,
\be
\langle \Delta S_{\rm external} \rangle
~ \approx ~\propto \, \left( \Delta t  \right)^2 \, . 
\ee 
However, in our set up time is only an average, ``large
scale'' concept, and only in the large scale, classical limit we
can write variations of entropy in terms of progress of a time coordinate
as in \ref{dst} and \ref{dstp}.
The fundamental transformation is the one given in expressions 
\ref{DeltaSS}, \ref{DeltaSSk}, and the term $g_{00}$ has only to be understood
in the sense of:
\be
\Delta S (A^{\prime}) \, \longrightarrow \,  
\langle \Delta S (A^{\prime}) \rangle \, \equiv \, 
\Delta t^{\prime} g_{00} \Delta t^{\prime} \, . 
\label{dsag}
\ee
The apparent vanishing of the metric~\ref{g00}
is due to the fact that we are subtracting contributions from the
first term of the r.h.s. of expression~\ref{DeltaSSk}, namely 
$\Delta S(A^{\prime})$,
and attributing them to the contribution of the
environment, the world external to the system of which we consider the proper 
time, the second term in the r.h.s. of \ref{DeltaSSk}, 
$\Delta S_{\rm external}(A)$. 
Any physical system is given by the superposition of an
infinite number of configurations, of which only the most entropic ones
(those with the highest weight in the phase space) build up
the classical physics, while the more remote ones contribute to 
what we globally call ``quantum effects''. Therefore,
taking out classical terms from
the first term, $\Delta S (A^{\prime})$, the ``proper frame'' term,
means transforming the system the more and more into a ``quantum
system''. In particular, this means that the mean value of whatever observable
of the system will receive the more and more contribution 
by less localized, more exotic, configurations, 
thereby showing an increasing quantum uncertainty.
In particular, the system moves toward configurations 
for which $\Delta x \rightarrow \gg 1 / \Delta p$.
Indeed, one never reaches the condition of vanishing of \ref{dsag}, because,
well before this limit is attained, also the notion itself of space, and time,
and three dimensions, localized object, geometry, etc..., are lost. The
most remote configurations in general do not
describe a universe in a three-dimensional space, and the ``energy'' 
distributions are not even interpretable in terms of ordinary observables.
At the limit in which we reach the surface of the horizon, the black hole will
therefore look like a completely delocalized object~\cite{blackholes-g}.

\section{String Theory}
\label{stringT}

In the previous sections we have derived what is in the average
the dominant geometry of the 
universe. To the whole resulting geometry contribute also
an infinite bunch of less entropic configurations, responsible for
``minor'' deviations, ``perturbations'' of the dominant configuration,
what we called ``inhomogeneities'' of the energy distribution.
In order to investigate what kind of perturbations do we have, it is 
convenient to map the combinatorial problem into a description of the
world in terms of propagating fields and particles. This is also a way to make
contact with the common approach to physics and observables.

Let us consider the geometry of the universe resulting from~\ref{ZPsi}. 
In its grounds, what 
we have is a distribution of amounts of energy along space, with the time 
ordering $E$. The rule of the dynamics is that the universe "evolves" 
in such a way that at any time what we have is basically the configuration of 
maximal entropy, plus ``quantum'' corrections which modulate the
smoothness of the dominant geometry into a bunch of propagating and
interacting ``wave packets''. 
For small intervals of "time" $\delta t$, 
the evolution can be approximated by a continuous evolution, parametrizable in 
terms of interactions and propagation of fields and particles. Indeed, a 
priori it is not at all obvious that such a description is possible or makes 
sense: if wave packets disperse too rapidly, such an approximation does not 
work. It makes sense only at a sufficiently large age of the universe
(large $N$). It is on the other hand 
important to notice that it remains just an approximation, and that 
a description in terms of 
fields, particles and their interactions, whatever this description may be,
is not necessarily the only possible description of the physical reality,
in the sense that, a priori, one may think to organize the universe
described by 
of \ref{ZPsi} also in another conceptual framework. 
Our mental organization in terms of the degrees of freedom of particles and 
fields is grounded on historical reasons, and finds here a justification
in the separation of the physics arising from \ref{ZPsi} 
into a dominant configuration plus corrections falling within the 
domain of the Heisenberg's Uncertainty. This means that particles
and fields that mediate their interactions will be quantum objects.
The Heisenberg Uncertainty allows us to
keep under control the approximation implicit in this organization
of physical phenomena, and account for its lack 
of precision.

In the world described by \ref{ZPsi}, a map to a description
in terms of quantum fields and particles makes only sense ``locally'', where
locality must be intended
not only in the sense of space but also in the sense of time. Any embedding
into a theory of time-dependent propagating fields and particles
framed in an infinite space-time is only formal: the map
must be ``updated'' at any time of the evolution of the universe, something
which can be kept into account by introducing time-dependent masses and
couplings.

We want now to see what are the conditions such a map
must satisfy in order to faithfully represent the physics of the combinatorial
framework. At the end of our discussion, we will conclude that such a map
not only exists, but that it is \emph{unique}, in the sense that the conditions
determine it uniquely.

\subsection{Mapping to quantum fields}
\label{mqf}

Let us summarize the key points at the base of the map to quantum fields
we are looking for.

$\bullet$ A first thing we want to show is that the ``canonical'' form
of the Uncertainty Principle, namely the inequality \ref{HUP} (i.e.
in the appropriate normalization $\Delta E \Delta t \geq 1/2$, 
which in a relativistic context
goes together with $\Delta P \Delta x \geq 1/2$),
implies, and is implied by, only one dimensionality of space, with
a well defined geometry.
In our combinatorial construction, we have seen that we obtain a "classical" 
D=3 dimensional space, \emph{plus} 
the Heisenberg Uncertainty.
The dimensionality of space becomes D=3+1 once we implement the "time" E
in a time coordinate suitable for a field theory description. 
Taking this into account, what we have seen is that:
\be
{\rm combinatorial ~ scenario}   ~ \Rightarrow ~ \left[ D = 3+1 \right]\, 
\cup \, 
\left[ \Delta E \Delta t \geq 1/2 \right] \,  .
\label{csimp}
\ee
This means also that:
\be 
\Delta E \Delta t \geq 1/2 ~ \Leftrightarrow ~ D = 3+1 \, .
\label{Hd3}
\ee 
Let us suppose in fact by absurd that   $\Delta E \Delta t \geq 1/2  \Leftrightarrow
D  \neq 3+1$. Then, in the sum of the rests considered to derive the 
uncertainty (see section~3), 
the ratio between weight of the classical and
weights of quantum configurations is different, something
that would lead to a different uncertainty. 
But there is more: $\Delta E \Delta t \geq 1/2$ 
not only is uniquely related to the dimensionality, but also to 
the geometry of space, 
because geometries different from the sphere have different entropy, and 
therefore different weight, leading to a different uncertainty. 
This means that the relation 
$\Delta E \Delta t \geq 1/2$  not only fixes dimension 
and main geometry, but also the spectrum of the theory.

$\bullet$ A second point to remark is that,
if we look for a description of this world in terms of quantum particles 
with wave-like behavior, their masses must correspond to momenta of a certain 
space, i.e. to the inverse of appropriate radii. In order to introduce higher 
masses than just the "cosmological mass", given by
the inverse of the radius of the 
universe, we must introduce an internal space. In this way, instead of  
$m = 1/R$ we can have $m = 1/\langle  R \rangle$, where 
$\langle  R \rangle = \sqrt[p]{R \times r_1 \times r_2 \cdots \times r_{p-1}}$,
\label{rri}
the $r_i$, $1 \leq i \leq p-1$ being internal radii. In the particular case in 
which all these turn out to be of elementary (= minimal) size, 
$r_i=1$ $\forall i$, the mass expression reduces to 
$m_{(p)} = 1 / \sqrt[p]{R}$. 
We need therefore a relativistic quantum field theory with internal dimensions.

\subsubsection{How many dimensions do we need?}
\label{ndim}

How many internal dimensions do we need?
We want to describe all the possible perturbations of the geometry
of a sphere in three dimensions, as due to fields and particles 
that propagate in it. Notice that it is not a matter of building a
set of fields \emph{framed} in a certain space, i.e. functions of
space-time coordinates. It is a matter of promoting the deformations
of the geometry themselves to the role of fields.
One may think at a description in terms of vector fields.
Once provided with a time
coordinate, the three-sphere $\times$ the time coordinate, 
which can be considered the 
D = 3+1 ``background'' space, corresponds to vector fields
possessing an $SO(3,1)$ symmetry. However, 
we must have both bosons and fermions. Fermions are needed because
we want a quantum relativistic description of fields. It is relativity what
leads to the introduction of spinorial representations of space.
This does not mean
we need bosons and fermions in equal number, nor even that they must
have the same mass (implying supersymmetry of the theory): supersymmetry
is not a symmetry of the real world (in the sense of an exact symmetry).
In terms of spinorial representations, $SO(3,1)$ is 
locally isomorphic to $SU(2) \times SU(2)$, a group
with 3+3 generators, which, once parametrized in terms
of bosonic fields, correspond to a space with six bosonic
coordinates. One would like to conclude that,
in order to have both a vectorial \emph{and} 
a spinorial representation of the background space with all its perturbations 
we need therefore the original 3+1 \emph{plus} 3+3 internal coordinates. 
With six internal dimensions it seems we are sure that whatever internal 
configuration can be mapped to a 
configuration of space-time, allowing for a non-trivial (and complete) 
mapping between the "fiber" and the "base" space, ensuring to
have a non-degenerate and complete description of all the perturbations. 
There is however a subtlety: our theory 
has also a minimal length, "1". This ``cut-off'' 
can be viewed as a specific value $g_0$ of a coupling $g$. Indeed, it is by definition the 
"gravitational" coupling of the theory (more on this point later). 
Unfreezing this one results in a new 
dimension, $g_0 = 1 \to g=x$. We need therefore at least 11 dimensions. 
Since this space is in general curved, there is one more parameter
of the theory, the curvature. This discriminates between geometries, and as
such is also part of the definition of the coupling of this geometric field
theory. It is not an independent parameter, being a function of all the
coordinates. However,
this coupling too can be viewed as a coordinate,
or, to better say, it may turn out convenient to embed it into a flat,
independent coordinate:
if we want to give a representation of an 
11-dimensional curved space in terms of \underline{flat} 
coordinates, we need 12 coordinates
\footnote{This is what gives the impression that the fundamental theory lives
in twelve dimensions (See for instance the works on F-theory, 
first proposed in~\cite{Vafa:1996xn}).}.

\subsubsection{T-duality}

$g = g_0 =1$ is a self-dual point. Any kind of 
"unfreezing", as a matter of fact simply "projects" onto 
one of the two possible 
decompactifications: either $x < 1$ or $x > 1$. In
the perturbative limit, $x \to 0$ or $x \to \infty$ respectively. 
This is a real projection, in the sense that at the limit 
part of the physical content is ``lost'':
it is in fact not possible to simply establish that a certain value of $g$,
say $g = g_0$, is the
cut off of the theory, the ``boundary'' value, above or below which the 
coordinate extends. Setting $g$ to a chosen, finite value $g_0$, 
to be the size in comparison to which all other ones are
measured, makes only sense if in the theory 
we have also non-trivial values of couplings \emph{above and below} $g_0$.
Otherwise, in practice $g_0$ would be a free, running parameter: its actual
value would be meaningless, being possible to scale it out by an
unphysical overall rescaling. As we will see, this in practice means that
the field theory representation will necessarily have strongly \emph{and}
weakly coupled sectors.
In principle, the two decompactification limits,
$g \to 0$ and $g \to \infty$, are equivalent. This does not mean
that the theory is necessarily invariant under T-duality, as is called 
the symmetry under inversion of the value of the coordinate (in the case of
a coupling, one speaks more properly of S-duality),
and indeed it must not be: if it was T-duality invariant,
then T-duality would be an unphysical transformation, mapping the
theory into itself. From the point of view of the phase space
(or of the configurations in the combinatorial scenario), this would mean
that all the configurations of the phase space possess this symmetry. The
latter could then be reabsorbed into a redefinition of
the elements of the phase space, not as single configurations
but as classes of configurations, given by the orbits of T-duality.
Although not an exact symmetry, it is however essential
that the complete theory must contain both the
``sectors'' related by T-duality, which are 
non-perturbative with respect to each other. 
Therefore, in general in a perturbative construction
not all the states are visible.

\
\\

To summarize, what we need is a quantum relativistic
field theory which is perturbative in ten dimensions,
is endowed with some kind of T-duality mechanism, 
and accounts for a vectorial and
spinorial realization of the fields

Any supersymmetric quantum string theory provides such
a realization because, in any of its perturbative realizations, 
it corresponds to all the above mentioned criteria,
and introduces spinors starting from the embedding 
of what, according to a result of Haag, Sohnius and Lopuszanski
(see~\cite{wessbagger}), is the most general graded Lie algebra compatible with 
a relativistic quantum field theory.
Supersymmetry is in general not a symmetry of the theory
derived from \ref{ZPsi}, which is always defined on a finite volume,
and is not a symmetry of every string vacuum either. 
However, perturbative field constructions are always realized in
a decompactification limit, and
every string vacuum can be constructed by starting from a 
vacuum which in this limit is supersymmetric, 
by reduction via projection/compactification on more
curved/singular spaces. Therefore, the supersymmetric one is the
most symmetric string configuration (supersymmetry is the most general 
extension of the Poincar\'e algebra). From the discussion we will
present in section~\ref{mimacro}, one derives that this is also
the less entropic in the string phase space among the
whole tower of derived conpactifications that descend from it. 
Supersymmetric string theory is therefore the ground construction
from which one starts to investigate the mapping of the combinatorial 
formulation in terms of a description based on quantum relativistic 
fields.

The existence of
a minimal length is naturally embedded in closed string theory
through its ``T-duality'' symmetry in any of its coodinates,
which shows up upon compactification onto circles
\footnote{Any 
perturbative string construction has its own proper length
$\ell_{s}$. It is only under string-string duality that,
in the most entropic string vacuum, one can match all these lengths
and identify them with the Planck length
$\ell_{\rm Pl}$.}. Type I string is obtained as an orientifold
of type II string \cite{Sagnotti:1987tw}, and as such doesn't possess
an explicit T-duality of target space coordinates upon toroidal
compactification. Nevertheless,
its physical content is equivalent to the one of the string
with T-duality, because one ensures to remain within 
the space of all the other string constructions
by enforcing extra sectors required by cancellation of anomalies.
These precisely correspond to what in the other constructions
are the ``T-dual'' or ``S-dual'' sectors.
We will discuss later the issue about uniqueness of string theory, 
namely the question whether all different types of string construction
are indeed slices of the same theory.

Perturbative superstring theory is realized precisely
in ten dimensions. The reason why this is the critical dimension
is apparently unrelated to the way we predicted this number of dimensions
in section \ref{ndim}. However, the fact that 10 dimensions arise
in string theory only once fermions are embedded, through the most general
extension of the Poincar\'{e} algebra, is a hint on the fact that there must be
a deep relation between this property, and the fact that quantum 
superstring theory bears already built-in in its construction 
the information that
it is the theory ``thought of'' in order to eventually describe the
vectorial and spinorial geometries of a four-dimensional quantum 
world.

Hidden behind the
technicalities of the properties of conformal field theory, there is
a deeper level of relations between
quantization and dimension of the most general vectorial and spinorial
realization of geometries, as discussed in section~\ref{ndim},
which ultimately relies on the logical 
structure we have introduced through our combinatorial approach.
String theory is built as a geometric theory endowed with a quantization
principle, which implements the Heisenberg's Uncertainty Relations, under which
all ``non-dominant'' configurations are ``covered''.
This procedure encodes therefore a choice of
``starting point'' for the approximation of the configurations, three
space dimensions, plus a rule implementing the ignorance due to
the fact of neglecting
the rest, the Uncertainty Relations.
In our set up, the space/momentum (and time/energy) relations appears
precisely in the form of the Heisenberg's inequality. 
This is related to a specific choice of the scaling of energy $\sim N$
as compared to the scaling of space, which appeared as the natural one.
We expect therefore that \emph{quantum}
string theory should be non-anomalous when built out of a number
of coordinates which allows a correspondence of these descriptions. 
Alternatively, one could think to choose a different
relation between ``energy'' $N$ and radius, or time. We would then get
a different uncertainty relation, of the type:
\be
(\Delta X)^{\alpha} \Delta p \, \geq \, {1 \over 2} \hbar \, ,
\ee    
where $\alpha$ is an exponent, $\alpha \neq 1$. In this case,
quantum string theory would be non-anomalous in a different number
of dimensions.

\subsection{Entropy in the string phase space}
\label{mimacro}

In order to
represent a mapping of our combinatorial problem at any finite time,
we must consider also string theory as
living on a \emph{compact space}.
This implies that it must be considered in
a \underline{non-perturbative} regime, where, in particular, owing
to compactness of the space, supersymmetry is broken. 
Considering string theory as \emph{defined} on a compact space, and
viewing infinitely extended space only as a limiting case of
a compact space, entails a \emph{deep change of perspective}, full of
consequencies for the interpretation of things that we compute 
in string theory. For instance, owing to the
always broken supersymmetry, 
the string partition function does not vanish. Owing to the lack of
translational invariance \emph{in the very definition} of the theory,
there is no inverse four-volume factor in the normalization of 
mode expansions and therefore on string amplitudes. As a consequence,
the vacuum energy one calculates on
the string vacuum with the usual techniques of taking the vacuum mean value
of the partition function does not compute an energy density but 
a global energy. In order to obtain the vacuum energy density one must
divide this quantity by a volume factor corresponding to
the actual radius of the horizon of the
universe in the dominant configuration. Precisely this volume factor
allows to obtain in a natural way and without fine tuning the correct 
value of the cosmological constant (see \cite{npstrings-2011},
cfr. also~\cite{spi}).

Consider now the collection of non-equivalent string constructions.
In its whole, this "world of string constructions" is the
string counterpart of the phase space of the combinatorial configurations
of the universe, of which it constitutes a representation. We will
call it the string phase space.
The physics of this "universe" will be obtained by the analogous 
of~\ref{ZPsi} for string constructions. This means that
we will have to take the sum over all string vacua, weighted 
by their volume of occupation in the string phase space.
First of all: what is now the counterpart of keeping fixed the
total energy $E$? This cannot be easily identified with the
time of the string target space, as in this description
time is not compact. In principle, it is possible to consider the
target space coordinate ``$X_0$'' as being compact, but this does not
solve our problem of correctly mapping the combinatorial problem of
summing configurations at fixed total $E$. Indeed, on the combinatorial
side $E$ is also the \emph{radius} of the classical space, 
i.e. the space on top
of which we build up the quantum theory. In that case, fixing the radius
means also fixing the volume of space, because the geometry is fixed to be
predominantly a three sphere. On the string side, the string target space
does not distinguish a priory between three selected space coordinates, the 
representatives of the three-dimensional physical space, and the coordinates
of an ``internal'' space: they start all on the same footing, and a possible
selection of just three as the physical space must come by itself,
in the string configuration that dominates in the string phase
space, not as an external input. 
This tells us that, when dealing with string vacua, instead of comparing
constructions at fixed $E$, we must compare constructions at fixed volume
of the target space, no matter how many dimensions this describes
(the concept itself of what are to be considered as dimensions of 
physical space, and what ``internal'' dimensions, will be clarified in the 
following). On the string space, \ref{ZPsi} becomes therefore:
\be
{\cal Z}_V \, = \, \int_V {\cal D} \psi \, {\rm e}^{S(\psi)} \, ,
\label{Zint}
\ee
where $\psi$ indicates now a whole, non-perturbative string configuration, 
and $V$ is the volume of the target space,
intended as ``measured'' in the duality-invariant Einstein frame. 
In order to understand what kind of ``universe'' comes out
of all the possible string configurations 
we must therefore find out those that correspond to the maximal entropy in
the phase space, at any fixed volume.

String configurations on a compact space are obtained by compactifying
the target space on certain spaces, which may be continuous and
differentiable, or even singular. In any case, a compactification
leads to a reduction of the symmetry of the initial theory.
This is obvious by construction in the case of orbifolds: the more
singular is the orbifold on which the string is compactified, the higher
is the amount of symmetry reduction.
But also the other kinds of compactifications
can be viewed as generalized cases of symmetry reduction, from
the most symmetric,
and therefore the very initial, geometry of the string target space.
In principle, this should
be a higher-dimensional sphere. However, owing to the
decompactification implicit in any perturbative construction, the sphere
unfolds to a flat space (one may think that the coupling coordinate
is precisely the curvature of the whole space), and flat is how the space appears perturbatively. In this limit, the
most symmetric compactifications are not spheres but tori.
The symmetries of the target space reflect then on the entire
string spectrum, in the sense that, if the initial and the derived
target space have symmetry
represented by the groups $G$ and $G^{\prime}$ respectively, such that
$G^{\prime} = G /H$, and the initial spectrum has
a symmetry $\tilde{G}$, the new spectrum will have a symmetry
$\tilde{G}^{\prime}$ such that
$\tilde{G}^{\prime} =  \tilde{G} / \tilde{H}$, where $\tilde{H} \cong H$.
We may say that both $H$ and $\tilde{H}$ are representations
of the same group, that for simplicity we call $H$~\footnote{Notice
that we are \underline{not} saying that $G \cong \tilde{G}$
nor $G^{\prime} \cong \tilde{G}^{\prime}$!}.
Let's consider the action of the group $H$ on the initial
string configuration, that we call $\Psi$, 
i.e. the action on its target space and on the spectrum. Consider then
the configuration $\Psi^{\prime}$ obtained by
modding by $H$. 
Elements $h \in H$ map $\Psi^{\prime}$ to
$\Psi^{\prime \prime} = h \Psi^{\prime}$, physically
equivalent to $\Psi^{\prime}$, in the sense that, by construction, there
is a one-to-one map between $\Psi^{\prime}$ and $\Psi^{\prime \prime}$
which simply re-arranges the degrees of freedom. 
From a physical point of view, there are therefore $|| H  ||$
ways of realizing this configuration. The occupation in the whole
phase space is therefore enhanced by a factor $|| H  ||$ as compared
to the one of $\Psi$. By reducing the symmetry of the target space, 
we have enhanced the
possibilities of realizing a configuration in equivalent ways
in the string phase space, in the same
sense as, in the two-cells example of section~\ref{setup}, 
by assigning different colours,
black and white, we have the possibility of realizing the configuration
``one-white/one-black'' in more ways than ``white-white'' or ``black-black''.
The string construction with the highest entropy will therefore be the
one obtained through the highest amount of symmetry reduction.

\subsection{Uniqueness of String Theory: summary of line of thinking}
\label{1stheory}

We have discussed how the combinatorial scenario introduced in section~\ref{setup}
constitutes the most general logical structure one may think about, and, 
as such, is therefore
unique. We have also seen that it implies a three-dimensional ``classical''
relativistic world plus quantum corrections, 
and that, if translated into a description
in terms of propagating fields (and particles) it implies a description in
terms of quantum strings. However, the uniqueness of the combinatorial
scenario does not imply a priori uniqueness of the string scenario.
That is, a priori the various types of superstring construction (heterotic,
type II, type I) could
correspond to separate sets of combinatorial configurations. 
Eleven dimensional supergravity too could be an a priori unrelated theory
(indeed, being gravity in itself not a quantum theory, 11-d supergravity
would be only a part of such a theory). Each one of the various string types 
would generate a chain of ``descendants'' obtained
via compactification and projection (orbifoldization or other kinds of
compactification), starting from the most symmetric (supersymmetric) one
and going toward the one compactified on the most singular 
(= less symmetric) space, something that would produce a configuration 
which would be the one of
highest entropy in the string phase space among the configurations 
of the particular subset corresponding the
specific chain of derived string vacua. A priori there could therefore be
several ``highest entropy'' configurations, one for each type of string. 
Indeed, the evaluation in itself 
of the weight of a certain configuration in the string phase space, obtained
as discussed in section \ref{mimacro} 
via comparison of the volumes of the symmetry groups, is something clearly
defined within each family chain. Comparing configurations belonging
to different chains of ``descendants'' is not obviously unambiguous in
itself: a priori there could be string vacua of different types possessing
the same degree of symmetry reduction. On the other hand, the
definition of weight, and entropy of a configuration, becomes unambiguous
also in the string phase space once this is put in relation with the
phase space of combinatorial configurations, namely once by symmetry of a
configuration one intends the symmetry of the full spectrum, i.e. the whole
physical content~\footnote{One must pay attention to not confuse the
symmetry of the string configuration, related to the symmetries of its 
target space, and the symmetry of a distribution of energy, as in the
combinatorial approach. Any string configuration corresponds to 
the superposition of an entire collection of combinatorial distributions, and 
the mapping between the two pictures implies a re-interpretation of 
superpositions of energy distributions in terms of quantum fields
and uncertainties. Nevertheless the two concepts are related, although in
a rather involved way.}. In this case, it appears clear that equal symmetry
means also equal physical content, i.e. complete equivalence of
configurations. It appears also clear that, working in a compact target space,
for each type of string construction one can reduce the 
symmetry only until one reaches a minimum of symmetry/maximum of 
entropy, because the combinatorial is finite, the volume of
the maximal symmetry is finite, and cannot be reduced ``ad libitum''. 

Let us consider the ``highest entropy'' configurations belonging
to two different chains of descendants, $X^{\rm i}_{\rm max}$ and
$X^{\rm j}_{\rm max}$, obtained by a reduction from the
most symmetric ``parent'' vacuum, $X^{\rm i}_0$ and $X^{\rm J}_0$
respectively, through a chain of projections /singularizations of the
target space:
\ba
{\rm type \, (i)}: \, X^{\rm i}_0 & \to &  X^{\rm i}_{\rm max} \nn \\
{\rm type \, (j)}: \, X^{\rm j}_0 & \to &  X^{\rm j}_{\rm max} \nn \, .
\ea 
A priori, $X^{\rm i}_{\rm max} \neq X^{\rm j}_{\rm max}$. The question is
to see whether $X^{\rm i}_{\rm max} = X^{\rm j}_{\rm max}$, and moreover
whether $X^{\rm i}_{\rm max} = X^{\rm j}_{\rm max} = X_{\rm max}$,
where $X_{\rm max}$ is
the representation in terms of fields of the universe that \ref{ZPsi}
pops out. If $X^{\rm i}_{\rm max} \neq X_{\rm max}$, it means that
$X^{\rm i}_{\rm max}$ corresponds to a superposition of combinatorial 
configurations whose entropy is such that, once summed up as in 
\ref{Zall1}--\ref{Zallint} in order to give
the size of the correction to the configuration of maximal entropy among
the set $(i)$, they do not reproduce the Heisenberg inequality. 
They correspond therefore to a \emph{different} uncertainty relation, 
giving rise to different quantization rules. That means: starting with 
a string configuration assumed to be a quantum representation of
the combinatorial problem, as implied by the properties of the universe
of combinatorial configurations discussed in section~\ref{mqf}, 
quantized according to the canonical quantization 
rules in order to realize an implementation
of the Heisenberg inequality, 
we end up with a vacuum that does not correspond to the
rules we imposed on it. This is a contradiction.
Therefore, $X^{\rm i}_{\rm max} = X^{\rm j}_{\rm max} = X_{\rm max}$,
and the most entropic string vacuum is unique.
Going backwards, through the chain of 
projections/compactifications, to lower entropy configurations of each 
family, we infer the equivalence of the various string types.
It is precisely by enforcing canonical quantization that we produce 
a string theory possessing uniqueness properties, and make
all the a priori different string types equivalent.   
The same quantization imposes also a well defined number of dimensions for
the (perturbative) string construction, precisely the number
of dimensions that leads to a 3+1 dimensional universe as the
most entropic configuration.

The fact that the most entropic configuration in \ref{ZPsi} is
a classical three-sphere, namely a $SO(4)/SO(3)$ space, characterized
by a symmetry which corresponds to a vectorial, not
a spinorial, representation of the group of rotations, tells us that
spinors, required by relativity in the basic definition of the 
quantum theory, in the most entropic configuration will eventually be
all paired. Indeed, interactions are only mediated by vectorial fields
(photon, graviton), which couple 
with pairs of fermions (think at the interaction with light, 
$\bar{\psi} \slash{\! \! \! \gamma} 
\slash{\! \! \! \! A} \psi $,
and gravity, through the metric which
multiplies the entire effective action).
In particular, this means that all fermions are massive
(the mass term is of the type $m \bar{\psi} \psi $).
This prediction is confirmed by a detailed analysis of the 
physics coming out from~\ref{Zint} (for
details, see~\cite{npstrings-2011}).

\subsection{A string path integral}
\label{aspi}

Proceeding as in section~\ref{UncP} 
it is possible to show that
the contribution to \ref{Zint} given by the configurations
of non-maximal entropy amounts to a quantity covered by the Heisenberg
Uncertainty. Although expressed in a different way, 
this somehow matches with the fact that, in the Feynman path integral,
the contribution of paths which do not extremize the classical action
corresponds to the quantum deviation from the classical physics.
Indeed, any configuration $\psi_V$ contributing to \ref{Zint}
describes a ``universe'' which, along the set of values of $V$, undergoes
a pressureless expansion.
In this case, the first law of thermodynamics:
\be
dQ \, = \, d U \, + \, P dV \, ,
\label{qup}
\ee
specializes to:
\be
dQ \, = \, dU \, .
\ee
Plugged in the second law:
\be
d S \, = \, {d Q \over T} \, ,
\label{sqt}
\ee 
it gives:
\be
d S \, = \, {d U \over T} \, .
\label{sut}
\ee
Here $T$ is the temperature of the universe, defined as the ratio
of its entropy to its energy. In the case of 
the configuration of maximal entropy, the universe behaves, 
from a classical point of view, as a 
an expanding, three-dimensional Schwarzschild
black hole, and the temperature is proportional to the
inverse of its total energy,
or equivalently, its radius: $T \, = \, \hbar c^3 / 8 \pi  G  M k $,
where $k$ is the Boltzmann constant and $M$ the mass of the universe,
proportional to its age according to $2 G M \, = \, {\cal T}$.
By substituting entropy by energy and temperature in \ref{Zint} according
to~\ref{sut}, we get:
\be
{\cal Z} \, \equiv \, \int {\cal D} \psi \; 
{\rm e}^{ \int {dU \over T}} \, ,
\label{zut}
\ee
where $U \, \equiv \, U ( \psi (T) )$.
If we write the energy in terms of the integral of a space density, and 
perform a Wick rotation from the real temperature axis to the imaginary one,
in order to properly embed the time coordinate in the space-time metric, 
we obtain:
\be
{\cal Z} \, \equiv \, \int {\cal D} \psi \; 
{\rm e}^{{\rm i} \int d^4 x \, E (x)} \, .
\label{zE}
\ee 
Let's now define:
\be
{\cal S} \, \equiv \, \int d^4 x \, E (x) \, .
\label{actiondef}
\ee
Although it doesn't exactly look like, 
${\cal S}$ is indeed the Lagrangian Action in the usual sense.
The point is that the density $E (x)$ here is a pure kinetic energy term:
$E (x) \, \equiv \, E_k $. In the definition of the action, we would 
like to see subtracted a potential term:
$E (x) \, = \, E_k \, - \, {\cal V}$. However, the ${\cal V}$ term that
normally  appears in the usual definition of the action, is in this 
framework a purely effective term, that accounts
for the boundary contribution.
Let's better explain this point. What one usually has in a quantum action
in the Lagrangian formulation, is an integrand:
\be
L \, = \, E_k \, - \, {\cal V} \, ,
\label{lev}
\ee 
where $E_k$, the kinetic term, accounts for the propagation
of the (massless) fields, and for their interactions. Were the fields 
to remain massless, this would be all the story. The reason why we usually
need to introduce a potential, the ${\cal V}$ term, is that we want 
to account for masses and the vacuum
energy (in other words, the Higgs potential, and the (super)gravity potential).
In our scenario, non-vanishing vacuum energy and non-vanishing masses 
are not produced, as in quantum field theory,
through a Higgs mechanism, but arise as momenta of a space of finite 
extension, acted on by a shift that lifts the zero mode 
(see Ref.~\cite{npstrings-2011}). When 
we minimize \ref{actiondef} through a variation of fields in a finite
space-time volume, we get a non-vanishing boundary term  due to the 
non-vanishing of the fields at the horizon of space-time (moreover, we obtain 
also that energy is not conserved). In a framework in which space-time is 
considered of infinite extension, as in the traditional field theory,
one mimics this term by introducing a potential term ${\cal V}$, which has to
be introduced and adjusted ``ad hoc'', with parameters whose
origin remains obscure~\footnote{Here we have another way to see why
the cosmological constant, accounting for the ``vacuum energy'' of the 
universe, as well as the other two contributions to the energy of the 
universe, correspond to densities $\rho_{\Lambda}$, $\rho_m$, $\rho_r$,
whose present values are of the order of the inverse square of the
age of the universe ${\cal T}$: 
\be
\rho \, \sim \, {1 \over {\cal T}^2} \, .
\label{rhoT}
\ee 
Were these ``true'' bulk densities, they should scale
as the inverse of the space volume, $\sim 1 / {\cal T}^3$.
They instead scale not as volume densities but
as surface densities: they are boundary terms, and as such they live on
a hypersurface of dimension $d \, = \,  dim [$space-time$] \, - \, 1$.
The Higgs mechanism of field theory itself can here be considered
a way of effectively parametrizing the contribution of the boundary    
to the effective action in a compact space-time.  
The Higgs mechanism, needed in ordinary
field theory on an extended space-time in order to cure the
breaking of gauge invariance introduced by mass terms, is somehow the
pull-back to the bulk, in terms of a density, i.e. a ``field'' depending on
the point $\vec{x}$, of a term which, once integrated, should reproduce
the global term produced by the existence of a boundary.}.

The passage from the entropy sum over configurations to the path integral
is not just a matter of mathematical trickery. It involves first of
all the \emph{reinterpretation} of amplitudes as \underline{probability}
amplitudes. This is on the other hand implemented in the string construction.
But besides this, there is something that may look odd at first sight.
In the usual quantum (field) theoretical approach, mean values as
computed from the Feynman path integral are in general 
complex numbers, as implied by 
the rotation on the complex plane leading to
a Minkowskian time, $1/T \to it$. Real (probability) amplitudes
are obtained by taking the modulus square of them. This means that what
we obtain from \ref{Zint}, \ref{zE}
is somehow the square of the traditional path integral.
This is related to the fact that, in order to build up the
fine inhomogeneities of a vectorial
representation of space, as implied by the combinatorials of energy 
distributions, we resort to 
a \emph{spinorial} representation of space-time.
Roughly speaking, spinors are ``square roots'' of vectors.
Indeed, as discussed in~\cite{npstrings-2011} and~\cite{spi},
masses are here originated by a $Z_2$ orbifold shift on the string
space. This shift gives rise to massive particles by pairing left and right
moving spinor modes (spinor mass terms in four dimensions are of the 
type $m \psi \bar{\psi}$). The $Z_2$ orbifold
projection halves the phase space by coupling two parts. In terms of the weight
in the entropy sum, we have at the exponent a pairing/projection
$(S+S) / Z_2$, what makes clear that the amplitudes of \ref{Zint}
are squares of those of the elementary fields
(with ``weight'' $\exp S$). Had we 
just a vectorial (bosonic) representation of space, this would not occur, 
because vectorial (spin 1, or scalar, spin 0) mass terms are of the type 
$m A^2$, $m \varphi^2$. 
That is, a mass pairs with \emph{one} boson (usually one sees this in terms of 
dimension of the field).

\section{The space-time, and what propagates in it}
\label{dimspt}

\subsection{Mean values and observables in the string picture}
\label{vevst}

Mean values of observables are computed through the analog of \ref{meanO}
for the string representation of the combinatorial problem, ~\ref{Zint}.
The quantity \ref{meanO} receives contributions mostly from
the configurations with the highest entropy.
One may ask what happens if ${\cal O}$ diverges on some non-extremal 
entropy configuration. In this case, the main 
contribution to the mean value of the observed quantity could come from 
seemingly negligible vacua. 
However, in this theoretical framework 
physical quantities such as masses, energies, couplings
have a value, a ``weight'' depending on their occupation in the phase
space. Therefore, observables
cannot ``blow up'' on rare configurations
on the ground of their very basic definition: they acquire a non-negligible
contribution from their being in correspondence with 
(relatively) often realized processes (see discussion in Ref.~\cite{npstrings-2011}).

\subsection{The scaling of energy and entropy}
\label{entrV}

We want now to compute the scaling of the total energy of the universe
(for a detailed computation of the various contributions to this
quantity we refer the reader to \cite{npstrings-2011}).
In a generic thermodynamical system, once the partition
function is known, energy is computed in a
simple way by taking its logarithmic derivative with respect to the
temperature. The point is however that, 
for a generic string configuration,
it is not obvious how a temperature should be defined, nor clear whether
it can be defined at all.
In (perturbative) string theory one can
consider the so-called "one-loop partition function", which is given as
the sum at genus one (one loop in string perturbation) over all
the states of the theory, weighted with a sign distinguishing their supersymmetry 
charge~\footnote{It is indeed the
partition function of the helicity supertraces, see for instance~\cite{Kiritsis:1997hj}.}. 
In the case of unbroken supersymmetry (and also in some special
cases of formally broken, but in which states with opposite charge exist at any
energy level in equal number), the string partition
function vanishes. Basically the reason is that the condition for
having preserved supersymmetry is the existence of a limit of
decompactification of the theory (i.e. the possibility of taking the limit in which
the parameter that tunes the breaking of supersymmetry vanishes).
This is always a limit of infinite volume, and therefore of vanishing 
energy density and temperature. In this situation also \emph{any} 
thermodynamical partition function vanishes.
In a situation in which instead
all the coordinates, including the one serving as coupling of the theory, are
compact, as is our case, supersymmetry is broken, and the string partition function
does not vanish. Its value gives the energy of the string vacuum, i.e. of the string configuration~\footnote{Differently from the
usual approach to string amplitudes, in our case they don't 
calculate densities, but \underline{global quantities}.  
The reason why in the traditional approach string computations produce 
densities, to be compared with the integrand appearing in the effective 
action, lies in the fact that space-time is assumed to be infinitely extended.
In an infinitely extended space-time, there is a 
``gauge'' freedom corresponding to the invariance under space-time
translations. In any calculation there is therefore a redundancy, 
related to the fact that any quantity computed at the point 
``$\vec{x}$'' is the same as at the point ``$\vec{x} + \vec{a}$''.
In order to get rid of the ``over-counting'' due to this symmetry,
one normalizes the computations by ``fixing the gauge'', i.e.
dividing by the volume of the ``orbit'' of this symmetry $\equiv$ the volume
of the space-time itself. Actually, since it is not possible to perform
computations with a strictly infinite space-time, multiplying and dividing
by infinity being a meaningless operation, the result
is normally obtained through a procedure of ``regularization'' of the
infinity: one works with a space-time of volume $V$, supposed to be
very big but anyway finite, and then takes the limit  $V \to \infty$.
In this kind of regularization, the volume of the space
of translations is assumed to be $V$, and it is precisely the fact of
dividing by $V$ what at the end tells us that we have computed a density.
In any such computation this normalization is implicitly assumed.
In our case however, there is never 
invariance under translations: a translation of a point
$\vec{x} \to \vec{x} + \vec{a}$ is not a symmetry,
being the boundary of space fixed. 
On the other hand, a "translation" of the 
boundary is an expansion of the volume and corresponds
to an evolution of the universe, it is not a symmetry of the present-day
effective theory.
In our framework, the volume of the group
of translations is not $V$. Simply, this symmetry does not exist at all.
There is therefore no over-counting, and what we compute is not a density,
but a global value. 
In our case, compactification of space to
a finite volume is not a computational trick as in
ordinary regularization of amplitudes, it is a physical condition.
In our interpretation of string coordinates, there is therefore
no ``good'' limit $V \to \infty$, if for ``$\infty$'' one intends the
ordinary situation in which there is invariance under translations.
In our case, this symmetry appears only strictly at that limit,
a point which falls out of the domain of our theory.}.
In our language, this is the energy of the universe. Indeed,
as we discuss in appendix~E of~\cite{npstrings-2011}, the string partition
function turns out to precisely correspond to the thermodynamical partition function.
What in the string approach serves as temperature, namely, a component of the world-sheet parameter~\footnote{This has a different definition on different perturbative
constructions, which can be of closed or open string.}, can be identified
with a string coordinate. The identification between
world-sheet and target space coordinates involves always two
string coordinates, namely a space and a time coordinate. The freedom
in the choice of the reparametrization of the map between
world-sheet and target space reflects a gauge invariance
of the theory, which is usually fixed for instance by choosing
a light-cone gauge, in which only the space transverse coordinates
explicitly appear in the construction. A condition for this
identification is that the target space coordinate is not twisted, otherwise
we don't have a string, but the mapping of a segment to a point.
Owing to this identification, the integration over the "temperature"
in the string partition function is always an integration
over a parameter of order one in string units. Indeed,
for the cases of relevance for us, it results to be of order one also
in Planck units (string coupling of order one).
As a consequence, the logarithmic derivative of the partition
function results to be of the same order as the partition function itself.
As long as one is just interested in the simple order
of magnitude and scaling law of
the vacuum energy, one can deal with the string partition
function as with the function which computes the vacuum energy, i.e.
like the derivative of the thermodynamical partition
function.
Through the approximation:
\be
{\partial {\cal Z} / \partial \beta \over {\cal Z}} ~ \approx ~
{{\cal Z} / \beta ~ (\sim 1) \over {\cal Z}} ~ \approx ~1 \, ,
\label{ZoverZ}
\ee
we obtain therefore that the vacuum energy is of order one. 
This is however not automatically the energy of the universe at time
${\cal T} > 1$: indeed, owing to the
identification of the time coordinate of the string target
space with the world-sheet time coordinate,
\ref{ZoverZ} is only the energy of the universe at time
${\cal T} = 1$.
In order to obtain the energy at a generic time ${\cal T} > 1$,
let us first compute the energy density. 
This is obtained by dividing the total energy by 
the volume of the target space. In doing this, one must bear in mind that
one coordinate of the space has been set to unit size by
the reparametrization which allows to trade it
for the coordinate of the string world-sheet. 
For simplicity, if we suppose that the space is $D$-dimensional, and that
all coordinates of the \emph{physical}
space are of length $R$, the energy density is:
\be
\rho(E) ~ \approx ~ {1 \over R^{D-1}} \, ,
\label{rhoE}
\ee
and not $1 / R^D$.
In this computation, $D$ is the number of \emph{all} the
non-twisted space coordinates, including the coupling.
In three dimensions (i.e. in four space-time dimensions) this
gives $\rho(E)_{D=3} = 1 / R^2$. However, this is not all
the story: as long as T-duality along these coordinates is not
broken, also energy densities are invariant under this symmetry.
Indeed, as long as T-duality is a symmetry, it is not even possible to
distinguish the universe at large volume from the one at small volume.
Expression~\ref{rhoE} is in general just the large volume limit of
a more complicate expression invariant under inversion of each radius.
In any case, since we are interested in the universe at large volume,
well beyond the Planck scale, this approximation is
justified. 
In this limit, for any $D$ the total energy is therefore:
\be
\langle E \rangle ~ = ~ \int_V \rho(E) ~ \approx ~ R \, .
\label{EpropR}
\ee 
If we start with a universe of Planck size
(${\cal T} = R_1 = \ldots = R_D = 1$)
and let the space to be stirred by propagating massless modes
(there exists always at least one such, the graviton)
we have at any time $R \sim {\cal T}$, and
the total energy can be identified with the age of the universe,
as in section~\ref{setup}.

The relation~\ref{EpropR} matches with those of section~\ref{eSp}.
This means that the string compactifications correspond to a
black hole-like universe only in three dimensions
(higher dimensional black holes have
energy and entropy scaling as higher powers of the radius,
see for instance~\cite{Rabinowitz:2001ag}). But, more importantly, it means that
the analysis of entropies carried out in section~\ref{eSp}, and the
conclusions we derived about the dimension of the space of the
configuration of highest entropy, carries over to the string representation.
The configuration of highest entropy corresponds therefore
to three space dimensions.
This comes not unexpected, because, as we pointed out in section~\ref{stringT}, quantum string theory is built
precisely with ingredients which make of it a representation
of the combinatorial problem, and the canonical quantization precisely imposes
the number of degrees of freedom leading to a consistent description
of the geometries of a three-dimensional space.
In~\cite{npstrings-2011} we will see that three
is precisely the number of space coordinates which are left untwisted
in the case of maximal reduction of symmetry, which, as discussed in 
section~\ref{mimacro}, corresponds
to the maximal entropy in the phase space.
We will therefore recover in another way the fact that
this scenario leads to automatically
to a four-dimensional space, without postulating this condition
a priori.

As we will discuss in~\cite{npstrings-2011}, in the string vacuum
of highest entropy T-duality is broken. This makes legal talking
of energy density of a large volume universe, as distinguished from small volume,
and allows to make contact with the geometric description of the universe
as implied by \ref{zsum1}, \ref{ZPsi}.

\subsection{``vectorial'' and ``spinorial'' metric}
\label{vecsp}

Till now, we have spoken of space-time as the space spanned by the
extended, expanding  coordinates, whereas the internal coordinates
are those which are frozen at a fixed value. This point deserves
to be considered more in detail. The question is about what is the meaning
of ``dimension'' of space-time. 
It can not be just the number of
coordinates we label as ``$X_{\mu}$''. 
Indeed, string theory implements coordinates and degrees of freedom 
in a framework of space-time-dependent fields. 
However, \emph {in the interpretation
of our theoretical framework, the dimension of space is given by the number
of independent massless fields}. 
In the dynamical representation of the combinatoric 
problem on the continuum, the expansion
of space-time is viewed as driven by the propagation of 
massless fields, which in some sense ``represent'' it. 
When we write type II string in its basic
formulation, namely in ten dimensions, at least perturbatively we write
the theory in ten dimensions, because we have ten bosonic fields free
to propagate with the speed of light.
But from inspection of the effective
action, and, after compactification and orbifolding, 
of the heterotic dual, we learn that the theory has much more
massless fields than just ten. It effectively describes a higher-dimensional
space, namely the ``fiber'' with
all gauge fields and tensors living on the ten-dimensional ``base'' space. 
The confusion about the counting of dimensions originates from
the fact that the space-time coordinates themselves are fields and not just
parameters. From a perturbative point of view,
at the origin or the existence of more massless fields than just
the number of dimensions is the nature of the bosonic coordinates as
functions of the coordinates of a string world-sheet. This allows the
separation into left and right movers, which can combine to give rise to
more degrees of freedom than just the number of coordinates. On the other 
hand, the existence of this possibility is also at the origin of the
existence of T-duality upon compactification, and therefore of an
effective minimal length as soon as the theory is targeted to a finite 
volume, a necessary condition to match with a representation of our
combinatorial scenario. We recover therefore the fact
that precisely the existence
of such a cut-off, which makes gravity consistent as a quantum theory,
produces the existence of a higher number of degrees of freedom that
just the counting of perturbative dimensions.

In order to properly represent the combinatorial problem,
the space-time coordinates must
correspond to massless fields, in the sense that there must be a 
non-degenerate mapping between space-time degrees of freedom and field
degrees of freedom. If, by absurd,
there were more massless degrees of freedom than space-time coordinates,
it would mean that we have wrongly computed the dimensionality of space-time:
the number of ``coordinates'' maximally extended, i.e. with the
minimal curvature, would be higher, and we would get into a contradiction. 
In the string configuration of highest entropy, i.e. of lowest symmetry, 
we must therefore have a correspondence:
\be
\left\{ X_0 = t, \vec{X}    \right\} \, \leftrightarrow \, 
\left\{ \phi_i (t, \vec{X})  \right\} \, , ~~~~~~~
\det \left[ \partial \phi_i(t, \vec{X}) \big/ \partial (t, \vec{X}) \right]
 \, \neq \, 0 \, ,
\label{detphi}
\ee  
for a set of fields $\phi_i$, $i = 1, \ldots, 4$. 
Here the role of time can be a bit misleading, being this coordinate
not precisely a field itself. On the other hand, it comes quite correctly in 
the counting, because the space we are describing is not flat. In other
words, instead of time one can speak of an additional coordinate, the fourth
space-coordinate, which is needed in order to described a curved
three dimensional space in terms of flat coordinates. This fourth
coordinate can be alternatively viewed as a curvature, $X_4 \sim 1/R$, 
anyway related to the time coordinate through the relation 
$R \sim E \sim {\cal T}$.  
As we are going to discuss in~\cite{npstrings-2011},
at the minimum of symmetry, the invariance of space under
rotations is broken. There are therefore
only ``diagonal'' degrees of freedom. Namely, in these configurations
the space built on the coordinates $x_1, \ldots, x_3$ 
doesn't possess a symmetry under 
rotations: $ x_i \, \to \, x^{\prime}_i = A_{ij} x_j$. 
After having gauged away the redundant degrees of freedom, for instance
in the light-cone gauge, where only transverse degrees of freedom 
appear, the graviton field has only the two ``diagonal'' entries:
\be
g_{\mu \nu} \, = \, \{ g_{11}, g_{22}  \} \, . 
\ee 
This makes up two field coordinates. We need two more to fill up
the space-time dimension and ensure that the map \ref{detphi} is 
non-degenerate. Being the speed
of expansion of the universe ``fixed'' to a finite constant $c$, 
the space-time is necessarily a relativistic space,
whose representations are built on spinors.
Vectorial representations can be built starting from spinorial ones, but
not the other way around.
The spin connection contains therefore a mixture of vectorial and purely 
spinorial components.

Consider now the role played by 
the field $g_{\mu \nu}$. It ``rotates'' two vectors, by contracting
their indices into a scalar, according to:
\be
V^{\mu }, \, V^{\nu} ~ \to ~ V^{\mu} g_{\mu \nu} V^{\nu} \, .
\label{vvg}
\ee
We expect that a ``purely spinorial'' spin connection in a similar way
rotates, and contracts, spinor indices.
\be
\psi^{\alpha}, \, \psi^{\beta} ~ \to ~ \psi^{\alpha} \tilde{A}_{\alpha \beta} 
\psi^{\beta} \, .
\label{psipsiA}
\ee 
Owing to the breaking of rotational symmetry, the only bi-spinors
present in the minimal string vacua
are those that pairwise build up diagonally vector coordinates. 
If we indicate the spinors associated to each bosonic coordinate as
$\phi^{\mu}_1$, $\phi^{\mu}_2$, this means that there are no
mixed states of the type:
\be
\phi^{\mu}_{\alpha} \otimes \phi^{\nu}_{\beta} \, , ~~~~~~~
\mu \neq \nu \, , ~~~~ \alpha, \beta \in \{1,2 \} \, ,
\ee
but only diagonal ones:
\be
x_{\mu} \, = \,
 \phi^{\mu}_{1} \otimes \phi^{\mu}_{2} \, .
\ee
We expect therefore the ``spinorial'' part of the spin connection to be
in bijection with a vectorial representation consisting of just two
transverse field degrees of freedom. This is actually the way the 
electromagnetic vector-potential field in these vacua works.
The field $A_{\mu}$ is a vectorial field, not a spinorial one.
On the other hand, the vector index $\mu$ must be somehow thought as 
a ``bi-spinor'':
\be
A_1 ~ \sim ~ A_{{1 \over 2} {1 \over 2}} \, .
\label{1-22}
\ee
Indeed, $A_{\mu}$, normally introduced through a gauge mechanism
applied to a scalar quantity built on a bi-spinor, somehow provides with field 
degrees of freedom a ``metric'' which contracts spinor indices to
a scalar:
\be
\overline{\psi} \partial {\! \! \! /} \psi ~ 
\stackrel{\rm gauge}{\longrightarrow} ~ 
\overline{\psi} A {\! \! \! /} \psi ~ = ~
\overline{\psi}^{\alpha} \gamma^{\mu}_{\alpha \beta} A_{\mu} \psi^{\beta} \, .
\label{ppgA}
\ee 
The gamma matrices, precisely introduced by Dirac 
in order to deal with ``square-roots'' of vectorial relations,
play the role of converter from bi-spinorial to vectorial indices.
The field $\gamma^{\mu}_{\alpha \beta} \equiv
{A {\! \! \! /}}_{\alpha \beta}$ corresponds therefore to
the ``spinorial spin connection'' $\tilde{A}_{\alpha \beta}$ introduced above.
This field provides the two missing degrees of freedom required in order
to complete the non-degenerate ``representation'' of space-time \ref{detphi}.

We stress that it is only in four dimensions that we can realize
such a non-degenerate map as \ref{detphi}, namely, with a consistent
correspondence between number of massless degrees of freedom and 
formal number of space-time dimensions.
Being a representation of space-time
means that graviton and photon propagate at the speed 
of space-time itself. As such, they  correspond to massless fields.
This proves that, \emph{in the configuration of highest entropy / minimal symmetry} 
there are \emph{two massless fields, the graviton and the photon}. 
As we will discuss in section~\ref{boundary}, their propagation
occurs at the speed of expansion of the horizon.
The constant of proportionality between the age and the radius
of the universe sets therefore what we call the "speed of light".

\subsection{Observations about a space-time \emph{built} by light rays}
\label{gr}

In this section we consider
the geometry of the classical space of our scenario, namely the space 
contained within the classical horizon of observation. We want to show that
a non-vanishing curvature is necessarily implied
if the space is considered as \emph{built}
by light rays propagating at finite speed. 
This curvature however does not correspond entirely to
a classical geometry, something that is signaled also by
a mismatch in the normalization.
The ``classical'' space we consider is such that: 
\begin{enumerate}
\item \emph{all} the points of the 
universe are causally connected to the observer. This means, not simply
they fall \emph{within} a space-like region, but \emph{are}
at a light-like distance, in space
and time, from the observer. 
\end{enumerate}
\noindent For the same reason,  
\begin{enumerate}
\item[2.] these
points are also light-connected to the origin, the ``Big Bang'' point.  
\end{enumerate}
These conditions are fully compatible with 
what we know about the universe. In practice they mean that
we consider a space-time corresponding to the region
causally connected to us. This space is bounded
by a horizon corresponding to the spheric surface, centered on our point
of observation, whose radius is given
by the maximal length stretched by light since the time of the Big Bang.
This region defines our ``universe'': there is no space-time outside this 
region. This implies that the entire space-time originates from a 
``point''.

At first look, the space included within the
horizon looks more like a ball than a curved surface.
If we set the origin of our system of coordinates
at the point we are sitting and making observations,
the universe up to the horizon is by definition the set of the points
satisfying the equation:
\be
x_1^2 \, + \, x_2^2 \, + \, x_3^2 \, \leq \, {\cal T}^2 \, .
\label{curv}
\ee 
However, owing to the finiteness of the speed of light,
the region close to the horizon corresponds to the early
universe, and the horizon ( i.e. for us the set of points 
$x_1^2 \, + \, x_2^2 \, + \, x_3^2 \, = \, {\cal T}^2$) 
effectively corresponds to the origin of space-time. 
This means that the points lying close to the
horizon are indeed also close in space. The set of points
$\vec{x} \, = \, (x_1,x_2,x_3):\sqrt{x_1^2 \, + \, x_2^2 \, + \, x_3^2} \in
[{\cal T}, {\cal T} - \epsilon]$ is, from an ``objective'' point of view,
a ball of radius $\epsilon$ centered at the origin. 
Let's here set the origin of the universe, i.e. of space-time,
at $(x_0=t,x_1,x_2,x_3)=(0,0,0,0)$. 
From a ``correct''
geometric point of view, we, namely the observers, are sitting at a point
on the hypersurface $x_1^2 + x_2^2 + x_3^2 \, = \, {\cal T}^2$.
This defines a 2-sphere of radius ${\cal T}$. 

\begin{figure}
\centerline{
\epsfxsize=14cm
\epsfbox{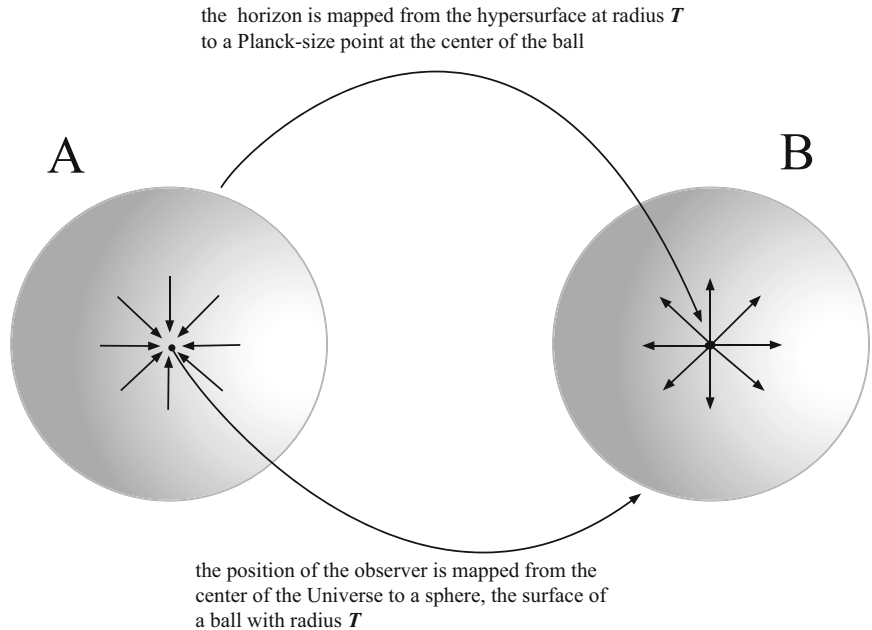}
}
\vspace{0.3cm}
\caption{The ball A represents the universe as it appears to us, 
located at the center and observing a space-time extended in any direction
(solid angle 4$\pi$) up to the horizon at distance $\cal T$. The arrows
show the direction of light from the horizon to us.
Ball B on the right represents instead the ``dual'' situation, with the 
horizon, corresponding to the origin of space-time, at the center,
and the light rays, indicated by the arrows, propagating this time outwards.
We are sitting on a point on the surface, a two-sphere oriented inwards.
The curvature therefore is negative.  This surface
has to be thought as an ``ideal'' surface: different points on the surface 
belong to different causal regions: there is no communication among them.  
The only point of it we know to 
really exist is the one at which we are located. From a ``real'' point of view,
the two-sphere boundary of ball B is therefore a ``class of surfaces'', in
each of which all the points of the surface have to be though to correspond
to a single point in the ``real'' space-time. The path of light rays
is therefore not straight but curved. Figure~\ref{fig2} helps to figure out
the situation. Both figures should be taken in any case only as a hint,
none of them being able to depict the exact situation.}     
\label{fig1}
\end{figure}

\begin{figure}

\centerline{
\epsfxsize=14cm
\epsfbox{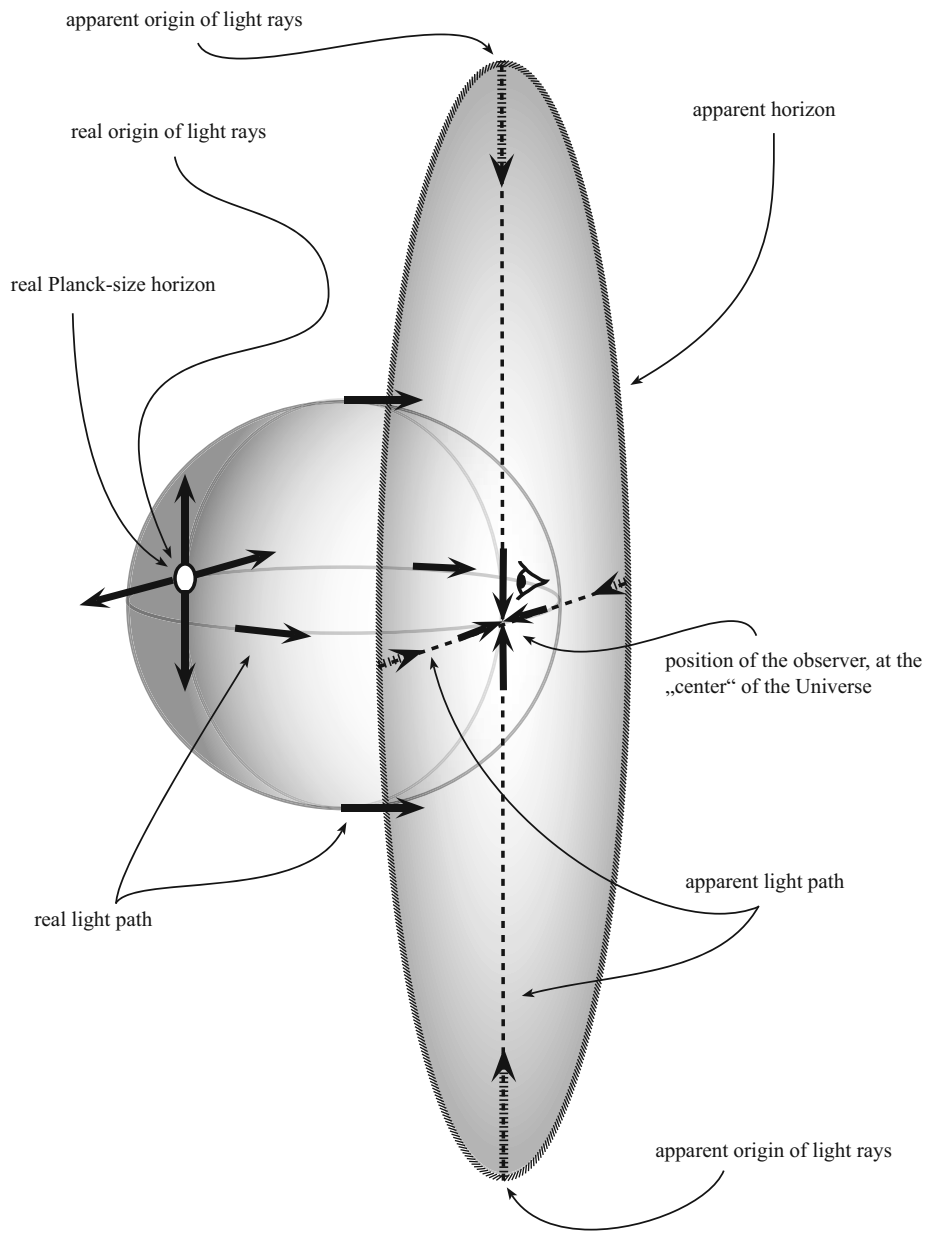}
}
\vspace{0.3cm}
\caption{The disc represents the space-time as it appears to the observer:
a flat, tangent space. The real space-time is however curved: the border of the
disc corresponds to a ``point'' on a 3-sphere, and is located
diametrically to the position of the observer.}
\label{fig2}
\end{figure}

The Ricci curvature scalar
for a 2-sphere is given by ${\cal R} \, = \, {2 \over r^2}$, where $r$
is the radius of the sphere, when thought of
as embedded in three dimensions.
In our case, $r \, = \, {\cal T}$. 
From the point of view of the \emph{local} physics
around the observer, the three-dimensional space is perceived 
as a staple of concentric two-spheres, like a kind of onion,
in which the local curvature. i.e. the amount of ground energy of space,
is basically the one corresponding to the tension of the two-sphere
with radius exactly corresponding to the age the universe 
has for the observer (different points are located 
at a space-time distance, and correspond to different ages
of the universe).
We can therefore consider the Einstein equations 
for a four-dimensional space-time:
\be
{\cal R}_{\mu \nu} \, - \, {1 \over 2} g_{\mu \nu} {\cal R} \, 
= \, 8 \pi G_N  T_{\mu \nu} \, + \, \Lambda g_{\mu \nu} \, ,
\label{einstein}
\ee
Let's also consider the space as ``empty'', and therefore
neglect the contribution of the stress-energy tensor. 
Contracting indexes we obtain: 
\be
- {\cal R} \, = \, 4 \Lambda \, .
\ee  
The
two-sphere we are considering is a surface oriented inwards; the observers sitting on this surface don't look
outside but toward the center of the sphere, toward the ``big-bang'' point.
At the point the observer is sitting, this surface has a metric with 
hyperbolic signature, and the Ricci curvature must be taken with
the opposite sign as compared to the one of the two-sphere.
Plugging this into the above equation, we obtain:
\be
\Lambda \, = \, {1 \over 2  r^2} \, = \, {1 \over 2 {\cal T}^2} \, .
\label{r2t}
\ee
The strange topology of this universe is illustrated in figure~\ref{fig1}. 
\noindent
In order to help the reader in visualizing the situation, we show
in figure~\ref{fig2} a two-dimensional, intuitive 
picture of the universe, illustrating the fact that, both for a flat
and a curved space-time, incident light rays arrive 
parallel to the hyperplane tangent to the observer, so that no difference is 
in practice locally observable (the only indication that the path 
of light is not straight but curved comes from a measurement of the 
cosmological constant or of a non-vanishing contribution
to the stress-energy tensor).
Owing to the curvature of space-time, the rays don't come from the apparent
horizon, the one obtained by straightly continuing the light paths along the 
tangent plane, but from a "point-like" horizon. 
\noindent
Although useful to the purpose of illustrating how things are going, 
both figures \ref{fig1} and \ref{fig2} are slightly misleading, 
none of them being able to account for the real situation. 
In particular, from figure \ref{fig1} we understand that there is
a symmetry between picture A on the left and picture B on the right.
Mapping from A to B, i.e. exchanging the origin with
the horizon, involves a ``time inversion''. This operation corresponds
to a \underline{duality} of the system. The configuration ``B'', associated
to the solution~\ref{r2t} of the Einstein's equations, corresponds to one
of the possible points of view, ``pictures'', from which to look at the 
problem. Had we looked from the seemingly rather unnatural
picture A, we would have concluded that the curvature is positive.
However, this too is a legitimate point of view.

\begin{figure}[h]
\centerline{
\epsfxsize=12cm
\epsfbox{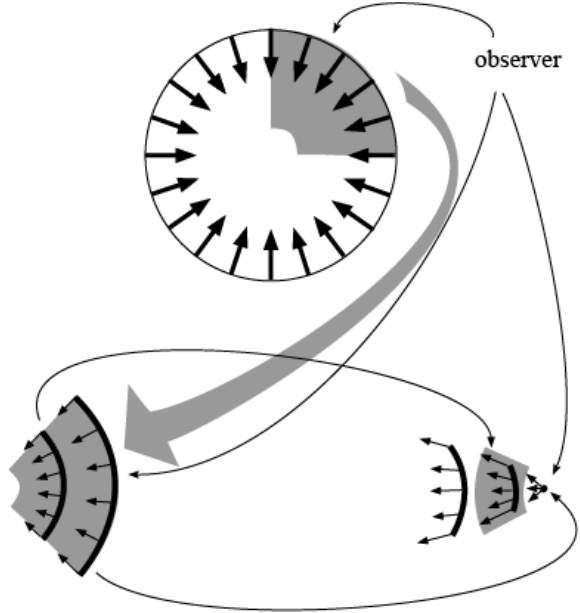}
}
\caption{Shrinking to a point the two-sphere on which the 
observer is located (here represented by the external boundary of the disk),
and considering the region causally connected to it,
leads to a 3-sphere geometry and to a change of sign of the curvature.}
\label{shrink}
\end{figure}

Let's have a closer look at the situation we are describing. As seen
from the point of view of the origin of the universe, the 
``surface'' given by the equation $x_1^2 + x_2^2 + x_3^2 = {\cal T}^2$
consists of points lying on non causally-related regions. We are sitting on
a point of this surface, which however is an ``ideal'' surface: the only 
point we know to exist is the one at which we are sitting, the hypothesis
of boundedness of space-time being precisely justified by the requirement
of describing only the region causally connected to us (alternatively,
we can think that this surface must be thought as equivalent to a 
point). In general, our universe is the set of points,
each one lying on a sphere $x_1^2+x_2^2+x_3^2 = r^2 \, < {\cal T}^2$,  
causally connected to us (notice that any 2-sphere with radius $r < {\cal T}$ 
contains several points causally connected to us). 
The curvature experienced
by the observer is positive: from the
point of view of the observer such a space made of 
a staple of two-spheric shells can be seen as obtained by
starting from two-sphere corresponding to the surface at the horizon,
and stapling two-spheres progressively more and more shrunk,
till the one he is sitting on, which is shrunk to a point. In other words,
the space ``opens up'' from the point of the observer, as roughly illustrated in figure~\ref{shrink}. If we consider the 
region causally connected to us as the only one we indeed know to exist (and 
therefore have the right to consider), and therefore
as the full existing space-time, it happens that
the light rays starting from a ``point'', the origin of the universe, 
end up also to a point, our point of observation.
It is not hard to realize that what we are describing is indeed the geometry 
of a 3-sphere. 
\noindent
The curvature of a 3-sphere is three times larger than the one of a two-sphere
with the same radius. This means that, from a classical point
of view, only one third of the whole curvature of the universe is
of purely "geometric" origin. Indeed, what we have investigated here 
is the pure "cosmological" contribution to
the energy density of the universe. The missing part can be explained
within the framework of a quantum scenario, such as the string 
scenario we discuss in Ref.~\cite{npstrings-2011} (see also Ref.~\cite{spi}).
We refer to \cite{npstrings-2011}, section~3.1,
for a discussion of the other contributions to the curvature of the universe.

\subsection{Closed geometry, horizon and boundary}
\label{boundary}

\begin{figure}
\centerline{
\epsfxsize=8cm
\epsfbox{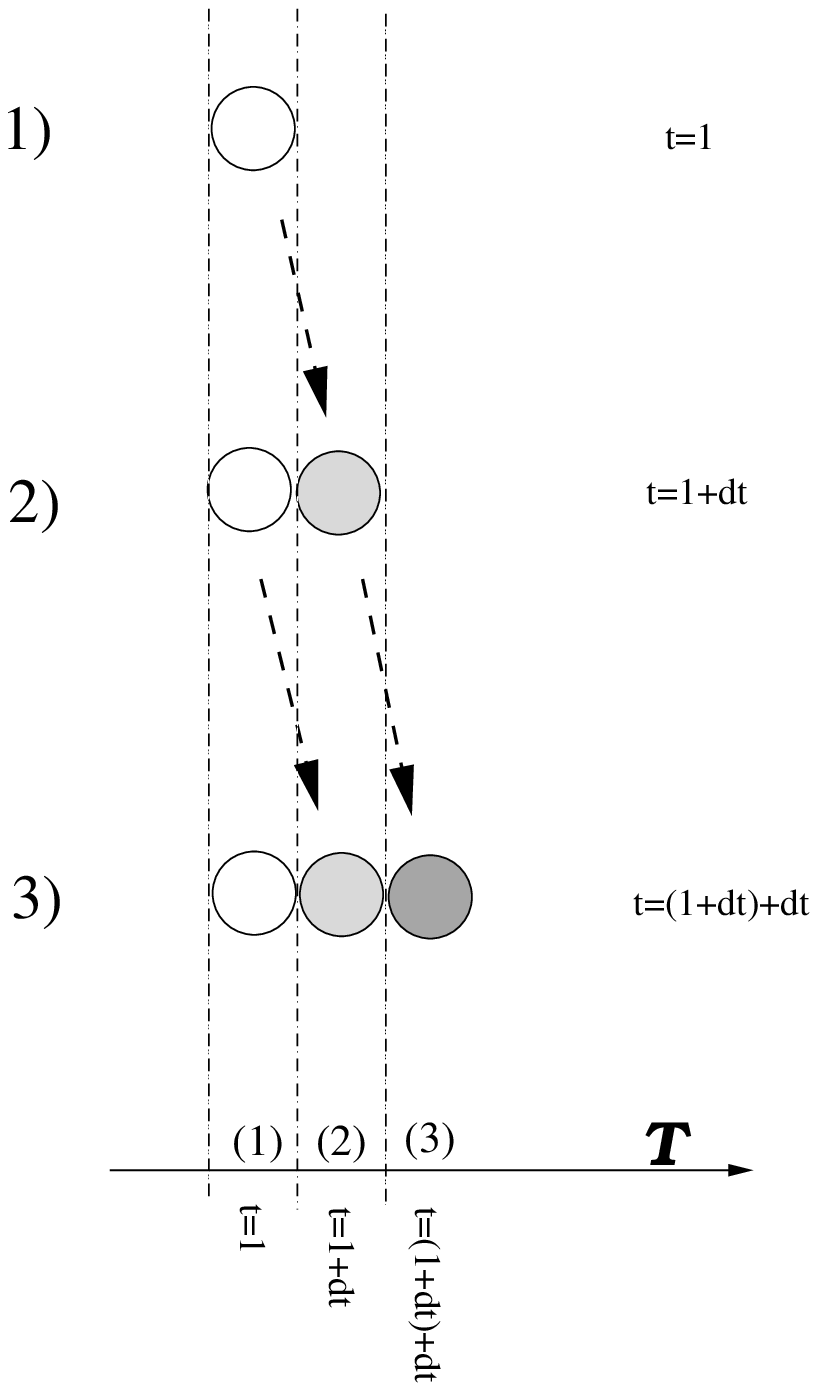}
}
\vspace{0.3cm}
\caption{The progress of the universe through increasing times $\sim$ number
of elementary cells. As the statistics grows, the configuration
gets more complex and differentiated. Configuration 3) does not ``include''
configuration 2), which in turn does not include number 1): with volume
increases also the radius, and the curvature of space decreases. 
The curvature and therefore the energy levels are different at 
different times. Configuration 1) ``flows'' to 2), and 2) to 3) only in the
sense that 2) is more similar to 3) than 1) or a configuration with
two equal balls are, and therefore this is the most realized ``evolutionary
path''.}     
\label{expansion2}
\end{figure}

\begin{figure}
\centerline{
\epsfxsize=8cm
\epsfbox{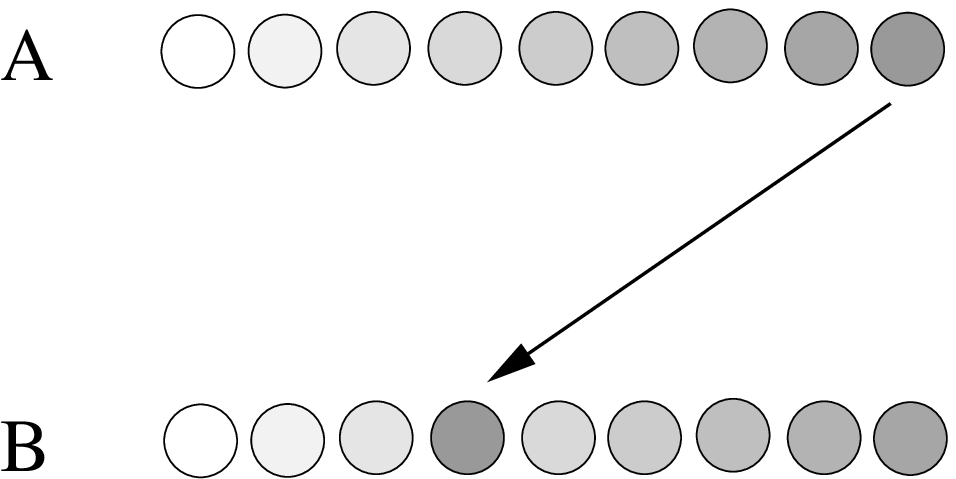}
}
\vspace{0.3cm}
\caption{Configuration B) is obtained by superposing to
the dominant configuration less entropic configurations than in A),
and therefore has a lower weight, the more and more negligible 
the higher is the ``jump'' between B) and A).}     
\label{discont-1}
\end{figure}

We want now to see how our universe builds up according to
the combinatorial scenario of section~\ref{setup}.
At $t = 1$ there is only one
possible configuration, that we illustrate in figure \ref{expansion2} with
a ball representing the unit cell.
Already at the next step, $N = 2$, we have many more (infinitely many)
possibilities, corresponding to any possible space ``dimension'',
being $N^p$ no more trivial, and the combinatorics increases very rapidly. 
For what discussed in section \ref{eSp}
we can concentrate the analysis to three dimensions,
which gives a contribution of the same amount as the sum of
all the neglected dimensionalities and configurations (Uncertainty Principle).
The dominant configuration
is the one that gives a ``homogeneous'' distribution of the occupied cells 
(what at large $N$ becomes a ``spheric'' geometry) . 
But already at $N = 2$ also non-extremal
entropy configurations start to exist and give non-trivial contributions.
The result is that, on the top of a homogeneous distributions, in the universe
start to show up inhomogeneities, of the order of the Uncertainty Relations.
We can represent this process by distinguishing the regions of the space
as balls with a different colour, figure~\ref{expansion2}.    
As time goes by (i.e. $N$ increases), we get new possibilities 
of differentiation from the basic homogeneity, in a sort of ``progressive
differentiation through steps of small perturbations''.
We indicate this with an increasingly darker coloration of the balls.
In principle, one could ask if there could be ``discontinuities''
in this progress, namely, whether there could be steps in which a 
darker ball falls between lighter balls, as illustrated 
in figure~\ref{discont-1}. 
Of course, nothing prevents
these configurations from appearing, and indeed they \emph{are present}.
However, they are less entropic than "smoother" ones:
the entropy of a configuration decreases the more and more,
the more it deviates from the "regular" geometry of the sphere. 
Configuration presenting discontinuities give therefore a minor contribution,
so that, in the average, the evolution takes place in the continuous
way we illustrate in figure~\ref{expansion3}.

The progress through times/values of the average curvature and its
inhomogeneities looks therefore like the propagation of a perturbation,
and is translated, in the continuous language, 
into a ``propagation'' of the information through an expanding universe.
As time goes by, the average curvature of space decreases,
and this takes place everywhere, throughout all the space. 
The ``perturbations'' too spread out: as $1 / N^2$ for those
that we interpret as ``massless fields'', or with a lower rate for
the ``massive'' ones. To an observer,
the local physics appears to be influenced, ``produced'' if one prefers,
by the sum of the information that propagated up to him from everywhere else
in the space. We already mentioned that a detector is a particular 
configuration of space, and an experiment is the detection of variations of 
this configuration during a certain interval of time.
By letting an experiment, or a detection, to take place
in a certain interval of time, the observer can resolve for different
``smearing rates'' of the inhomogeneities, and consequently organize
the interpretation of
what happens in terms of massless and/or massive objects. 
Of course, this is only an approximation, because any possible
configuration contributes, although to an uncertainty expressed by
the Uncertainty Principle. The ``experimental
observation'' of the universe surrounding an observer indeed carries
the information about the geometry of the full universe,
the progress toward an increasing ``local'' differentiation from the
average geometry being interpretable as a continuous, ``jump-less''
evolution that propagates the perturbation of the spheric geometry.
What \underline{we observe} is indeed \underline{always just our 
local physics}, but we disentangle the jungle of data by organizing
them as a superposition of informations coming from different places,
and therefore ``originating'' at different times in the past.
Notice that, although
we see regions of the universe corresponding to past times, 
we don't see ourselves in our past times: neighboring shells
are not derived the one from the other through time evolution.

\emph{At any time ${\cal T}$ the universe is given by the
(weighted) superposition of the configurations of the phase space 
\underline{at present time}}. 

In this sense, the actual configuration 
depends only on the present phase space, not on the past:
the evolution belongs to our interpretation.
For the observer the universe turns out to mainly consist
of a progression from the farthest configuration, the ``initial'' one, 
to the nearest, representing the physics at present time.
At any time, the dominant configuration is however not derived
by a process of evolution of the one at previous time. 
\begin{figure}
\centerline{
\epsfxsize=8cm
\epsfbox{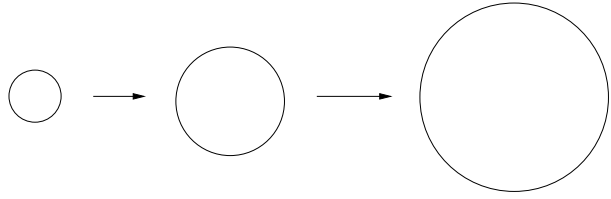}
}
\vspace{0.3cm}
\caption{During the time evolution, the three-dimensional space
build up of an increasing number of elementary cells 
expands with the geometry of a three-sphere of growing radius.}     
\label{expansion1}
\end{figure}
To an observer the space appears built
as an ``onion'': the observer is surrounded by shells, 
two-spheres corresponding to farther and older phases of the universe, 
up to the horizon, that correspond to the ``big bang'',
as illustrated in figure~\ref{expansion3}.
\begin{figure}
\centerline{
\epsfxsize=8cm
\epsfbox{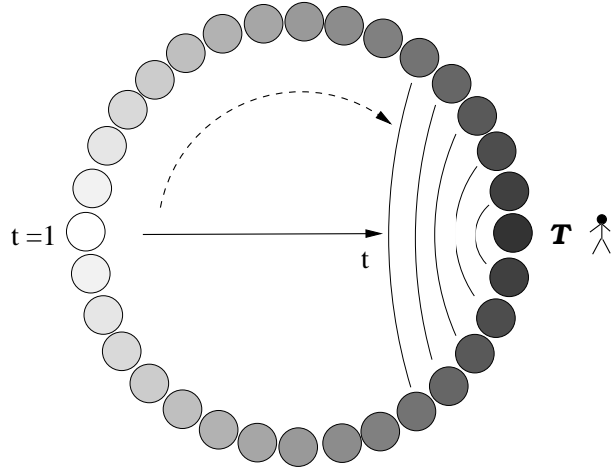}
}
\vspace{0.3cm}
\caption{The universe at time ${\cal T}$ appears to the observer as
a set of surrounding shells made out of elementary cells.
These shells, here represented only through sections constituted
by two linked antipodal cells, go from the one closest to the
observer (the darkest one), 
which is also the closets in time, up to the farthest, the horizon,
that corresponds to the ``big-bang'' cell. The curvature of space
is on the other hand everywhere the one of the age corresponding to the
observer, ${\cal R} \sim 1 / {\cal T}^2$. Notice that the two-sphere 
corresponding to the horizon appears to ``conserve'' the total energy flux, 
$1$, of the initial unit cell.}     
\label{expansion3}
\end{figure}
However, in itself there is no real ``big-bang point'', located somewhere
in this space: the argument leading to this description of the observed 
universe is not related to a particular choice of the point of observation.
This does not mean that the universe looks absolutely identical
for any choice of this point inside the whole space; simply, it means that
from any point the universe appears built as an ``onion'', with informations
coming from everywhere around, going backwards in time, and space, up to
the horizon, which is an ``apparent horizon'', with no real physical location, 
but always at distance $N \sim {\cal T}$ from the observer.

Wherever the location of the observer in the space is,
what he will see is only the ``tangent'' space around him,
experienced through the modifications it produces in himself.
In this way, he will have access to the knowledge of the
average energy density of the universe, and indeed an indirect experience
of the whole universe: the ``local physics'' is the one
specific of the actual time, and as such ``knows'' about the full
extension of the space. Owing to the special ``boundary conditions''
that ``sew'' the borders of space into a sphere,
the latter will look in the average homogeneous all around in every direction:
the observer will always have the impression of being located ``at the center''
of the universe.
Through the modification of his configuration, the ``local physics'',
(i.e. the set of all what happens to him, light rays hitting from the various
directions, gravitational fields etc..., that he will \emph{interpret}
as coming from all over the space around), he will then ``measure''
the energy density through all the space, concluding that it is 
$\rho(E) \sim 1 / R^2$. 
On the other hand, knowing that such an 
energy-density scaling law is the one of a sphere, he will deduce
that the horizon surface at a distance $R$ from himself, and with
area $\sim R^2$, has boundary conditions such that the space 
closes-up to a sphere.
The observer will then \emph{interpret}
the set of cells spread out along the horizon,
a surface with area $\sim R^2$ and
energy density $\sim 1 / R^2$, as corresponding to a point, a 
unique cell of unit volume and unit energy, ``smeared'' over
something that appears like a two-sphere.
He will therefore refer to this as to the ``Big Bang'' point, 
the initial configuration, energy one at volume one,
and he will say that what he sees by looking at the horizon, is indeed
the beginning of everything. We repeat however that
this point, or surface, namely, the horizon, does not really exist as a
special point located somewhere in space: the interpretation would be the same
for any observer, located at whatever point in this space.

\subsection{Quantum geometry}
\label{qgeom}

If one divides the total energy of a Schwarzschild black hole,
as given by the relation $2M = R$, by the volume of the three-ball
enclosed by the two-sphere surface of the horizon, i.e. by 
${4 \over 3} \pi R^3$, one finds an energy density which,
once inserted into the Einstein's equation relating the curvature of
space to the energy-momentum tensor,
precisely corresponds to the curvature of a three-sphere. 
So, one starts with an object that from outside looks like a ball,
i.e. a flat space,
and finds out that "as seen from inside" looks like a sphere, a curved space.
This is the situation of our universe:
despite the fact that a sphere doesn't have boundary, as observers
located inside we have the impression that there indeed
is a surface, a two-sphere at radius $R$, the natural boundary of the universe,
which works as the horizon of a black hole. 
In the discussion of the previous sections, we have seen that
the topology of this boundary is however very special:
as a matter of fact, it is the expanded representation of just a point.
From the point of view of classical geometry, being the universe a sphere
and being the interior of a black hole are conditions that cannot be both
consistent at the same time. The point is that the true geometry
of the universe is a quantum geometry. From the discussion of 
section~\ref{mbkhl}.
One can see that, close to the horizon of a black hole,
the metric is heavily affected by the contribution of configurations 
of lower entropy. These are the less classical ones,
the ones with a higher quantum delocalization. Going closer and closer to the horizon, 
the quantum non-localization increases, and at the limit of the "point"
of big bang, the non-localization is the maximal one: this point covers the entire
horizon of the universe (see discussion in Ref.~\cite{blackholes-g}).

\subsection{Non-locality and quantum paradoxes}
\label{nonloc}

The uncertainty encoded in 
the Uncertainty Principle, lifting up the 
predictive power of classical mechanics to a 
probabilistic one, leads also to
non-locality, possibly violating the bound on the speed of transfer
of information set by the speed of light $c$.

It has been a long debated question, whether this 
had to be considered
as something really built-in in natural phenomena, or simply an 
effect due to our ignorance of all the degrees of freedom involved, something
that could be explained through the introduction of ``hidden variables''
\cite{EPR}.
It seems that indeed the physical world lies on the side of true quantum 
interpretation, which, as shown by Bell \cite{bell}, is
not reproducible with hidden variables. 
At the quantum level, the bound on the speed of information, $c$,
can be violated by non-locality of wave-functions.  
How can we understand all that in the light of our framework? 
In our framework, the uncertainty of the Uncertainty Principle, 
at the base of quantum mechanics, is due to the fact that what we observe
and measure is the sum of an infinite number of configurations, among which
also tachyonic ones contribute. This is 
in agreement with the fact that non-locality
implies somehow a propagation at speed higher than $c$, thereby
violating causality. And indeed,
the ``dynamics'' described in the previous section, resulting
from the fact that at any time the universe is
the sum of all configurations at the actual time, 
implies in some sense an instantaneous transfer of 
information: although the classical geometric deformations propagate
at maximal speed $c$,
the local configuration around the observer is uniquely related
to the whole configuration of the universe.
At any time, an electron, or any other particle, ``knows'' how large
is the universe up to the horizon. How would such a boundary 
information determine the properties of each particle, 
even those locally produced in laboratory, if any information could only be
transferred at maximal speed $c$? This non-local, basically instantaneous
knowledge implied in the combinatoric scenario is reproduced in string theory,
a relativistic theory, essentially in two ways: 

1) through an explicit 
quantization, imposed in order to reproduce the ordinary 
quantum effects, which
include non-local, tachyon-like effects such as experiment correlations
violating Bell's inequalities and so on, and 

2) through the very basic properties of the string construction, in which,
as we discuss in~\cite{npstrings-2011} (see also Ref.~\cite{spi}),
massive particles live partly in the extended space, identified with the
ordinary space-time, the space along which they propagate,
and partly have a foot in the internal string space.

\begin{figure}
\centerline{
\epsfxsize=8cm
\epsfbox{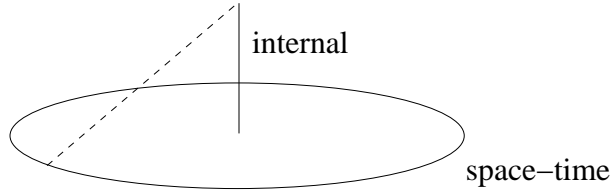}
}
\vspace{0.3cm}
\caption{A pictorial illustration of how can massive particle/fields have
knowledge of the full space-time. Here the extended but compact space-time
is represented as a disc on the horizontal plane. The internal dimension 
is represented by the vertical segment.
A field/particle confined to live tangent to the space-time
cannot ``see'' the horizon, but just the neighborhoods of the point where it is
sitting. A particle/field
with a foot on the internal space can lift-up its point of view
and see the horizon, having from above a global view of the disc on the 
plane.}     
\label{int-ext}
\end{figure}

Roughly speaking,
being extended also in the internal space provides massive states an
extra dimension from which they can ``look'' at the entire space-time, anyway
a compact space, of which they can ``see'' the boundary, because 
the internal coordinate is non-local with respect to the external ones
(see figure~\ref{int-ext}).
This is not true for the massless fields (photon and graviton), that live
entirely in the extended space-time. They therefore feel only the local
physics.

\subsection{The role of T-duality and 
the gravitational coupling as the self-dual scale}

\
\\
T-duality acts in string theory in such a way
to ensure the existence
of extra matter and fields,
in the right amount to imply the presence of both a strongly
and a weakly coupled sector. 
It is the presence of both these sectors what stabilizes the scale:
if there was just one of the two, we would be no scale at all,
because both the gravitational and the interaction 
scale would become "running" scales". 
In our framework, gravitational
interactions are the characterization at the
level of vectorial representation of space of the evolution through ``geometries''
of energies. 
When introducing the investigation of the
deformations of the dominant geometry in terms of propagating fields and
particles, we said that the finest description of the three-dimensional
sphere is not vectorial, but spinorial. One would have the
impression that a spinorial space is a better environment in which
to investigate the geometries of the distribution of energies, already at the 
level of the discussion of section~\ref{eSp}. 
Indeed, deformations of the main
geometry as given by spinor fields weight less, and as such can be
treated as perturbations, ``quantum perturbations''.
This is basically the reason why we perceive a space which is 
classically a vector space. Spinors (which in our framework
are only matter particles) turn out
to come in the game with the role of describing objects, ``energy packets''
which move at lower speed than the expansion of the classical space, 
and the propagation of information, itself. 
Gravity is the "average" interaction, 
the one related to the measure, and scale, of space, 
intended as the average geometry 
resulting from all the contributions to energy given also by the other
interactions. Its coupling is therefore "in between" strong and weak coupling, 
and is set to 1 by convention (one may take this as a renormalization 
prescription). If all the interactions were "weak", the space would be only 
spinorial (we would perceive the space as spinorial, therefore we would not 
see propagation of information through vectorial bosons). It is the existence 
of both weakly and strongly coupled sectors (consequence of, even softly 
broken, T-duality) what forces spinors to behave also as vectors, and 
results in an impression of vectorial geometry of space.

\subsection{Natural or real numbers?}
\label{NorR}

In this work, we have introduced an interpretation of the
real world as the collection of all possible ``binary codes''. 
The universe mostly appears as the average of these configurations, 
that we interpret as assignments of energy amounts along a space.
Reducing everything to a collection of binary codes means reducing
everything to a discrete description in terms of natural numbers.
One may ask the question: are natural numbers enough to encode \emph{all} 
the information of the universe, or is there a finer description, for instance
in terms of real numbers? Are we sure that with natural numbers we
can catch, and express, all the information?
At first sight, one would say that real numbers say ``more'', allow to express
more information. Moreover, they appear to be ``real'' in the true sense
of something existing in nature. For instance, one can think to draw with
the pencil a circle and a diameter. Then, one has \emph{physically} realized
two lines of lengths that don't stay in a ratio expressible as a rational 
number. However here the point is: what is really about the microscopical 
nature of these two drawings? At the microscopical level, at the
scale of the Planck length, the notion of space itself is so fuzzy 
to be practically lost. In our scenario, 
an analysis of the superposition
of configurations tells us that, before reaching this scale, remote
configurations, whose contribution is usually collected 
under the Heisenberg's Uncertainty, count more and more. In other terms,
the world is no more classical but deeply quantum mechanical, to the point
that the uncertainty in the length of the two lines doesn't allow us
to know whether their ratio is a real or a rational number. Moreover,
real numbers are introduced in mathematics through procedures, such as
limits, Dedekind sections etc., whose informational content
can be ``written'' as a text with a computer program, such as the one 
with which I am writing here these words. This means that, 
\emph{as a matter of pure information}, real numbers can be introduced via
natural numbers. Perhaps, all the information of the universe is expressible
through natural numbers, and, as a consequence, the discrete description
of the universe, and in particular of space-time, is not just
a simplification, an approximation, but indeed the most
fundamental one can think about.

\newpage

\providecommand{\href}[2]{#2}\begingroup\raggedright\endgroup

\end{document}